\documentclass[times,11pt]{elsarticle}
\usepackage{graphicx}
\usepackage{subfig}
\usepackage{fullpage}
\usepackage{multirow}
\usepackage{url}
\usepackage{amsmath,amssymb,amsthm,subfig,bm,units,tikz,booktabs}
\usetikzlibrary{arrows}
\usepackage[mathscr]{euscript}
\usepackage{mdframed,booktabs,rotating}
\usepackage[title,toc,titletoc,page]{appendix}
\usepackage[ruled,algo2e]{algorithm2e}
\newtheorem{conclusion}{Conclusion}

\newcommand{\hms}{\textcolor{black}} 
\newcommand{\rewb}{\textcolor{black}} 

\DeclareMathOperator{\sgn}{sgn}

\theoremstyle{definition}

\newtheorem{remark}{Remark}

\usepackage{algorithm, algorithmic}

\newcommand{\kaparef}{$\kappa_{\text{ref }}$}
\title{{\bf A low-dissipation reconstruction scheme for compressible single- and multi-phase flows based on artificial neural networks}}
\DeclareUnicodeCharacter{2212}{-}

\author[sjtu]{Minsheng Huang}
\ead{mingo.stemon@sjtu.edu.cn}
\author[aero]{Lidong Cheng \corref{cor}}
\ead{critters@sjtu.edu.cn}
\author[sjtu,ins]{Wenjun Ying \corref{cor}}
\ead{wying@sjtu.edu.cn}
\author[icl]{Xi Deng}
\author[ist]{Feng Xiao}
\cortext[cor]{Corresponding author}
\address[sjtu]{School of Mathematical Sciences, Shanghai Jiao Tong University, Shanghai, P.R.China.}
\address[aero]{School of Aeronautics and Astronautics, Shanghai Jiao Tong University, Shanghai, P.R.China.}
\address[ins]{MOE-LSC and Institute of Natural Sciences,
Shanghai Jiao Tong University, Shanghai, P.R.China.}
\address[icl]{Department of Chemical Engineering, Imperial College London, SW7 2AZ, United Kingdom.}
\address[ist]{Department of Mechanical Engineering, Institute of Science Tokyo, 2-12-1 Ookayama, Meguro-ku, Tokyo, Japan.}

\begin{document}

\begin{abstract}
Solving compressible flows containing both smooth and discontinuous flow structures remains a significant challenge for finite volume methods. Godunov-type finite volume methods are commonly used for numerical simulations of compressible flows. One of the key factors in obtaining high-quality solutions is high-fidelity spatial reconstruction. 
In this work, we introduce a new paradigm for constructing high-resolution hybrid reconstruction schemes for compressible flows. This approach generates training data based on BVD (Boundary Variation Diminishing) schemes for supervised learning and employs ANN (Artificial Neural Networks) to create an indicator that pre-selects the most suitable reconstruction scheme for each cell, achieving the lowest global numerical dissipation. 
The numerical schemes under this paradigm are more computationally efficient than similar schemes within the BVD framework, as each cell only requires constructing a single interpolation function. Following this paradigm, a novel low-dissipation reconstruction scheme based on the MUSCL-THINC-BVD scheme, named the deepMTBVD scheme, is proposed for compressible single- and multi-phase flows. 
The polynomial function used is the MUSCL (Monotone Upstream-centered Schemes for Conservation Laws) scheme with the van Leer limiter, while the THINC (Tangent of Hyperbola for INterface Capturing) function, a sigmoid function, is employed to mimic the discontinuous solution structure. A multilayer perceptron culminating in a simple two-hidden-layer neural network is designed and trained offline. This network identifies the most suitable candidate interpolation to be employed within the target cell.
The performance of the proposed scheme has been extensively verified through benchmark tests of single- and multi-phase compressible flows, where discontinuous and vortical flow structures, like shock waves, contact discontinuities, and material interfaces, as well as vortices and shear instabilities of different scales, coexist simultaneously. \hms{Numerical results indicate that the new deepMTBVD scheme performs as well as the original MUSCL-THINC-BVD scheme for numerical simulations of compressible flows while reducing computational time by up to 40\%. 
Additionally, the deepMTBVD scheme attains higher accuracy and more flexibility than the MUSCL-THINC-BVD scheme. Thus, the proposed scheme shows highly comparable results with the MUSCL-THINC-BVD scheme and has the potential to be extended into a series of low-dissipation numerical schemes.}
\end{abstract}


\maketitle
\section{Introduction}

Compressible fluid dynamics is one of the active and challenging research areas, attracting considerable interest in theoretical research and practical industrial applications. Discontinuities like shock waves and contact discontinuities are common features in compressible flows. They can lead to intricate flow structures, especially in situations involving multi-phase fluids with material interfaces.
Such complex flows are far beyond theoretical studies and difficult to experiment with in many situations. In some cases, numerical simulation has proven to be the most effective or even the only feasible approach to provide the basic mechanisms behind the complex phenomena of compressible flow. 

The finite volume method (FVM) is a common framework for numerically solving compressible flows. While the first-order FVM is simple to implement, it suffers from excessive dissipation, making it unsuitable for practical engineering and scientific applications. To tackle this issue, the Total Variation Diminishing (TVD) concept led to the development of TVD schemes \cite{HARTEN1997260}, such as the MUSCL (Monotone Upstream-centered Schemes for Conservation Law) scheme \cite{vanLeer-344}. TVD schemes achieve second-order accuracy in smooth regions and effectively handle discontinuities without oscillation. However, they still experience excessive numerical dissipation, particularly around discontinuities and extrema.
To further improve order of accuracy, uniformly high-order reconstruction schemes such as Essentially Non-Oscillatory (ENO) \cite{HartenEngquist-204,ShuOsher-199,ShuOsher-847}, Weighted ENO (WENO) \cite{LiuOsher-80,JiangShu-79}, and Targeted ENO (TENO) \cite{FuHu-399,FuHu-400} have been proposed. These schemes use polynomial interpolation for smooth regions and specially designed limiters for discontinuities, theoretically achieving arbitrarily high-order accuracy.
In contrast to conventional FVM, methods like the Multi-moment Constrained finite Volume method (MCV) \cite{RN70, RN46}, high order Discontinuous Galerkin (DG) \cite{Cockburn1990545},  Flux Reconstruction(FR)/Correction Procedure via Reconstruction (CPR) \cite{RN82, RN83, huynh2007} and their variants \cite{RN86, RN87, RN38, RN89}, use locally defined degrees of freedom (DOFs) to construct high-order reconstructions. These methods feature more compact cell stencils but require particular strategies, such as artificial viscosity or local limiting processes, to handle discontinuities without oscillations \cite{DUMBSER201447}. 
Despite the aforementioned efforts, limiting processes or artificial viscosity methods for high-order polynomial-based schemes often introduce excessive numerical dissipation, which can smear and blur flow structures, especially at discontinuities like shock waves and interfaces \cite{RN108, RN110}.

Due to the presence of the Gibbs phenomenon \cite{Hewitt1979TheGP}, high-order polynomials are not optimal for reconstructing discontinuous flow fields. In contrast, sigmoid functions perform better in such scenarios. For example, the THINC (Tangent Hyperbolic Interface Capturing) scheme utilizes the hyperbolic tangent function for reconstruction \cite{XiaoHonma-13}. At the early stage, this scheme was applied to solve the VOF (Volume of Fluid) equation for multiphase flow interfaces \cite{ShyueXiao-35}. With the introduction of the BVD (Boundary Variation Diminishing) principle and BVD schemes, it has been widely used to solve compressible single-phase and multiphase flows \cite{RN8, RN2}.
The fundamental idea of the BVD scheme is to apply polynomial-based schemes in smooth regions and use the THINC scheme to capture discontinuous solutions. Developed BVD schemes include MUSCL-THINC-BVD \cite{RN12,ChengDeng-692}, THINC-THINC-BVD \cite{DengXie-99}, (W/T)ENO-THINC-BVD \cite{RN8,HouZhao-1564,huang2024tenothincbvd}, and Pn-Tm-BVD \cite{DengShimizu-142,DengJiang-848}. These schemes can capture delicate flow structures with high fidelity, including shock waves, interfaces, contact discontinuities, and multi-scale vortices. However, BVD schemes necessitate preparing all candidate reconstruction functions before selecting the final one within a given cell, suggesting that there is potential for improving computational efficiency.

Recently, the deep learning method has become a novel tool in solving Partial Differential Equations (PDEs), with approaches generally falling into three categories. The first involves using deep neural networks to simulate PDEs, directly adhering to boundary and initial conditions. Physics-Informed Neural Networks (PINNs), as discussed in references \cite{RN98, RN99, RN100}, are designed to solve PDEs by integrating physical principles and initial boundary conditions. Operator neural networks, like DeepOnet \cite{RN95, RN101}, FNO \cite{li2021fourier}, treat differentials or integrals as operators, learning to manipulate these through neural networks. The second approach applies deep neural networks within a variational or weak formulation context. The Deep Ritz Method \cite{RN106} addresses PDEs convertible into energy minimization problems. The Weak Adversarial Network \cite{RN107} is tailored for high-dimensional PDEs in their weak form. The final approach involves training surrogate models offline to augment traditional numerical methods. This includes integrating discontinuity or trouble-cell detectors \cite{RN102, RN105, ray2018troublecell}, enhancing shock-capturing techniques on Cartesian or unstructured grids \cite{RN14, RN15}, and refining conventional limiters \cite{zhu2021siam, RN104}. 
Salazar et al. \cite{salazar2020hybrid} developed a hybrid scheme by machine learning method of WENO and THINC scheme. The proper weights of stencils in the WENO-type schemes are obtained through the deep learning method \cite{bezgin2022weno3nn, nogueira2024weno5nn, liu2020weno}. Tompson et al. \cite{tompson2022accelerating} proposed a data-driven method to expedite solving incompressible Euler equations, and Feng et al. \cite{Feng2023deepreinforcement} applied a deep reinforcement learning framework for compressible flow simulations.

Inspired by the aforementioned works, we introduce a new paradigm for constructing high-resolution hybrid schemes for compressible flows. This approach generates training data based on BVD schemes for supervised learning and employs artificial neural networks to create an indicator that pre-selects the most suitable reconstruction scheme for each cell. Based on the MUSCL-THINC-BVD scheme, a novel low-dissipation reconstruction scheme is proposed for compressible single- and multi-phase flows. The fundamental idea aligns with the BVD scheme, which applies polynomial-based schemes for reconstructing smooth regions and the THINC scheme for discontinuous regions to minimize global numerical dissipation. The key difference lies in the method of determining the final reconstruction scheme for each cell. The BVD scheme constructs all candidate interpolation functions before selecting the optimal one based on the BVD principle. In contrast, the new method uses a pre-trained artificial neural network to predict the appropriate reconstruction scheme for each cell and directly constructs the corresponding interpolation function. Compared to the original BVD scheme, the new approach offers the following advantages: 1)  only one interpolation function needs to be created per cell, leading to a substantial reduction in computational cost during the reconstruction loop; 2) the flexibility of artificial neural networks presents significant potential for algorithm optimization. Numerical test results indicate that the new low-dissipation scheme surpasses the original BVD scheme in both computational efficiency and the accuracy of numerical solutions. 

The remainder of this paper is organized as follows. Mathematical models of single and multi-phase compressible flows are described in section \ref{mathematical_model}. In section \ref{numerical_method}, after a brief review of the MUSCL-THINC-BVD scheme, the details of constructing the deepMTBVD scheme are presented. Numerical results and discussion are presented compared to other methods in section \ref{numerical_result}, followed by some concluding remarks in section \ref{conclusion_remarks}.

\section{Mathematical model} \label{mathematical_model}
\subsection{Governing equations}
In this work, we consider the inviscid single and two-phase compressible flows. 
The Euler equation governs single-phase compressible flow, while two-phase compressible flow is described by the five-equation model developed in \cite{RN49}. 
For simplicity, a general form of conservation laws can be written as:
\begin{equation}
    \frac{\partial }{\partial t} \bm{U} + \nabla \cdot \bm{F}(\bm{U}) = \bm{S}.
    \label{model:general}
\end{equation}
which can be instantiated as follows:

\begin{itemize}
    \item Euler equations\\
    Inviscid single-phase compressible flow can be described by Euler equations, which is reformulated by $\eqref{model:general}$:
\begin{equation}
    \bm{U} = \begin{pmatrix}
    \rho \\
    \rho \bm{V} \\
    E
    \end{pmatrix}, \quad \bm{F}(\bm{U}) = \begin{pmatrix}
    \rho \bm{V} \\
    \rho \bm{V} \otimes \bm{V} + p \bm{I}\\
    (E+p)\bm{V}
    \end{pmatrix}, \quad \bm{S} = \begin{pmatrix}
    0\\
    \bm{0}\\
    0
    \end{pmatrix},
    \label{model:euler}
\end{equation}
where $\rho, \bm{V}, p$ denote density, velocity, pressure respectively. Here $E$ is the total energy per unit volume
\begin{equation}
    E = \rho(\frac{1}{2}|\bm{V}|^{2} + e),
\end{equation}
where $e$ is the specific internal energy determined by the equation of state (EOS) discussed in the next subsection.
    \item The five-equation model for two-phase inviscid compressible flows\\
    Assuming that the pressure and velocity fields at the material interface reach equilibrium at the moment of interaction, the five-equation model comprises two-phase mass equations, a mixture momentum equation, a mixture total energy equation and a volume fraction transport equation, which is given as follows
\begin{equation}
    \bm{U} = \begin{pmatrix}
    \rho_{1}\alpha_{1}\\
    \rho_{2}\alpha_{2}\\
    \rho \bm{V}\\
    E\\
    \alpha_{1}\\
    \end{pmatrix}, \quad \bm{F(\bm{U})} = \begin{pmatrix}
    \rho_{1}\alpha_{1}\bm{V}\\
    \rho_{2}\alpha_{2}\bm{V}\\
    \rho \bm{V} \otimes \bm{V} + p\bm{I}\\
    (E + p)\bm{V}\\
    \alpha_{1}\bm{V} 
    \end{pmatrix},  \quad \bm{S} = \begin{pmatrix}
    0\\
    0\\
    \bm{0}\\
    0\\
    \alpha_{1} \nabla \cdot \bm{V}
    \end{pmatrix},
    \label{model:five_equation}
\end{equation}
where $\rho_{k}, \alpha_{k}, k \in \left\{1, 2\right\}$ are the density and volume of fraction of $k^{th}$ fluid respectively. 
The volume fraction $\alpha_{k}$ is always non-negative and ranges between 0 and 1.
Different from \eqref{model:euler}, here $\bm{V}$ is the velocity field, $p$ is mixture pressure and $E$ is total mixture energy per unit volume. One can extend the two-phase model to the arbitrary number of phases by adding a continuity equation and a new advection equation of volume fraction of additional phase \cite{RN2}.
\end{itemize}

\subsection{The closure strategy}
To close the above systems, the fluid of a single phase is assumed to satisfy the stiffened gas EOS as follows
\begin{equation}
    p_{k}(\rho_{k}, e_{k}) = (\gamma_{k} - 1)\rho(e_{k} - e_{\infty,k}) - \gamma_{k} p_{\infty,k},
    \label{model:stiffened}
\end{equation}
where $k$ denotes the $k^{th}$ phase, $k=1$ for single phase, and $k \in \{1,2\}$ for two-phase flow. 
$\gamma_{k}$ is the ratio of the specific heats, and $p_{\infty,k}$ and $e_{\infty,k}$ are the prescribed parameters chosen to describe the condensed material property of interest and could be determined by empirical fitting with the experimental data \cite{RN50}. The sound speed of the single phase by \eqref{model:stiffened} is defined as 
\begin{equation}
    c_{k}^{2} = \frac{\gamma_{k}(p_{k}+p_{\infty,k})}{\rho_{k}}.
\end{equation}
The ideal-gas law is recovered by $p_{\infty,k} = 0$. The stiffened EOS is widely used in numerical simulation based on diffuse interface methods because of its ability to reproduce the shock relationships and saturation curves of liquid-vapor mixtures when simulating phase transitions. 

For two-phase flow, the volume fraction of each phase satisfies the saturation condition:
\begin{equation}
    \alpha_{1} + \alpha_{2} = 1.
\end{equation}
Conservative constraints define mixture density and mixture internal energy:
\begin{equation}
    \begin{split}
        & \rho = \alpha_{1}\rho_{1} + \alpha_{2}\rho_{2}, \\
        & \rho e = \alpha_{1}\rho_{1}e_{1} + \alpha_{2}\rho_{2}e_{2}.
    \end{split}
\end{equation}
Derived from \cite{RN52, RN53}, the mixed ratio of the specific heats $\gamma$, pressure parameter $p_{\infty}$ and internal energy parameter $e_{\infty}$ are given by:
\begin{equation}
    \begin{split}
        & \frac{1}{\gamma - 1} = \frac{\alpha_{1}}{\gamma_{1}-1} + \frac{\alpha_{2}}{\gamma_{2}-1},\\
        & \frac{p_{\infty}}{\gamma-1} = \frac{\alpha_{1}p_{\infty,1}}{\gamma_{1}-1} + \frac{\alpha_{2}p_{\infty,2}}{\gamma_{2}-1},\\  
        & \rho e_{\infty} = \alpha_{1}\rho_{1}e_{\infty,1} + \alpha_{2}\rho_{2}e_{\infty,2},
    \end{split}
    \label{model:thermodynamic}
\end{equation}
the resulting mixture EOS reads,
\begin{equation}
    p(\rho, e, \alpha_{1}, \alpha_{2}) = \left(\rho e - \sum_{k=1}^{2}\alpha_{k}\rho_{k}e_{\infty,k} + \sum_{k=1}^{2}\frac{\alpha_{k}p_{\infty,k}}{\gamma_{k}-1}\right) \bigg{/} \sum_{k=1}^{2}\frac{\alpha_{k}}{\gamma_{k}-1}.
    \label{model:mixeos}
\end{equation}
Mixed thermodynamic variables, defined as \eqref{model:thermodynamic} and \eqref{model:mixeos}, are free of spurious oscillations across the material interface \cite{RN56}. The sound speed of the mixture is calculated by Wood's formula \cite{RN54}:
\begin{equation}
    \rho c^{2} = \frac{\alpha_{1}}{\rho_{1} c_{1}^{2}} + \frac{\alpha_{2}}{\rho_{2} c_{2}^{2}}.
\end{equation}

\section{Numerical methods}\label{numerical_method}
To simplify the explanation, we introduce the numerical method in one dimension. Our work focuses on Cartesian grids, which can be extended to multiple dimensions using dimension-splitting techniques. We will first review the basic framework of the Godunov-type finite volume method, followed by the MUSCL-THINC-BVD scheme. Finally, we will provide details on the new reconstruction scheme, the deepMTBVD, which is based on an artificial neural network.

\subsection{The Godunov-type finite volume method}
In the Cartesian grid, we consider the discrete method of Eqn.~\eqref{model:general} in x-dimension only, y-dimension and z-dimension are similar. The computational domain is divided into $N$ cells, and denotes $i^{th} $cell $I_{i} = \left[x_{i-1/2}, x_{i+1/2}\right], i \in \{1, \cdots, N\}$. 
Assume that grid size is $\Delta x_{i} = x_{i+1/2}-x_{i-1/2}$ and cell center locates at $x_{i+1/2}$, the general integral formulation of the conservative variable $\bm{U}$ over a finite volume cell $I_{i}$, which follows the cell-average values, is given by:
\begin{equation}
    \Bar{\bm{U}}_{i} = \frac{1}{\Delta x_{i}} \int_{x_{i-1/2}}^{x_{i+1/2}}\bm{U} dx.
\end{equation}
To solve Euler equation \eqref{model:euler} and five-equation model
\eqref{model:five_equation}, we apply wave propagation 
method \cite{RN11, RN2} for the spatial discretization of cell $I_{i}$.

\begin{equation}
    \frac{d}{dt} \Bar{\bm{U}}_{i} = -\frac{1}{\Delta x_{i}}(\mathcal{A}^{+}\Delta\bm{U}_{i-1/2} + \mathcal{A}^{-}\Delta\bm{U}_{i+1/2} + \mathcal{A}\Delta\bm{U}_{i}),
    \label{numerical:fvm}
\end{equation}
where $\mathcal{A}^{+}\bm{U}_{i-1/2}$ and $\mathcal{A}^{-}\bm{U}_{i+1/2}$ are right-moving and left-moving fluctuations and $\mathcal{A}\bm{U}_{i}$ is the total fluctuation of cell $I_{i}$. In detail, right-moving and left-moving fluctuations can be determined by corresponding Riemann problems:
\begin{equation}
    \mathcal{A}^{\pm}\Delta\bm{U}_{i\mp1/2} = \sum_{k=1}^{K}\left[s^{k}(\bm{U}_{i-1/2}^{L}, \bm{U}_{i+1/2}^{R})\right]^{\pm} \bm{W}^{k}(\bm{U}_{i-1/2}^{L}, \bm{U}_{i+1/2}^{R}),
    \label{numerical:wave_propagation}
\end{equation}
where $K$, $s^{k}$ and $\bm{W}^{k}$ are the number of waves, moving speeds, and jumps of three propagation discontinuities solved by Riemann solvers such as HLL/HLLC Riemann solver \cite{torobook}, respectively. $\bm{U}_{i-1/2}^{L}$ and $\bm{U}_{i-1/2}^{R}$ are reconstruction values from neighboring cells by specific schemes at the interface $x = x_{i-1/2}$. The reconstruction method is the core of this paper, which will be discussed in the next section. Similarly, the total fluctuation can be calculated by 
\begin{equation}
    \mathcal{A}\Delta\bm{U}_{i} = \sum_{k=1}^{K}\left[s^{k}(\bm{U}_{i-1/2}^{R}, \bm{U}_{i+1/2}^{L})\right]^{\pm} \bm{W}^{k}(\bm{U}_{i-1/2}^{R}, \bm{U}_{i+1/2}^{L}),
    \label{numerical:total_fluctuation}
\end{equation}
Once given the spatial discretization, we employ the two-stage second-order SSP (Strong Stability-Preserving) Runge–Kutta scheme \cite{RN59} for numerical tests:
\begin{equation}
    \begin{aligned}
    \Bar{\bm{U}}^{*} & = \Bar{\bm{U}}^{n} + \Delta t \mathcal{L}(\Bar{\bm{U}}^{n}), \\
    \Bar{\bm{U}}^{n+1} & = \frac{1}{2}\Bar{\bm{U}}^{n} + \frac{1}{2}(\Bar{\bm{U}}^{*}+\Delta t \mathcal{L}(\Bar{\bm{U}}^{*})).
    \end{aligned}
    \label{numerical:timeintegral}
\end{equation}
Here $\Bar{\bm{U}}^{*}$ is the intermediate value at the time sub-step. We also apply three-stage third-order SSP Runge-Kutta for generating the dataset, which will be discussed in section \ref{numerical:method:deepmtbvd:section}:
\begin{equation}
    \begin{aligned}
    \Bar{\bm{U}}^{*} & = \Bar{\bm{U}}^{n} + \Delta t \mathcal{L}(\Bar{\bm{U}}^{n}), \\
    \Bar{\bm{U}}^{**} & = \frac{3}{4}\Bar{\bm{U}}^{n} + \frac{1}{4}(\Bar{\bm{U}}^{*}+\Delta t \mathcal{L}(\Bar{\bm{U}}^{*})), \\
    \Bar{\bm{U}}^{n+1} & = \frac{1}{3}\Bar{\bm{U}}^{n} + \frac{2}{3}(\Bar{\bm{U}}^{**}+\Delta t \mathcal{L}(\Bar{\bm{U}}^{**})).
    \end{aligned}
    \label{numerical:timeintegral2}
\end{equation}


\subsection{The MUSCL-THINC-BVD reconstruction scheme}
In this part, we briefly review the MUSCL-THINC-BVD scheme in the one-dimensional Cartesian grid. Denoting the left- and right-side values of an arbitrary cell boundary as $\bm{U}^{L}$ and $\bm{U}^{R}$, the approximate Riemann solver in the canonical formulation is given as follows \cite{RN8}:
\begin{equation}
    \bm{F}(\bm{U}^{L}, \bm{U}^{R}) = \frac{1}{2}\left(\bm{F}(\bm{U}^{L})+\bm{F}(\bm{U}^{R})\right)-\bm{A}(\bm{U}^{L}, \bm{U}^{R})(\bm{U}^{R}-\bm{U}^{L}).
    \label{numerical:canonical}
\end{equation}
The right-hand side of Eqn.~\eqref{numerical:canonical} is divided into a central part and a dissipative part. Here, $\bm{A}(\bm{U}^{L}, \bm{U}^{R})$ is a system matrix computed from $\bm{U}^{L}$ and $\bm{U}^{R}$, and $\left|\bm{U}^{R}-\bm{U}^{L}\right|$ is referred to as the boundary variation. This dissipative part can introduce excessive diffusion in the numerical solution. The fundamental idea of the BVD principle is that minimizing boundary variation can reduce numerical dissipation in flux calculations.

\begin{figure}[htbp]
    \centering
    \includegraphics[width=0.8\textwidth]{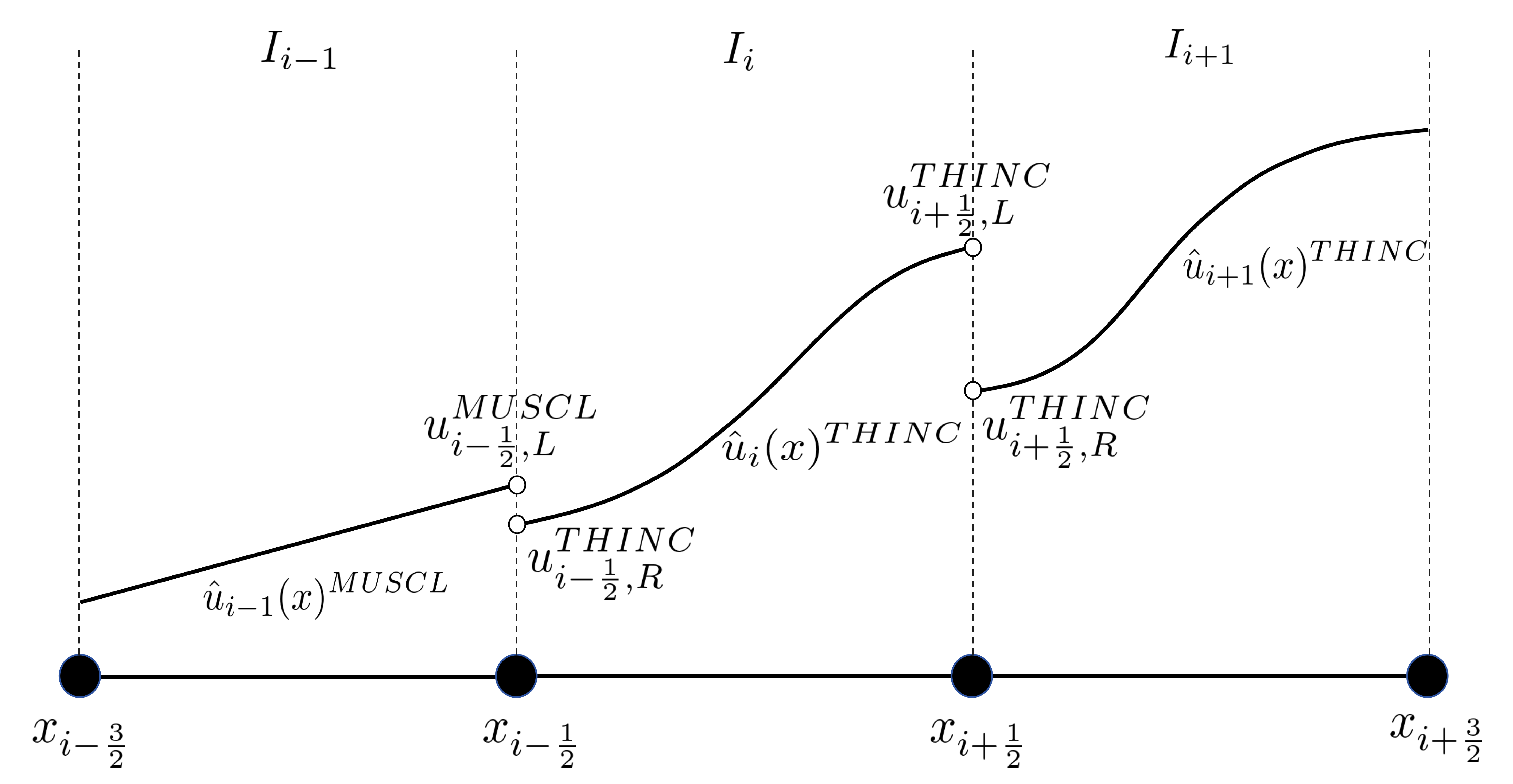}
    \caption{Schematic diagram of BVD scheme}
    \label{numerical:muscl_thinc_bvd}
\end{figure}

For the sake of simplicity, we denote a single variable as $u$ for reconstruction, which can be a primitive, conservative, or characteristic variable.
In the MUSCL scheme, a piece-wise linear function is constructed with cell averaged value $\Bar{u}_{i}$:
\begin{equation}
    \hat{u}(x)_{i}^{MUSCL} = \Bar{u}_{i} + \sigma_{i}\frac{x - x_{i}}{x_{i+1/2}-x_{i-1/2}}, \quad x\in [x_{i-1/2}, x_{i+1/2}].
    \label{numerical:muscl}
\end{equation}
The $\sigma_{i}$ is the slope defined in cell $I_{i}$, which is followed
\begin{equation}
    \sigma_{i} = \Tilde{\Delta}\left(\Bar{u}_{i} -  \Bar{u}_{i-1}, \Bar{u}_{i+1} -  \Bar{u}_{i}\right).
\end{equation}
Here $\Tilde{\Delta}$ involves a kind of slope limiter to prevent non-physical oscillation. This paper uses the Vanleer slope limiter \cite{RN43} for all of the numerical tests. \rewb{The corresponding boundary values reconstructed by the MUSCL scheme on $I_{i}$ are computed as
$$
\hat{u}^{MUSCL}_{i-1/2,R} = \Bar{u}_{i}- \frac{x_{i-1/2}-x_{i}}{x_{i+1/2}-x_{i-1/2}}\sigma_{i}, \quad \hat{u}^{MUSCL}_{i+1/2,L} = \Bar{u}_{i}- \frac{x_{i+1/2}-x_{i}}{x_{i+1/2}-x_{i-1/2}}\sigma_{i}.
$$
}
As illustrated in Fig.~\ref{numerical:muscl_thinc_bvd}, $u_{i-1/2, L}^{MUSCL}$ represents the left-side reconstructed value at the right boundary of cell $I_{i-1}$. Although the MUSCL scheme is renowned for numerical simulations, it generates excessive numerical dissipation and smears out the discontinuities. This issue is even more pronounced in the five-equation model, where the MUSCL scheme captures the material interface with greater dissipative error as discussed in section $\ref{numerical_result}$.

Another candidate scheme, THINC scheme \cite{XiaoHonma-13} is a monotone and \rewb{differentiable} hyperbolic tangent function, which suits well monotone step-like discontinuity. The THINC reconstruction is given as 
\begin{equation}
    \hat{u}(x)_{i}^{THINC}=\Bar{u}_{\min} + \frac{\Bar{u}_{\max} - \Bar{u}_{\min}}{2}\left(1 + \theta  \tanh\left(\beta\frac{x-x_{i-1/2}}{x_{i+1/2}-x_{i-1/2}} - \Tilde{x}_{i}\right) \right),
    \label{numerical:thinc}
\end{equation}
where $x \in [x_{i-1/2}, x_{i+1/2}]$. $\Bar{u}_{\min} = \min\{\Bar{u}_{i-1}, \Bar{u}_{i+1}\}, \Bar{u}_{\max} = \max\{\Bar{u}_{i-1}, \Bar{u}_{i+1}\}$ and $\theta = \sgn(\Bar{u}_{i+1} - \Bar{u}_{i-1})$. $\beta$ is a parameter to control the steepness of the jump. 
A small $\beta$ results in a diffusive solution and performs as similarly as TVD schemes while a larger one enforces the anti-diffusion effect and tends to capture the discontinuity sharply \cite{XiaoHonma-13, XiaoIi-46}.
To resolve sharp discontinuities like shocks and material interfaces, a $\beta$ valued from 1.6 to 2.2 produces good or acceptable results \cite{DengXie-99}.
In our numerical simulation, we set $\beta=1.6$ for all numerical tests presented in this paper. Following the work \cite{XiaoIi-46}, the parameter $\beta$ depends on the unit normal of discontinuity for multidimensional cases. For example, $\beta_{x}=\beta|n_{x}|$ where $n_{x}$ is the x-component of the unit normal of discontinuity. Similarly, $\beta_{y}$ and $\beta_{z}$ can be determined similarly. 

The unknown variable $\Tilde{x}_{i}$, representing the location of the jump center, can be determined from cell-average value $\Bar{u}_{i}$. We present the reconstructed values of the THINC scheme on the cell boundary in $I_i$ as
\begin{align}
    \Hat{u}_{i+1/2,L}^{THINC} & = \Bar{u}_{\min} + \frac{\Bar{u}_{\max} - \Bar{u}_{\min}}{2}\left(1+\theta \frac{\tanh{\beta}+A}{1+A\tanh{\beta}}\right), \label{thinc:left} \\
    \Hat{u}_{i-1/2,R}^{THINC} & = \Bar{u}_{\min} + \frac{\Bar{u}_{\max} - \Bar{u}_{\min}}{2}(1 + \theta A), \label{thinc:right}
\end{align}
where 
$$
    A = \frac{\frac{B}{\cosh{\beta}} - 1}{\tanh{\beta}}, \quad
    B = e^{\theta \beta(2C-1)}, \quad
    C = \frac{\Bar{u}_{i} - \Bar{u}_{\min} + \epsilon}{\Bar{u}_{\max}-\Bar{u}_{\min}+\epsilon}.
$$
Here, C is a projecting mapping function onto $[0,1]$, and $\epsilon = 10^{-20}$ is introduced to prevent arithmetic failure. 

The reconstruction strategy for each finite-volume cell relies on the BVD algorithm \cite{RN8}, which minimizes the boundary variation of reconstructed states at cell boundaries, as stated in Eqn.~\eqref{numerical:canonical}. When a discontinuity is detected in the target cell, the BVD algorithm favors the THINC scheme while another high-order scheme for the smooth region. BVD algorithms for selecting reconstruction schemes have been implemented on both Cartesian and unstructured grids. For further details, interested readers can refer to \cite{RN12,ChengDeng-692, DengShimizu-142}. \rewb{In this study, we adapt the MUSCL-THINC-BVD algorithm proposed by Deng \cite{RN2}, which is defined as follows:}
\begin{equation} 
\hat{u}_i(x)^{BVD}=\left\{
\begin{aligned}
\hat{u}_i(x)^{THINC}, & \text { if } \delta<C<1-\delta, \text { and }\left(\bar{u}_{i+1}-\bar{u}_i\right)\left(\bar{u}_i-\bar{u}_{i-1}\right)>0, \\
\quad & \text { and } TBV_{i,\min}^{THINC}<TBV_{i, \min}^{MUSCL}, \\
\hat{u}_i(x)^{MUSCL}, & \text { otherwise }.
\end{aligned}
\right. 
\label{numerical:bvd:algorithm}
\end{equation}
\rewb{Here $\delta$ is a small positive value, which is set to $10^{-8}$ in the present paper. The total boundary variation $TBV_{i,\min}^{s}$ for reconstruction function $\hat{u}_{i}^{s}(x)$ at the cell $I_{i}$ is given by 
\begin{equation}
TBV_{i,\min}^{s} = \min \left\{BV(M, s, M), BV(T, s, T), BV(M, s, T), BV(T, s, M)\right\}, s\in\left\{M, T\right\}. \label{bvd:tbv}
\end{equation}
Here, $M$ stands for the MUSCL scheme and $T$ for the THINC scheme. $BV(s_{1}, s, s_{2})$ is the function that evaluates the boundary variation of the $I_{i}$, which is defined as 
\begin{equation}
    BV(s_{1}, s, s_{2}) = \left|\hat{u}_{i-1/2,L}^{s_{1}}-\hat{u}_{i-1/2,R}^{s}\right|+\left|\hat{u}_{i+1/2,L}^{s}-\hat{u}_{i+1/2,R}^{s_{2}}\right|, \quad s_{1}, s_{2} \in \left\{M, T\right\}. \label{bvd:mtbvd}
\end{equation}
$s_{1}, s$ and $s_{2}$ denotes the reconstruction scheme selected on cell $I_{i-1}, I_{i}$ and $I_{i+1}$, respectively. From \eqref{bvd:tbv} and \eqref{bvd:mtbvd}, the present BVD algorithm minimizes boundary variations at the two ends of the target cell, which reduces the numerical dissipation.}

\subsection{The artificial neural network-based deepMTBVD scheme}\label{numerical:method:deepmtbvd:section}
The BVD algorithm minimizes the total variation of the boundary, and Eqn. \eqref{numerical:bvd:algorithm} gives the reconstruction scheme for a local minimum. It is noted that the BVD algorithm requires pre-reconstructing all candidate schemes at each sub-timestep. Our goal is to train an artificial neural network to pre-determine the most suitable reconstruction scheme for each cell, which allows us to construct only one interpolation function within each cell, thereby improving computational efficiency.
Artificial neural networks such as Recurrent Neural Network (RNN), Convolutional Neural Network (CNN), and fully-connected neural networks (FNNs) include an input layer, an output layer, and at least one hidden layer. In this paper, we consider an ANN with fully connected layers, which means each node in one layer connects to every node in the next layer with a certain weight and bias. It should be pointed out that we have tried other NNs and found that ANNs with fully connected layers are convenient to train and efficient. 

\subsubsection{Multi-layer perceptron architecture}
\begin{figure}[htbp]
    \centering
    \includegraphics[width=0.6\textwidth]{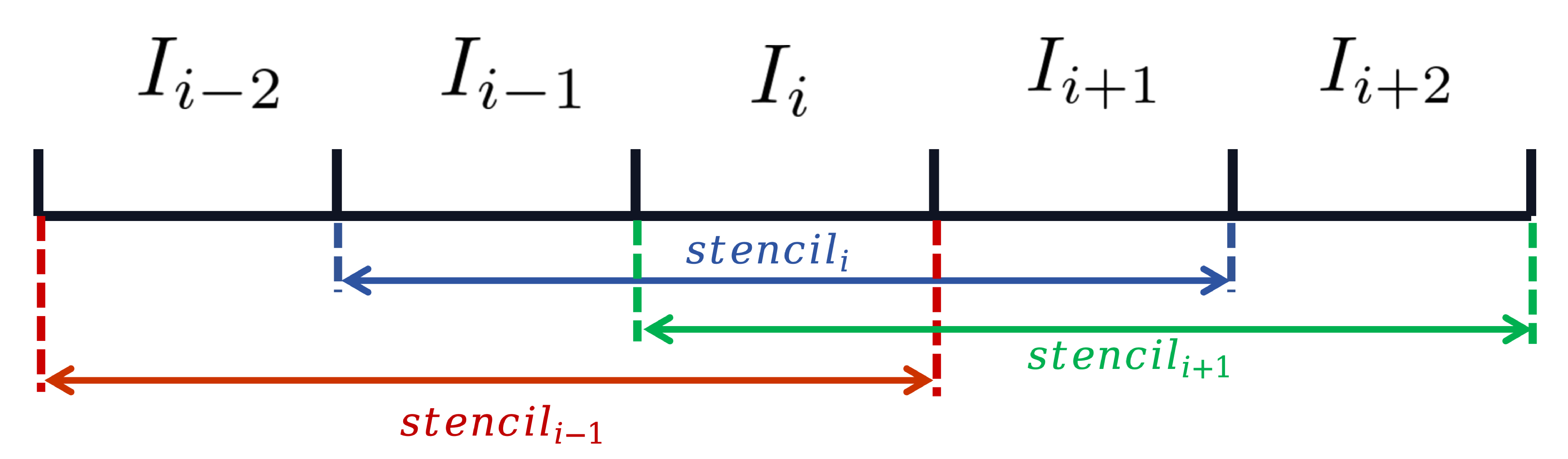}
    \caption{Three sub-stencils of reconstruction for MUSCL-THINC-BVD scheme}
    \label{numerical:stencil}
\end{figure}

As shown in Fig.~\ref{numerical:stencil}, the MUSCL scheme and THINC scheme both consist of three sub-cells $I_{i-1}, I_{i}, I_{i+1}$ for reconstructing. Deciding the reconstruction scheme of cell $I_{i}$, the BVD algorithm need to consider three overlapped stencils $stencil_{i-1}, stencil_{i}$ and $stencil_{i+1}$, which is exactly five-cells stencil from $I_{i-2}$ to $I_{i+2}$.  
Recall that the formula of the MUSCL scheme Eqn. \eqref{numerical:muscl} and the THINC scheme Eqn. \eqref{numerical:thinc}, we find that both schemes depend on the difference between the target cell $I_{i}$ and neighbor cells. Besides, different physical quantities will differ by large orders of magnitude, which makes it hard to train neural networks. 
\rewb{Therefore, we first propose two conclusions and then give the rule of the data normalization.}

\begin{conclusion}\label{conclusion:1}
    \rewb{Assume that $\bar{u}_{i-1}, \bar{u}_{i}, \bar{u}_{i+1}$ are strictly monotone increasing (decreasing), 
    then the reconstruction function $\hat{u}^{s}(x)$ in cell $I_{i}$ satisfies
    \begin{equation}\label{conclusion:normal}
        \frac{\hat{u}^{s}(x) - \bar{u}_{\min}}{\bar{u}_{\max}-\bar{u}_{\min}} \in (0, 1), \quad s \in \{M, T\}, \quad x \in [x_{i-1/2}, x_{i+1/2}],
    \end{equation}
    where $\bar{u}_{\min} = \min\left\{\Bar{u}_{i-1}, \Bar{u}_{i+1}\right\}, \bar{u}_{\max} = \max\left\{\Bar{u}_{i-1}, \Bar{u}_{i+1}\right\}$.}
\end{conclusion}
\rewb{The detail of the proof is shown in the Appendix.~A. The conclusion gives a reasonable way to normalize stencil data of the MUSCL scheme or the THINC scheme. Meanwhile, different from the MUSCL or THINC scheme depends on the compact stencil among the cells $I_{i-1}, I_{i}, I_{i+1}$, the MUSCL-THINC-BVD scheme in \eqref{numerical:bvd:algorithm} relies on additional one cell in both sides, since the left-side and right-side values on the interface at $x_{i-1/2}$ and $x_{i+1/2}$ are required. We donate $S_{i} = \left\{I_{i-2}, I_{i-1}, I_{i}, I_{i+1}, I_{i+2}\right\}$ as the the stencil of target cell $I_{i}$ in the MUSCL-THINC-BVD scheme, then we have the following conclusion below}
\begin{conclusion}\label{conclusion:2}
    \rewb{Assume the individual cell average values in stencil $S_{i}$ are not identical, then the target $I_{i}$ chooses the same scheme as the MUSCL-THINC-BVD scheme under the normalization as follows
    \begin{equation}\label{conclusion2:normalize}
        \Tilde{u}_{i} = \frac{\bar{u}_{i}-\Tilde{u}_{\min}}{\Tilde{u}_{\max}-\Tilde{u}_{\min}},
    \end{equation}
    where $\Tilde{u}_{\max} = \max\left\{\bar{u}_{i-2}, \bar{u}_{i-1}, \bar{u}_{i}, \bar{u}_{i+1}, \bar{u}_{i+1} \right\}$, $\Tilde{u}_{\min} = \min\left\{\bar{u}_{i-2}, \bar{u}_{i-1}, \bar{u}_{i}, \bar{u}_{i+1}, \bar{u}_{i+1} \right\}.$}
\end{conclusion}
\rewb{The proof is also omitted here and will be presented in Appendix.~B for interested readers. According to the conclusion above, we confirm that the original choice of the BVD algorithm remains unaffected under the normalization procedure \eqref{conclusion2:normalize}. Therefore, we consider the transformation on the stencil data $D_{i} = \left\{\Bar{u}_{i-2}, \Bar{u}_{i-1}, \Bar{u}_{i}, \Bar{u}_{i+1}, \Bar{u}_{i+2}\right\}$ as follows}
\begin{equation}
    \rewb{
    \Tilde{u}_{i} = \left\{
    \begin{aligned}
    \frac{\Bar{u}_{i} - \Tilde{u}_{\min}}{\Tilde{u}_{\max} - \Tilde{u}_{\min}}, & \quad |\Tilde{u}_{\max} - \Tilde{u}_{\min}| \ge \zeta \text{ and } \chi_{i} = 1,\\
    0, & \quad |\Tilde{u}_{\max} - \Tilde{u}_{\min}| < \zeta \text{ or } \chi_{i} = 0.
    \end{aligned}
    \label{numerical:reformulate}
    \right.}
\end{equation}
Here $\Tilde{u}_{\max} = \max\{\bar{u}_{i-2}, \bar{u}_{i-1}, \bar{u}_{i}, \bar{u}_{i+1}, \bar{u}_{i+2}\} $ and $\Tilde{u}_{\min} = \min\{\bar{u}_{i-2}, \bar{u}_{i-1}, \bar{u}_{i}, \bar{u}_{i+1}, \bar{u}_{i+2}\}$, which is the maximum and minimum of stencil $D_{i}$, respectively. $\zeta = 10^{-15}$ is a small positive number. $\chi_{i}$ denotes the monotone indicator, which is defined as follows:

\begin{equation}
    \chi_{i} = \left\{
    \begin{aligned}
        0, & \quad (\bar{u}_{i} - \bar{u}_{i-1})(\bar{u}_{i+1}-\bar{u}_{i}) < 0, \\
        1, & \quad \text{otherwise}.
    \end{aligned}
    \right.
\end{equation}
It makes sense that the discontinuity and material interface are described as relative values ranging from 0 to 1, which is more suitable for the neural network.

\begin{figure}[htbp]
    \centering
    \includegraphics[width=0.8\textwidth]{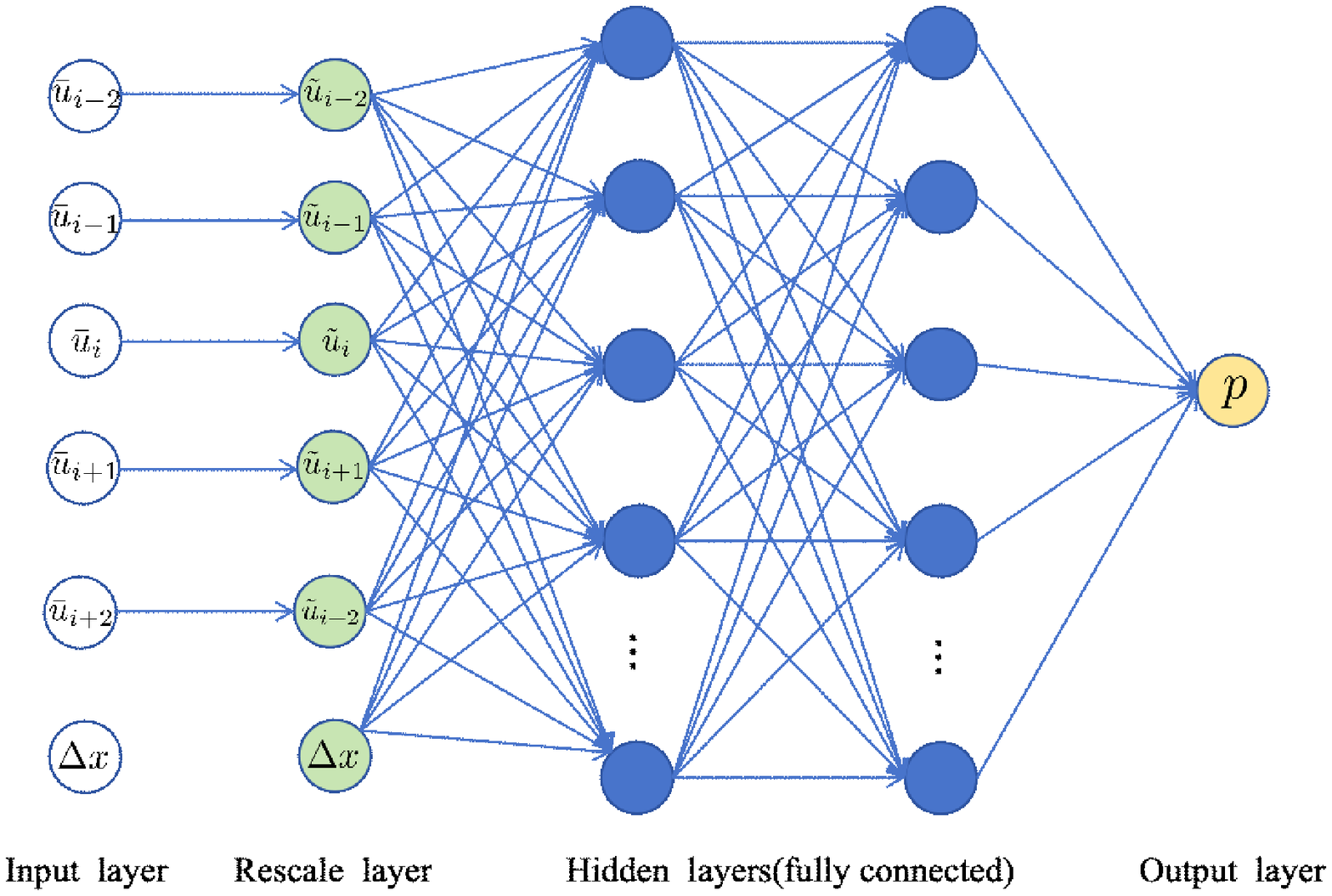}
    \caption{The multilayer perceptron architecture}
    \label{numerical:MLP}
\end{figure}

Here we introduce a multi-layer perceptron architecture as in Fig.~\ref{numerical:MLP}, a fully connected neural network with six inputs, several hidden layers, and one output. Followed by discussion above, we choose $\Bar{u}_{i-2}, \Bar{u}_{i-1}, \Bar{u}_{i}, \Bar{u}_{i+1}, \Bar{u}_{i+2}$ and monotone indicator $\chi_{i}$ of cell $I_{i}$ as preliminary inputs before reformulating by \eqref{numerical:reformulate}. Sigmoid function $\left(\sigma(x) = \frac{1}{1+e^{-x}}\right)$ is chosen as the activation function to achieve nonlinearity. 
Different from normal binary classification by using cross entropy function as loss function, we use weighted focal loss \cite{lin2018focal} as loss function, which reads
\begin{equation}
    FL(\kappa_{t}) =  -\omega_{t} (1 - \kappa_{t})^{\eta} \log(\kappa_{t}),
    \label{numerical:focal:loss}
\end{equation}
where 
\begin{equation}
    \kappa_{t} = \left\{
    \begin{aligned}
        \kappa, & \quad y = 1,\\
        1-\kappa, & \quad y = 0.
    \end{aligned}
    \right. \quad
    y = \left\{
    \begin{aligned}
        1, & \quad \text{THINC},\\
        0, & \quad \text{MUSCL}.
    \end{aligned}
    \right.
\end{equation}
In the above, $y\in \left\{0, 1\right\}$ specifies the ground-truth class where ``1" stands for THINC scheme and ``0" for MUSCL scheme. The output $\kappa\in [0,1]$ represents the probability of a monotone discontinuity inside the cell and prefers to be reconstructed by the THINC scheme. $\omega_{t},\eta$ are the weight for different classes and the positive tunable focusing parameter, respectively. The small $\eta$ adjusts the rate where easy samples are down-weighted and does not affect misclassified loss. The parameter $\omega_{t}$ initially represents the unbalance of positive and negative samples. If positive samples are more than negative, $\omega_{t}$ is supposed to be smaller than $\frac{1}{2}$. 
As cells containing discontinuities tend to select the THINC function, and the number of these cells is always less than those selecting MUSCL, imbalanced data will occur during training.
Additionally, although the THINC function with $\beta=1.6$ captures sharp discontinuities, it introduces excessive anti-diffusion errors and contaminates the smooth region if mistakenly applied there.
Based on this observation, we would rather accept that THINC is mistakenly classified as MUSCL, but we cannot accept that MUSCL is mistakenly classified as THINC, which leads to larger weight $\omega_{t}$ for $y=1$ and smaller for $y=0$. 
\rewb{In the practice experiment, we set $\omega_{t} = 0.1$ and $0.5$ for each classification, respectively, and set $\eta = 2$ to get satisfying results. The chosen weights ratios show the best performance of the correct classification of two schemes and misclassification of the MUSCL scheme, as demonstrated in Appendix.~C.}

\subsubsection{Construction of training set and validation set}
Different from Feng \cite{RN15} for shock indicator and Xue \cite{RN68} for WENO indicator, we generate data set from numerical simulation in different test cases. As shown in Algorithm.\ref{numerical:method:algorithm}, we first implement the MUSCL-THINC-BVD algorithm in the 1D Euler equation. Assume computation domain is $[0, 1]$, then we set up initial data by 
\begin{equation*}
    (\rho^{(0)}_{i}, u^{(0)}_{i}, p^{(0)}_{i}) = \left\{ 
    \begin{aligned}
        (\rho^{L}_{i}, u^{L}_{i}, p^{L}_{i}), \quad x \leq 0.5,\\
        (\rho^{R}_{i}, u^{R}_{i}, p^{R}_{i}), \quad x \ge 0.5.
    \end{aligned}
    \right.
\end{equation*}
We prefer to choose those initial data that generate shock waves, contact discontinuity, and rarefaction waves over a long simulation time. \rewb{The Appendix.~D shows the Rimemann initial state used in this work.} In practice, we will switch the left and right states at $x=0.5$. In Algorithm.~\ref{numerical:method:algorithm},  we use Eqn. \eqref{numerical:timeintegral2} for generating more data since it produces three sub-time profiles. In order to improve the generalization and to be problem-independent, we only advance one initial condition to 10 time steps(30 sub-time steps) with a small Courant number with $cfl=0.1$. Furthermore, we take mesh size $\Delta x$ range from $\frac{1}{100}, \frac{1}{200}, \frac{1}{300}, \frac{1}{400}$ for each test cases. 

\IncMargin{1em}
\begin{algorithm}[!ht] 
\SetKwData{Left}{left}\SetKwData{This}{this}
\SetKwData{Up}{up} \SetKwFunction{Union}{Union}
\SetKwFunction{MTBVD}{$\text{MUSCL-THINC-BVD}$} 
\SetKwFunction{thirdssp}{$\text{third-order SSP Runge-Kutta}$}
\SetKwInOut{Input}{input}
\SetKwInOut{Output}{output}
    \Input{Set of initial profile $\left\{\rho^{L}_{i}, u^{L}_{i}, p^{L}_{i}\right\}, \left\{\rho^{R}_{i}, u^{R}_{i}, p^{R}_{i}\right\}, i \in \left\{1, 2\cdots, N\right\}$, mesh size $M = \left\{\frac{1}{100}, \frac{1}{200}, \frac{1}{300}, \frac{1}{400}\right\}$ .} 
    \Output{The training and validation set $\mathcal{R}$.}
    \BlankLine
    \emph{// Step1: Generate the unprocessed set $\mathcal{R}_{0}$}\label{numerical:method:algorithm:substep1}
    
    \For{$i\leftarrow 1$ \KwTo $N$} { 
        \For {$j \in M $ }{
            1. Set up initial data $\left\{\rho^{L}_{i}, u^{L}_{i}, p^{L}_{i}\right\}, \left\{\rho^{R}_{i}, u^{R}_{i}, p^{R}_{i}\right\}$, located at $x = 0.5$ with mesh grid size $M[j]$.
            
            \For {$k \leftarrow 1$ \KwTo 10}{
                2. Calculate $k^{th}$ sub-time step by $\thirdssp$ by $\MTBVD$ scheme.\label{numerical:method:ssp}
                
                3. Add the stencil values and label of cell $I_{i}$, $\left\{\Bar{u}_{i-2}, \Bar{u}_{i-1}, \Bar{u}_{i}, \Bar{u}_{i+1}, \Bar{u}_{i+2}, \chi_{i}, y_{i}\right\}$\ into $\mathcal{R}_{0}$.
            }
        }
    } 
    \BlankLine
    \emph{// Step2: Generate $\mathcal{R}$ from the set $\mathcal{R}_{0}$}\label{numerical:method:algorithm:substep2}
    \For{$r_{0} = [\Bar{u}^{0}_{i-2}, \Bar{u}^{0}_{i-1}, \Bar{u}^{0}_{i}, \Bar{u}^{0}_{i+1}, \Bar{u}^{0}_{i+2}]^{T} \in \mathcal{R}_{0}$} {
        \vspace{.1cm}
        distance = 1, flag = 0
        
        \For {$r = [\Bar{u}_{i-2}, \Bar{u}_{i-1}, \Bar{u}_{i}, \Bar{u}_{i+1}, \Bar{u}_{i+2}]^{T} \in \mathcal{R}$} {
            \vspace{.1cm}
            $\text{distance} = \min(\text{distance},  ||r_{0} - r||_{\infty})$

            \If{$\text{distance} < \delta$} {
                \emph{// $\delta$ is a small positive number, we set it as $10^{-3}$}
                
                flag = 1, break
            }
        }
        \If{$\text{flag} == 0$} {
            Add $r$ into $\mathcal{R}$;
        }
    }
    \caption{Generate training and validation data set.}
    \label{numerical:method:algorithm}  
\end{algorithm}
\DecMargin{1em}

\rewb{In step.~(1) of Algorithm.~\ref{numerical:method:algorithm}, unprocessed set $\mathcal{R}_{0}$ needs to be generated firstly. Given a target cell $I_{i}$ and its input $\left\{\Bar{u}_{i-2}, \Bar{u}_{i-1}, \Bar{u}_{i}, \Bar{u}_{i+1}, \Bar{u}_{i+2}, \chi_{i} \right\}$, we label it by MUSCL-THINC-BVD algorithm. If the target cell is reconstructed using the THINC scheme, the label is set to ``1". Otherwise, the label is set to ``0", representing the MUSCL scheme. The set $\mathcal{R}_{0}$  is mostly composed of ``0" values, with fewer ``1"s, since the discontinuities, which are the target of the THINC reconstruction scheme, are primarily located at $x = 0.5$, while other regions remain smooth or constant. In step. (2) of Algorithm.~\ref{numerical:method:algorithm}, deriving $\mathcal{R}$ from $\mathcal{R}_{0}$ involves manually eliminating duplicate elements, ensuring that only representative samples are maintained.}

Following the above-mentioned procedure, we generate 8000 supervised data for training and 2000 for validation. 
We set up two hidden layers, each with eight neurons. The Optimizer is Adam optimizer \cite{kingma2017adam} with a bath size of 128 and a learning rate of 0.001 with 2000 iterations. Finally, we find that the training model achieves an accuracy rate of over $98\%$ on both the training and validation sets.
\begin{remark}
Even though the data is trained from the one-dimensional Euler equation, the trained neural network overcomes the limitations of governing equations and variables through normalization in Eqn. \eqref{numerical:reformulate}. \rewb{First, the trained neural network serves as the predictor in the deepMTBVD scheme. No matter the one- or high-dimensional case, as long as the input is arranged as the vectors of the form $S_{i}$, the trained neural network is able to infer the results. Second, the MUSCL-THINC-BVD scheme, as described in \eqref{bvd:mtbvd}, relies on the boundary variation of physical variables between cell boundaries and can be consistently applied to all physical quantities without modification. Similarly, the deepMTBVD scheme is directly applicable to all physical variables in both single-phase and two-phase flows, in one-dimensional and two-dimensional settings, using the dimensional splitting method. Following the normalization procedure outlined in \eqref{conclusion2:normalize}, the physical variables are scaled to a range between 0 and 1, effectively mitigating issues arising from significant differences in magnitudes across physical variables. Moreover, as demonstrated in Conclusion.~\ref{conclusion:2}, this normalization process preserves the results chosen by the original BVD algorithm. Third, The numerical results presented in Section~\ref{numerical_result} further illustrate the broad applicability and robustness of the deepMTBVD scheme.}
\end{remark}

\subsubsection{Framework of the deepMTBVD method in FVM}
In this subsection, we briefly show the framework of applying the offline-trained indicator deepMTBVD into Eqn. \eqref{numerical:fvm}. The implementation of deepMTBVD into FVM is detailed in the Algorithm. 2.
\IncMargin{1em}
\begin{algorithm}[!ht] 
\SetKwData{Left}{left}\SetKwData{This}{this}
\SetKwData{Up}{up} \SetKwFunction{Union}{Union}
\SetKwFunction{deepmtbvd}{$\text{deepMTBVD}$} 
\SetKwFunction{secondssp}{$\text{second-order SSP Runge-Kutta}$}
\SetKwInOut{Input}{input}
\SetKwInOut{Output}{output}
    \Input{Physical variable $\Bar{u}_{i}^{n}$ of $I_{i}$ at time $t^{n}$, including density $\rho$, velocity $\bm{V}$,pressure $p$, volume fraction $\alpha_{k}$.} 
    \Output{$\Bar{u}_{i}^{n+1}$ of each $I_{i}$ at time $t^{n+1}$}
    1. Select input stencil $\left\{\Bar{u}^{n}_{i-2}, \Bar{u}^{n}_{i-1}, \Bar{u}^{n}_{i}, \Bar{u}^{n}_{i+1}, \Bar{u}^{n}_{i+2}, \chi_{i}\right\}$ for each cell $I_{i}$;
    
    2. Calculate output $\kappa_{i}$ by the deepMTBVD and select the reconstruction scheme by comparing it with reference $\kappa_{ref}$;
    
    3. Obtain the reconstruction value on the cell boundary for each $I_{i}$ and calculate flux by Riemann solver;
    
    4. Advance by \eqref{numerical:timeintegral} and \eqref{numerical:fvm} and repeat steps (1)-(3) in each sub-time step. 
    \caption{Application of deepMTBVD method in \eqref{numerical:fvm}}
    \label{numerical:method:deepmtbvd}  
\end{algorithm}
\DecMargin{1em}
Unlike the original MUSCL-THINC-BVD scheme, which requires pre-reconstruction of all schemes for each cell, the deepMTBVD method necessitates reconstruction only once in each sub-time step.
In the next section, we will compare numerical cases under the MUSCL and MUSCL-THINC-BVD scheme framework to the present indicator deepMTBVD. 

\begin{remark} Here are some remarks for training: \\
$\bullet$ Different from previous work\cite{RN15, RN68, salazar2020hybrid}, the network is more complex and we implement it into C++ code by ONNX file, which has the capability to perform matrix operation efficiently.  \\
$\bullet$ We have tested different network structures for training, such as using each stencil to correspond to a sub-network and then training jointly in the next hidden layer. The numerical results are similar, but the efficiency will be reduced. \\
$\bullet$ \hms{The \kaparef value serves as a threshold for controlling the results of binary classification. A higher \kaparef (larger than 0.5) leads to an increased number of MUSCL cells, which exhibit more diffusive behavior. In contrast, a lower \kaparef results in more THINC cells, which causes numerical instability and requires higher computational costs. This issue is discussed further in Appendix.~E. In this study, we have chosen a uniform value of \kaparef=0.45 for all numerical simulations.}

\end{remark}
\section{Numerical results}\label{numerical_result}
This section will evaluate the performance of the deepMTBVD scheme in numerical simulations of compressible single-phase and two-phase flows using benchmark tests of the Euler equations and the five-equation model. In all cases, the \kaparef is taken as the uniform value of 0.45.

\hms{\subsection{Accuracy test for linear advection equation}
In this section, we illustrate the accuracy test for the linear advection equation, which is carried out on refined grids to evaluate the convergence rate of the proposed scheme. The one-dimensional advection equation is given as:
\begin{equation}
    \frac{\partial u}{\partial t} + \frac{\partial u}{\partial x} = 0.\label{result:adv:eqn}
\end{equation}
The initial condition is defined as follows:
\begin{equation*}
    u(x,0) = sin(\pi x), x \in [-1,1].
\end{equation*}
We define the $L_{1}$ norm error as:
$$
L_{1, \text{error}} = \frac{\sum_{i=1}^{N} |\Bar{u}_{i}^{num}-\Bar{u}_{i}^{exa}|}{N},
$$
where $\Bar{u}_{i}^{num}$ is the numerical solution of the average value in cell $i$ and the $\Bar{u}_{i}^{exa}$ refers to the exact solution of the average value in cell $i$. The velocity in \eqref{result:adv:eqn} is set to 1, and the periodical boundary condition is employed on both sides in this example. 
\begin{figure}[ht!]
    \centering
    \subfloat[$L_{1,\text{error}}$ with mesh refinement]{\includegraphics[width=0.49\linewidth]{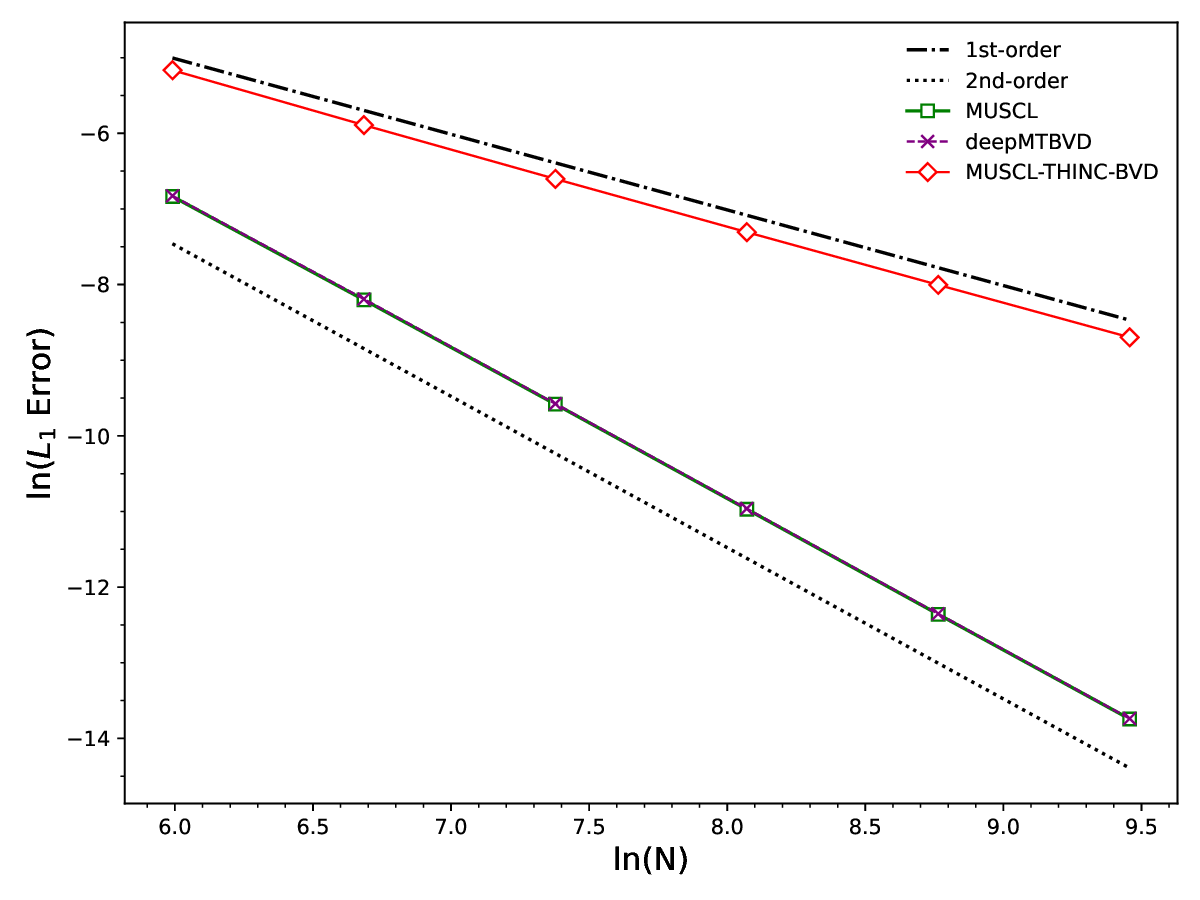}} 
    \subfloat[Time (s) with mesh refinement]{\includegraphics[width=0.49\linewidth]{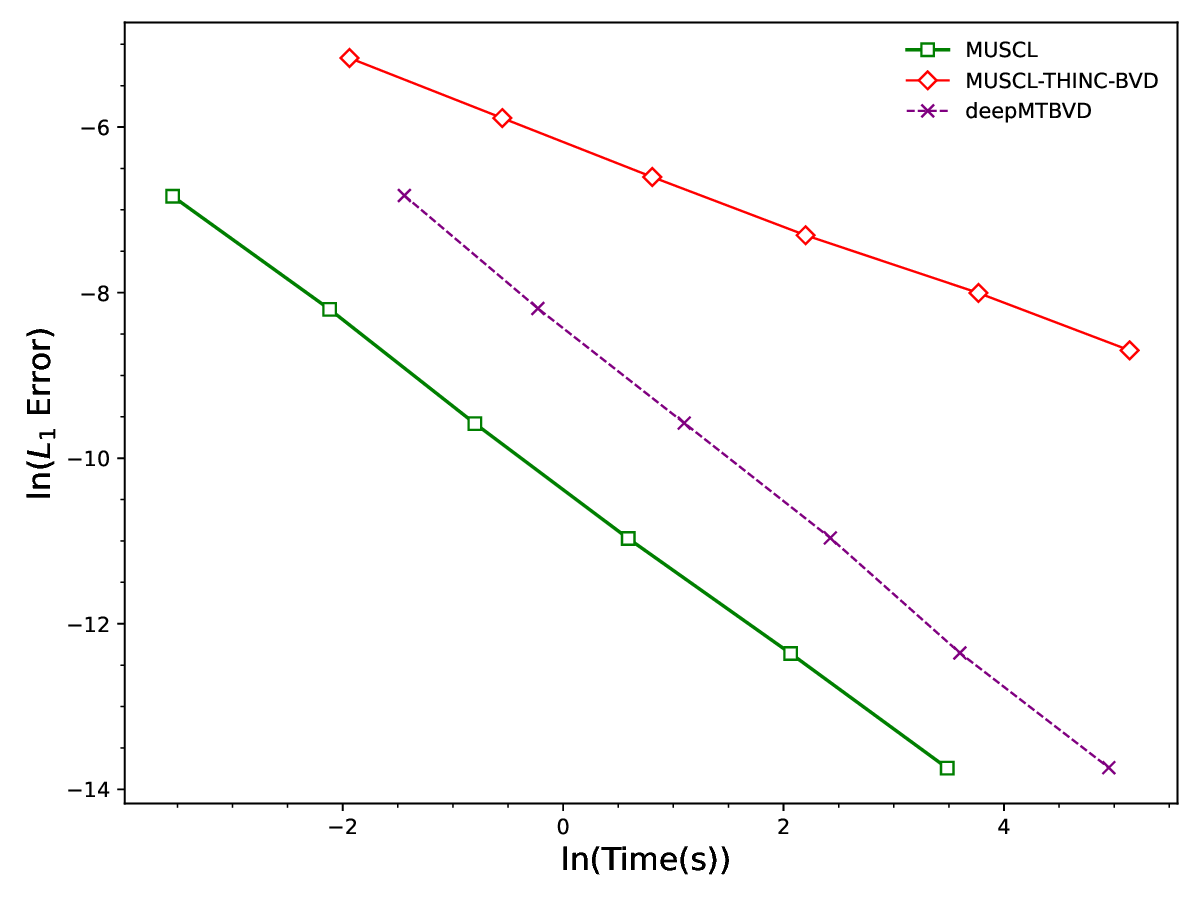}}
    \caption{Numerical results of the linear advection equation with 5 periods. Comparisons are made among the MUSCL scheme, MUSCL-THINC-BVD scheme, and deepMTBVD scheme.}
    \label{result:advection}
\end{figure}
We ran the computation for the 5 periods (time $t = 10$) with different mesh sizes: 400, 800, 1600, 3200, 6400, and 12800. The summarized numerical errors in $L_{1}$ norm and the cost time are presented in Fig.~\ref{result:advection}. It can be seen that the MUSCL scheme is second-order in $L_{1}$ norm and is the most efficient in the smooth profile. However, the MUSCL scheme will suffer more diffusion errors, as demonstrated in the following examples. The MUSCL-THINC-BVD scheme is only first order and costs most of the time as the mesh size is larger than 6400. From the figure above, we conclude that deepMTBVD is threefold: 1) Same with the MUSCL scheme in the smooth region, the deepMTBVD is also a second-order scheme in $L_{1}$ norm error while the MUSCL-THINC-BVD scheme is only first-order. 2) The deepMTBVD scheme costs more time than the MUSCL scheme since it needs an additional procedure for inference by the neural network. 3) As mesh refinement, the deepMTBVD is a more accurate and more efficient scheme compared to the MUSCL-THINC-BVD scheme. }

\subsection{Euler equations}
Classical one-dimensional shock tube problems and two-dimensional shock reflection problems were used to test the performance of the deepMTBVD scheme in solving the Euler equations. Numerical results show that the deepMTBVD scheme performs comparably or even better than the MUSCL-THINC-BVD scheme in the numerical simulation of compressible single-phase flows, and it achieves higher computational efficiency.
The ideal gas law is used with the ratio of specific heats of $\gamma = 1.4$. The HLLC Riemann solver \cite{torobook} is used for computing the numerical flux across cell interfaces. \hms{The Courant number is set to 0.4.}

\subsubsection{Sod's Problem}
The Sod's problem is one of the shock-capturing tests, which is applied here to test the ability of numerical schemes to resolve shock waves and contact discontinuities. The computational domain is a unit-long tube and the initial condition is given by 
\begin{equation*}
\left(\rho, v, p\right)= \begin{cases}(1,0,1), & 0 \leqslant x \leqslant 0.5, \\ (0.125,0,0.1), & \text { otherwise }.\end{cases}
\end{equation*}
\begin{figure}[ht!]
    \centering
    \includegraphics[width=\linewidth]{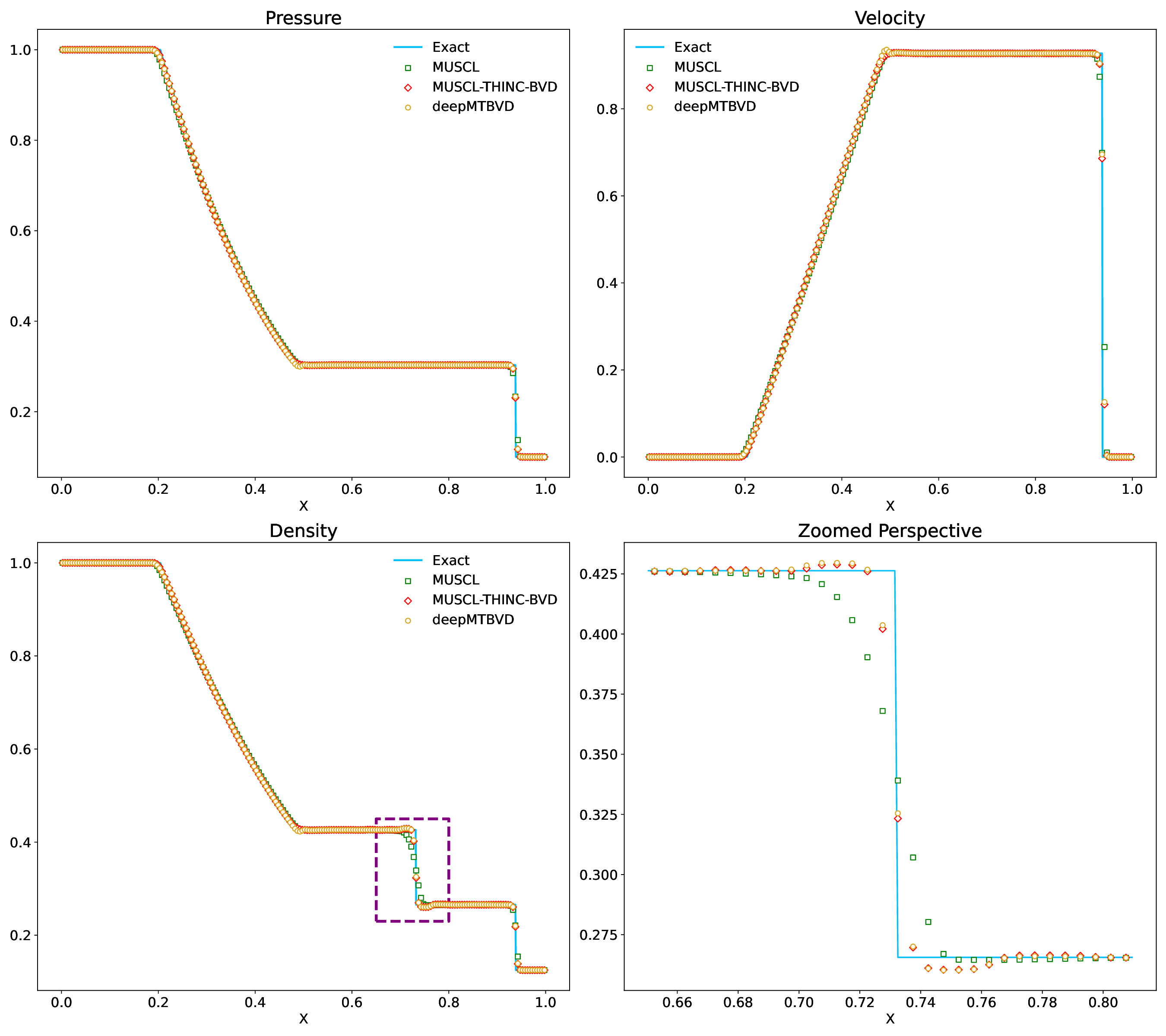}
    \caption{\hms{Numerical results of Sod's problem for velocity, pressure, and density at time $t=0.25$ with 200 mesh cells. The solid blue line is the exact solution, the green square symbol denotes the MUSCL scheme, the red diamond denotes the MUSCL-THINC-BVD scheme and the golden circle denotes deepMTBVD. The right bottom is a zoomed perspective of the dashed square.}} 
    \label{result:example1}
\end{figure}
The mesh number is set to 200 in this test. The results from MUSCL, MUSCH-THINC-BVD, and deepMTBVD are presented in Fig.~\ref{result:example1} for comparison. It obviously shows that deepMTBVD resolves the shock wave as well as MUSCL-THINC-BVD and MUSCL schemes.
It is obvious that the deepMTBVD scheme performs consistently with the MUSCL scheme, capturing sharp shocks and contact discontinuities. This indicates that the ANN accurately selected the THINC scheme at these locations.

\subsubsection{Lax's Problem}
\begin{figure}[htbp!]
    \centering
    \includegraphics[width=\linewidth]{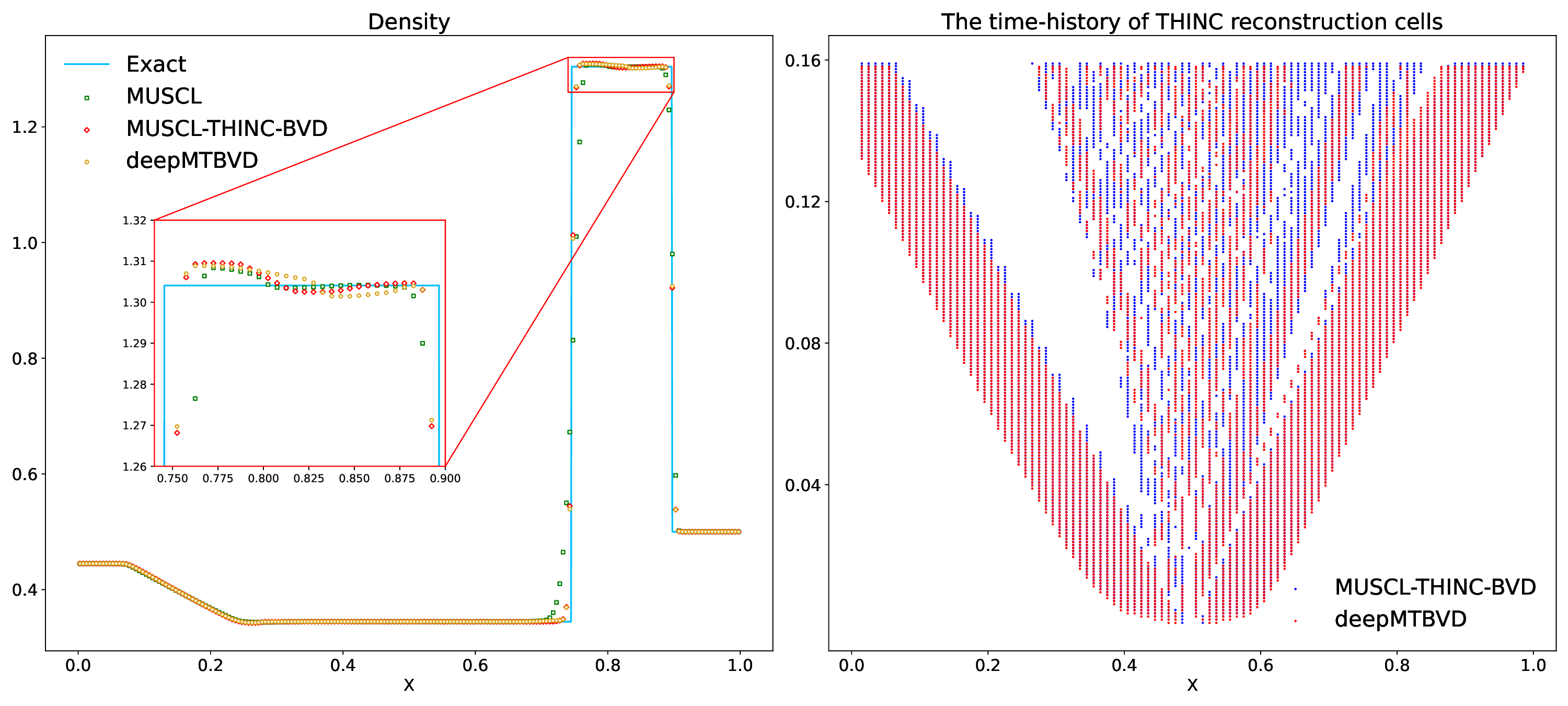}
    \caption{\hms{Numerical results of Lax's problem at $t=0.16$ with 200 uniform cells. The left panel represents the density profiles computed by three schemes. A zoomed perspective is added in the figure, which refers to the less numerical oscillation introduced by the deepMTBVD scheme compared to the MUSCL-THINC-BVD scheme. The right panel indicates whether the cell is selected as the THINC scheme by MUSCL-THINC-BVD (blue) or deepMTBVD (light red).}}
    \label{result:example3}
\end{figure}

This is a widely used benchmark test with the initial condition given by \cite{DengShimizu-142}:
\begin{equation*}
    \left(\rho, v, p\right)= \begin{cases}(0.445,0.698,3.528), & 0 \leqslant x \leqslant 0.5, \\ (0.5,0,0.571), & \text { otherwise }.\end{cases}
\end{equation*}
We present numerical results at time $t = 0.16$ with the mesh number of 200 cells. The exact solution consists of a left-moving rarefaction, a contact discontinuity, and a right-moving shock wave. The left panel Fig.~\ref{result:example3} shows the density profile of numerical results. DeepMTBVD and MUSCL-THINC-BVD reproduce the contact discontinuity and shock wave better than the MUSCL scheme while maintaining high accuracy for a smooth profile. 
At the end of the rarefaction wave, the results of the deepMTBVD scheme are closer to the exact solution compared to the MUSCL scheme while exhibiting weaker local numerical oscillations than the MUSCL-THINC-BVD scheme between the shock wave and contact wave in the enlarging perspective.

We also plot the $t-x$ diagram in the right of Fig.~\ref{result:example3} to demonstrate where the THINC reconstruction is implemented. 
The light red symbols represent the cell where the THINC function is selected by deepMTBVD but not by MUSCL-THINC-BVD at time $t$.
Otherwise, it is blue. If both schemes implement THINC reconstruction, we denote that case as dark red. It shows that deepMTBVD correctly identified contact discontinuity and shock waves. The BVD criterion learned by the neural network is nearly identical to the MUSCL-THINC-BVD scheme, especially for contact discontinuity and shock waves. 


\subsubsection{Strong Lax's problem}\label{strong_lax}
\begin{figure}[ht!]
    \centering
    \includegraphics[width=\textwidth]{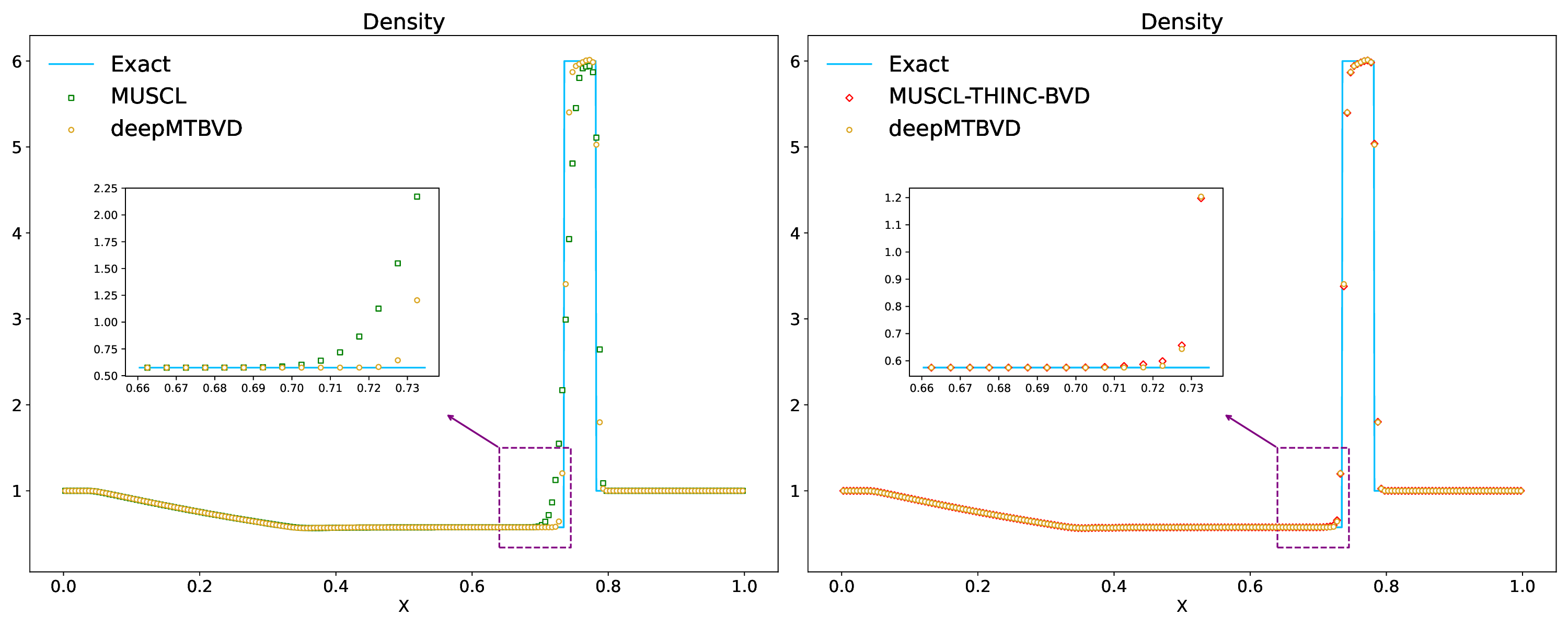}
    \caption{Comparison of numerical results with the exact solution of density at $t=0.012$ with 100 mesh cells. Left: MUSCL (green square) and deepMTBVD (golden circle); Right: MUSCL-THINC-BVD (red diamond) and deepMTBVD (golden circle). The solid line and square denote the exact solution in the instant time. }
    \label{result:example2}
\end{figure}

This benchmark is half of the blast wave of Woodward and Celella \cite{RN36}, whose solution contains a left rarefaction, a contact wave, and a right-moving shock wave. We consider the following conditions:
\begin{equation*}
    (\rho, u, p)= \begin{cases}(1.0,0.0,1000.0), & \text { if } x<0.5, \\ (1.0,0.0,0.01),& \text { otherwise. }\end{cases}
\end{equation*}
The computation domain is $[0, 1]$ with uniform 100 mesh cells. The density of exact and numerical results at time $t = 0.012$ are shown in Fig.~\ref{result:example2}. 
The left part shows that deepMTBVD resolves the shock wave sharper than the MUSCL scheme. Moreover, deepMTBVD produces a much less dissipative solution across the contact discontinuity than MUSCL, as shown in the zoomed dash square.
In the right part of Fig.~\ref{result:example2}, it should be noted that the numerical dissipation of deepMTBVD near strong contact discontinuities is lower than that of the MUSCL-THINC-BVD scheme.

\begin{remark}
    \begin{table}[ht!]
        \centering
        \caption{Time cost in previous three examples (N=10000).}
        \begin{tabular}{|c|rr|rr|rr|}
            \hline
            \multirow{2}{*}{Time(s)} &  \multicolumn{2}{c|}{Example1} &  \multicolumn{2}{c|}{Example2}& \multicolumn{2}{c|}{Example3}\\
             & \tiny{MUSCL-THINC-BVD} & \tiny{deepMTBVD} & \tiny{MUSCL-THINC-BVD} & \tiny{deepMTBVD} & \tiny{MUSCL-THINC-BVD} & \tiny{deepMTBVD} \\
             \hline
             Reconstruction Time & 37.04& 9.11& 51.20& 13.41& 42.17& 8.93\\
             Selection Time      & 2.99& 23.56& 4.10& 31.60& 3.18& 25.58\\
             Total Time          & 40.04& 32.67& 55.30& 45.01& 45.36& 34.51\\
             \hline
        \end{tabular}
        \label{time cost1}
    \end{table}

       \begin{table}[ht!]
        \centering
        \caption{Time cost in previous three examples(N=100000).}
        \begin{tabular}{|c|rr|rr|rr|}
            \hline
             \multirow{2}{*}{Time(s)} &  \multicolumn{2}{c|}{Example1} &  \multicolumn{2}{c|}{Example2}& \multicolumn{2}{c|}{Example3}\\
             & \tiny{MUSCL-THINC-BVD} & \tiny{deepMTBVD} & \tiny{MUSCL-THINC-BVD} & \tiny{deepMTBVD} & \tiny{MUSCL-THINC-BVD} & \tiny{deepMTBVD} \\  
                \hline
             Reconstruction Time & 5193.18& 569.11& 6071.26& 889.25& 4966.18& 862.73\\
             Selection Time      & 1072.32& 1996.91& 1197.09& 3037.32& 1009.79& 2772.40\\
             Total Time          & 6265.50& 2566.02& 7268.35& 3926.57& 5975.97& 3635.13\\
             \hline
        \end{tabular}
        \label{time cost2}
        \end{table}
    The above three examples compare the computational efficiency between the conventional MUSCL-THINC-BVD scheme and deepMTBVD. The mesh number in Table.\ref{time cost1} is set to 10000 while in Table.\ref{time cost2} is set to 100000. 
    The MUSCL-THINC-BVD scheme reconstructs all admissible reconstruction functions, such as MUSCL and THINC, and then chooses one of these, which leads to local minimal boundary variation. Unlike MUSCL-THINC-BVD, the deepMTBVD can pre-determine the proper admissible function by neural networks and thus reconstructs only once.
    Due to the complexity of the neural network, the deepMTBVD takes more time than the MUSCL-THINC-BVD scheme in the selection procedure but saves more time in the reconstruction procedure. Although the amount of saved time varies in different examples, our experiments show that it can improve performance by $10\%$ to $40\%$.
\end{remark}

\begin{figure}[ht!]
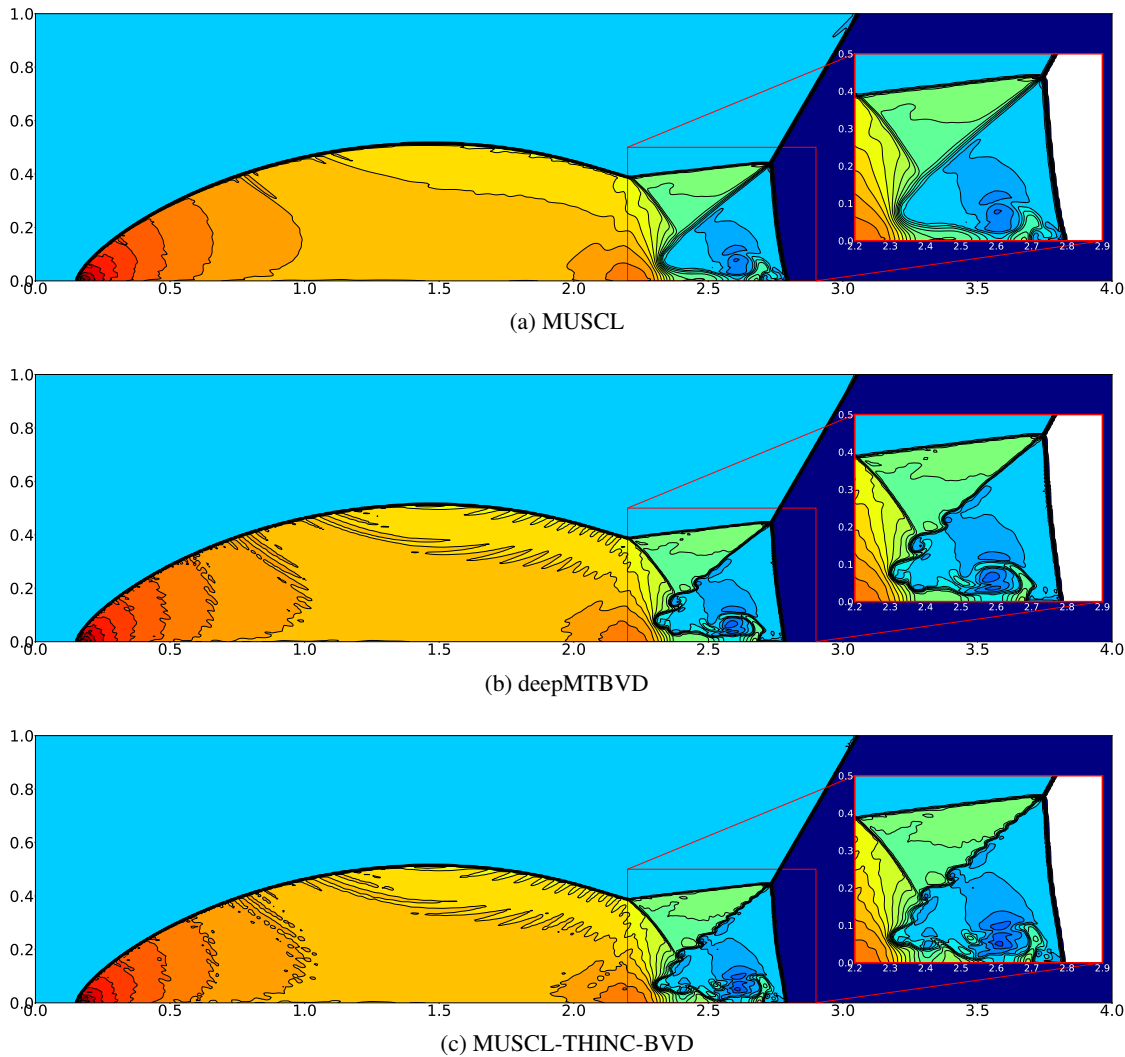

    \centering
    \subfloat[MUSCL]{\includegraphics[width=0.9\linewidth]{doublemach_200_muscl.eps}} \\
    \subfloat[deepMTBVD]{\includegraphics[width=0.9\linewidth]{doublemach_200_deepmtbvd_45.eps}} \\
    \subfloat[MUSCL-THINC-BVD]{\includegraphics[width=0.9\linewidth]{doublemach_200_mtbvd2.eps}}
    \caption{Density contours ranging from 1.5 to 21.5 with 30 levels at time $t = 0.2$. The sub-figure in each figure is the enlarged view within the red box.}
    \label{result:doulemach}
\end{figure}
\subsubsection{Double Mach Reflection}
Originally proposed in \cite{RN36}, a Mach 10 propagating planar shock with the 30-degree ramp is simulated in this example. This test has been adopted to simulate the strong reflected and refracted shocks as well as the richer vortical structures, which are sensitive to the dissipation of the numerical schemes. 
The effectiveness of a numerical scheme in reducing dissipation can be assessed by observing vortical structures formed due to Kelvin–Helmholtz instabilities along the slip line in the recirculation zone. Numerical schemes with much numerical dissipation often result in less refined structures.
The computation domain is $[0,4] \times [0,1]$. The reflecting wall lies along the bottom boundary, beginning at $x = \frac{1}{6}$. A right-going shock is imposed with 60 degrees relative to the x-axis and extends to the top $y=1$. The short region from $x \in [0, \frac{1}{6}]$ along the bottom boundary and the whole left-hand boundary is assumed to be the initial post-shock flow. The right boundary condition is given by zero-gradient.
lThe numerical solution is computed up to the time $t = 0.2$ with uniform mesh size $\Delta x = \Delta y = \frac{1}{200}$. 

 \begin{table}[htb]
    \centering
    \caption{Time cost in Double Mach Reflection.}
    \begin{tabular}{|c|rr|}
        \hline
        Time(s) &  \tiny{MUSCL-THINC-BVD} & \tiny{deepMTBVD} \\
         \hline
         Reconstruction Time &\hms{192.21}  & \hms{88.25}\\
         Selection Time &\hms{68.25}  & \hms{136.62}  \\
         Total Time &\hms{260.46} & \hms{223.87} \\
         \hline
    \end{tabular}
    \label{time:double_mach}
\end{table}
The result of density contours is shown in Fig.~\ref{result:doulemach}, which ranges from 1.5 to 21.5 with 30 equidistant levels. Both the MUSCL-THINC-BVD and deepMTBVD schemes capture shock waves with less dissipation compared to the traditional MUSCL scheme.
From the zooming perspective of each scheme, both deepMTBVD and MUSCL-THINC-BVD can reproduce fine vortical structures along the slip line, whereas the MUSCL scheme fails due to excessive numerical dissipation.
Compared with MUSCL-THINC-BVD, deepMTBVD resolves the compatible numerical flow structure of vortex chains in Fig.~\ref{result:doulemach} and requires less reconstruction time in Table. \ref{time:double_mach}.

\subsubsection{A Mach 3 Wind Tunnel With a Step}
This benchmark also tests the capability of the numerical scheme to capture the strong shock and vortical structures. The problem begins with the uniform Mach 3 flow in a wind tunnel with a step. The wind tunnel is 1 unit wide and 3 units long. A forward step is located at 0.6 units from the left-hand boundary to the end of the tunnel. The left boundary condition is specified as an inflow, the right boundary condition is assumed to have zero gradients, and the remaining boundaries are treated as reflective walls.
he tunnel is filled with the ideal gas, with Mach=3.0,$\gamma = 1.4$, density 1.4, pressure 1.0, and velocity 3.0, respectively. The computation region is divided into uniform mesh with $\Delta x = \Delta y = \frac{1}{160}$. 
\begin{figure}[ht!]
    \centering
    \subfloat[MUSCL]{\includegraphics[width=0.8\linewidth]{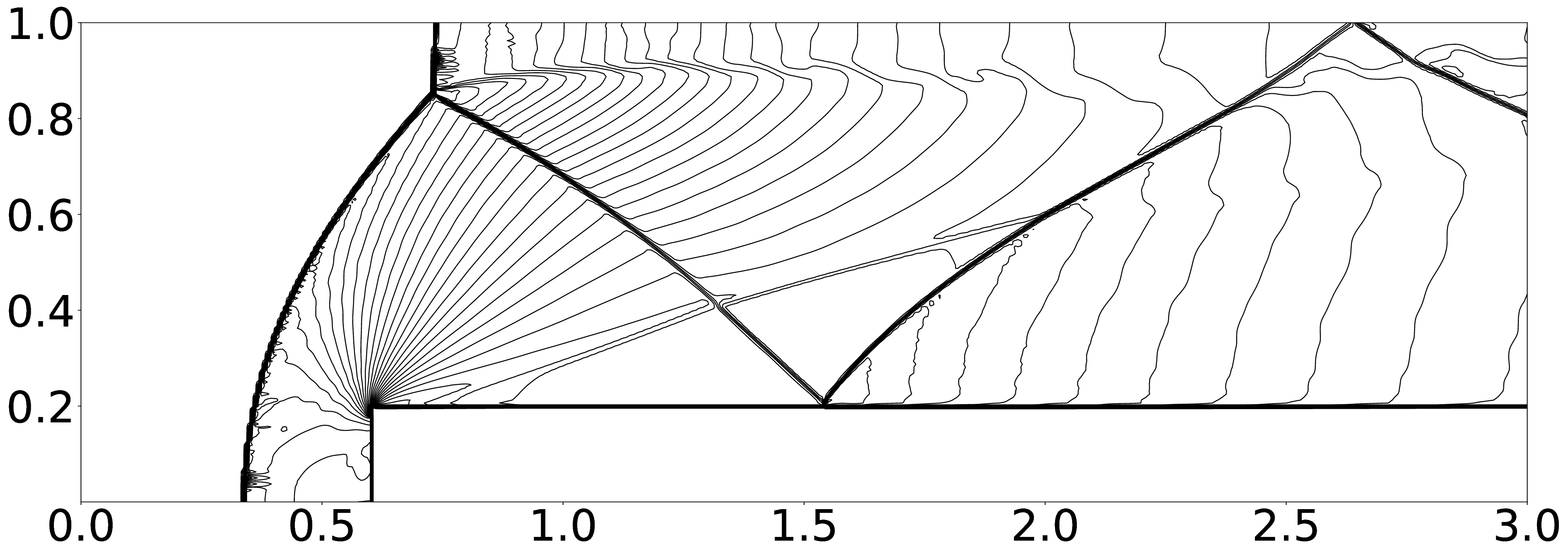}}\\
    \subfloat[deepMTBVD]{\includegraphics[width=0.8\linewidth]{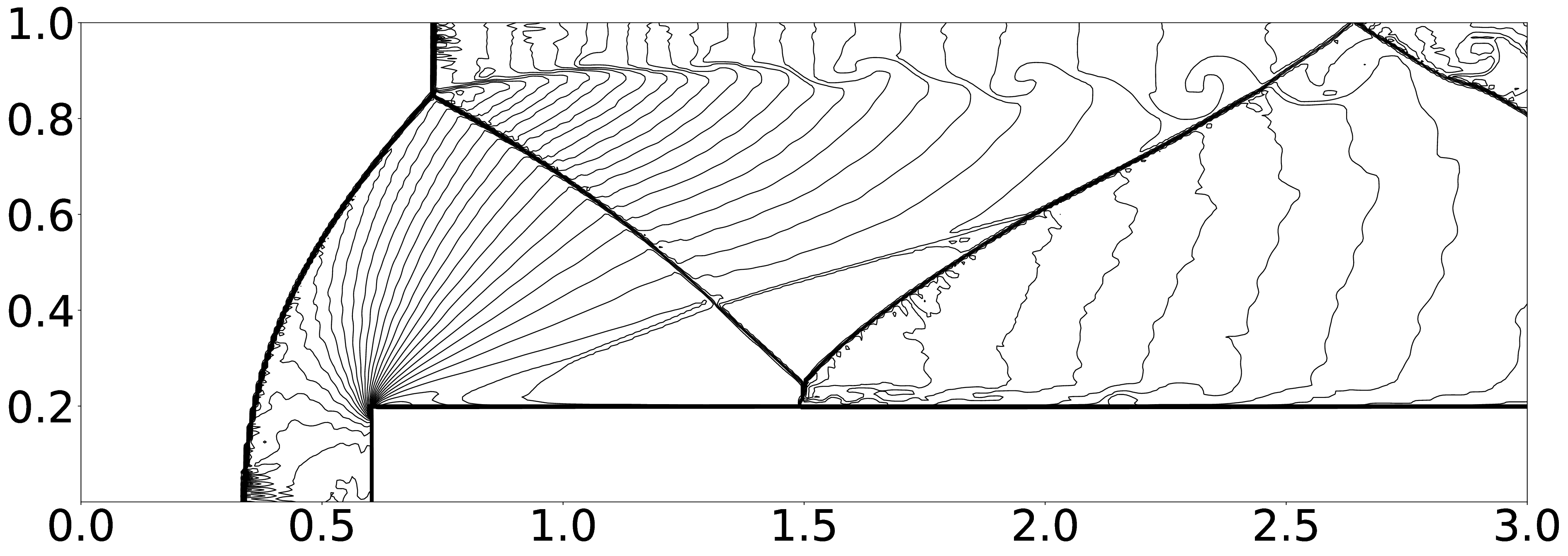}}\label{frontstep:deepmtbvd}\\
    \subfloat[MUSCL-THINC-BVD]{\includegraphics[width=0.8\linewidth]{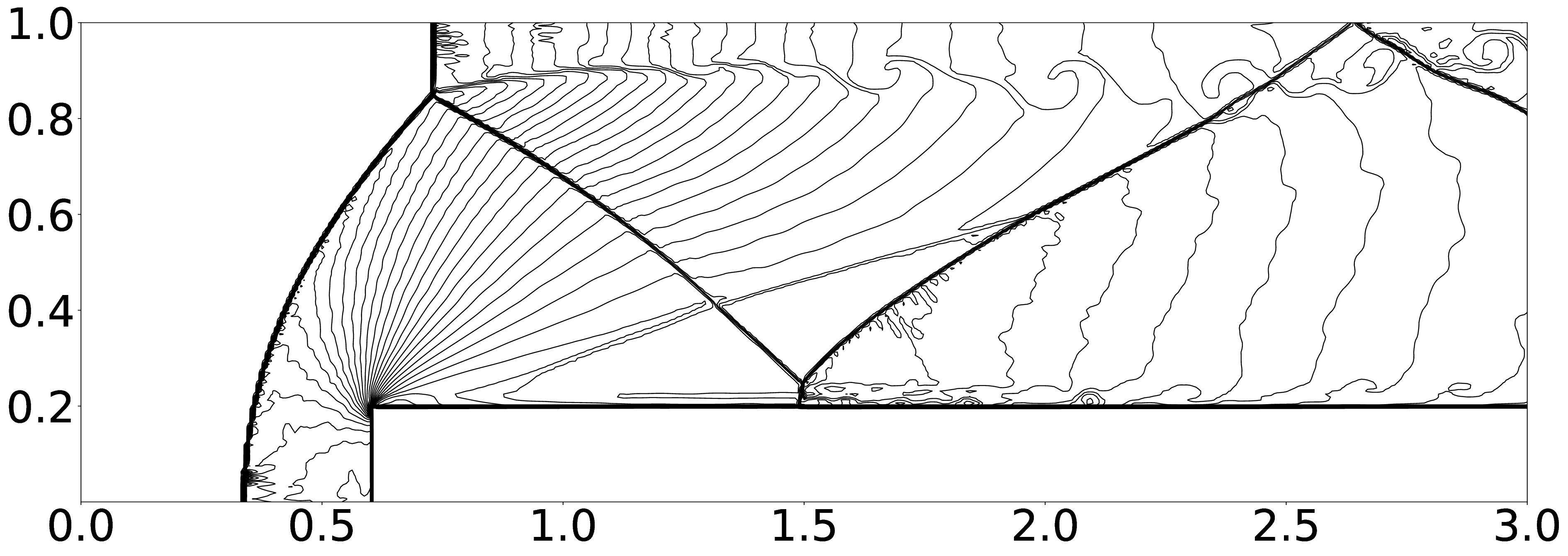}}\label{frontstep:mtbvd}
    \caption{Density contours range from 1.5 to 21.5 with 30 levels at time $t=4.0$.}
    \label{result:frontstep}
\end{figure}
\begin{figure}[ht!]
    \centering
    \subfloat[MUSCL-THINC-BVD]{\includegraphics[width=0.49\textwidth]{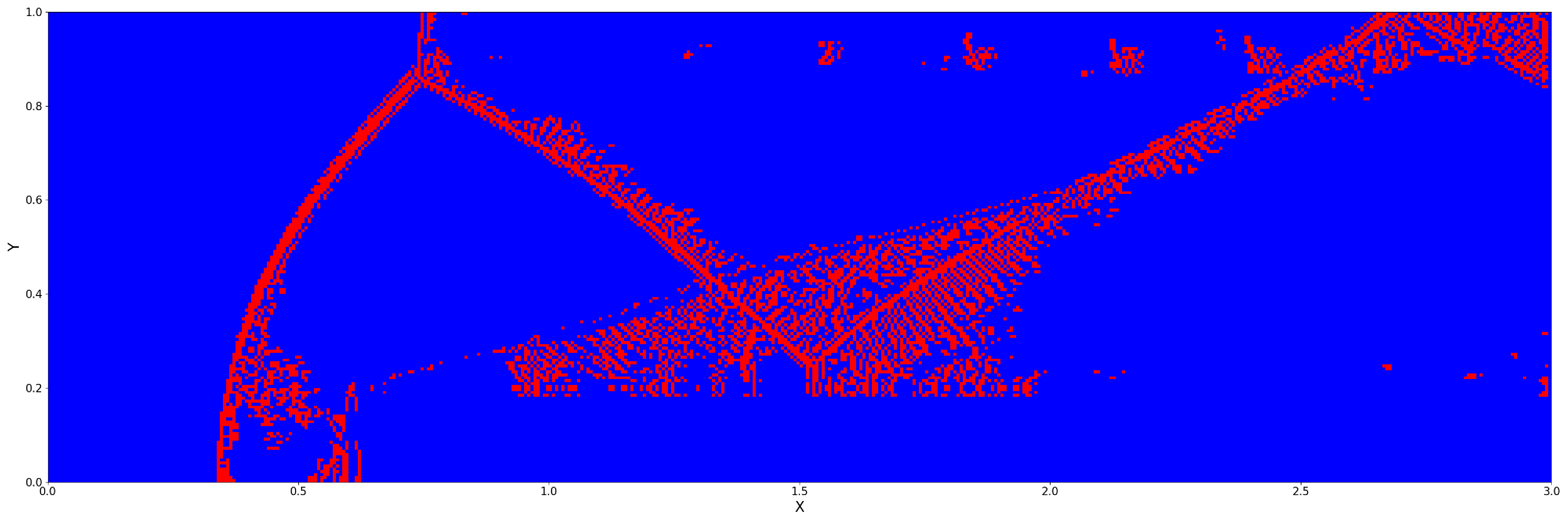}}
    \subfloat[deepMTBVD]{\includegraphics[width=0.49\textwidth]{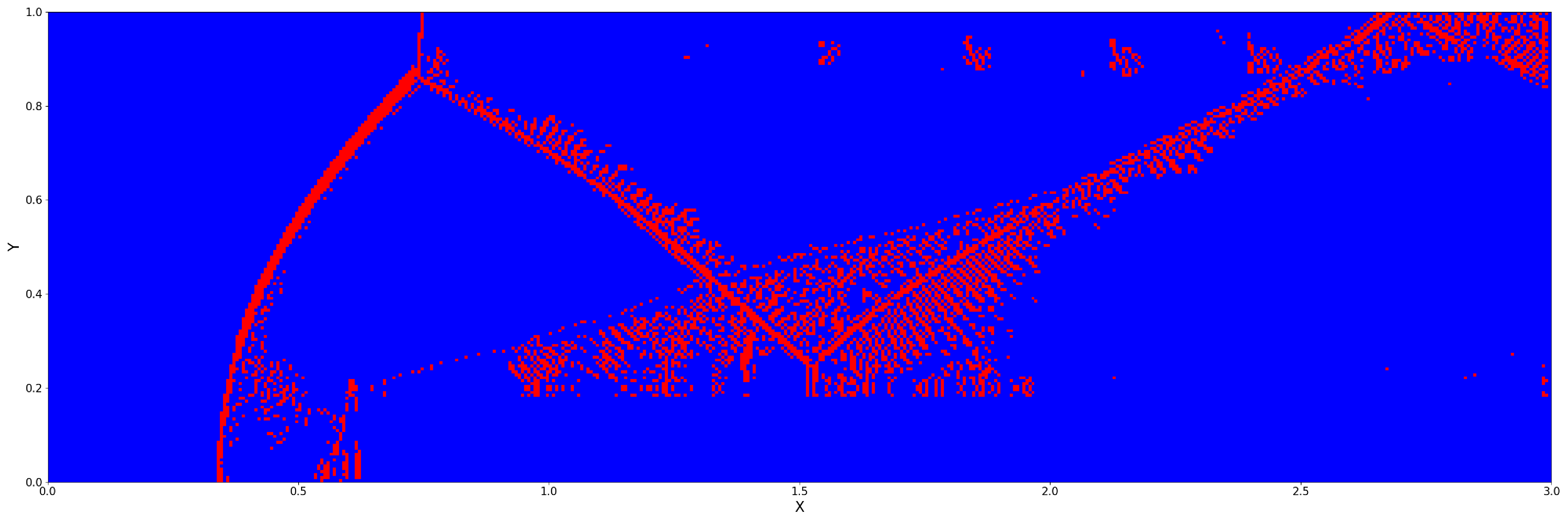}}
    \caption{\hms{The cell where is selected as THINC scheme of pressure $p$ in the x-axis through the MUSCL-THINC-BVD scheme (left) and the deepMTBVD scheme (right). ``Red" denotes the cell reconstructed by the THINC scheme, while the ``blue" represents the MUSCL scheme.}}
    \label{result:frontstep:celltype}
\end{figure}
 \begin{table}[htbp]
    \centering
    \caption{Time cost in Mach 3 wind tunnel with a step.}
    \begin{tabular}{|c|rr|}
        \hline
        Time(s) &  \tiny{MUSCL-THINC-BVD} & \tiny{deepMTBVD} \\
         \hline
         Reconstruction Time &212.22  &83.98\\
         Selection Time &50.32  &131.68  \\
         Total Time &242.54 &215.66 \\
         \hline
    \end{tabular}
    \label{time:frontstep}
\end{table}

The density contours range from 0.28 to 6.15 at time $t = 4.0$ are shown in Fig.~\ref{result:frontstep}. From the numerical results, the BVD schemes resolve shocks better than the MUSCL scheme. Furthermore, the BVD schemes can reproduce the vortices that cannot be resolved by the MUSCL scheme, showing the characteristics of the BVD scheme. The deepMTBVD scheme successfully captures the shock wave and resolves the vortical structure, which is similar to the MUSCL-THINC-BVD scheme. \hms{Fig.~\ref{result:frontstep:celltype} demonstrates the cells that are reconstructed by the THINC scheme through the MUSCL-THINC-BVD scheme (left) and the deepMTBVD scheme (right). From it, we find that both two schemes capture the shock waves, and the results are consistent with each other. Compared with Cheng's results in \cite{RN12}, an excessive number of THINC cells were selected around the shock wave at the top of the step due to the dimension-splitting process.} Furthermore, the deepMTBVD scheme requires less total reconstruction time than the MUSCL-THINC-BVD scheme, which is shown in Table. \ref{time:frontstep}.

\subsection{Five equation model of two-component flow}
We consider the five-equation model to simulate the two-component flow. Capturing moving immiscible interfaces with minimal diffusion is essential yet challenging for numerical schemes.
The MUSCL-THINC-BVD scheme shows that the THINC reconstruction scheme can effectively prevent smearing-out of moving interface. Furthermore, the transition layer of two material interfaces can be maintained within a few mesh cells in long-time simulation \cite{RN2, DengShimizu-142, RN12, RN11}. Deng \cite{RN2} pointed out that the MUSCL-THINC-BVD scheme limits the interface to a narrow thickness and captures more accurate structures compared to higher-order polynomial reconstruction algorithms, such as WENO-JS.
Cheng \cite{RN12} extends this algorithm to the unstructured grid and expands the size of the waiting-function set, which shows excellent performance for two-phase compressible flow simulations.

In this section, we show that the proposed deepMTBVD scheme has very similar performance to the MUSCL-THINC-BVD scheme and is far better than the MUSCL scheme in resolving the sharp interface of two-phase flow. Besides, the numerical stability of deepMTBVD in multi-phase flow is better than that of the BVD scheme. \hms{Same to single-fluid flow, we set \kaparef = 0.45 for following cases.}

\begin{figure}[htbp!]
    \centering
    \includegraphics[width=0.4\textwidth]{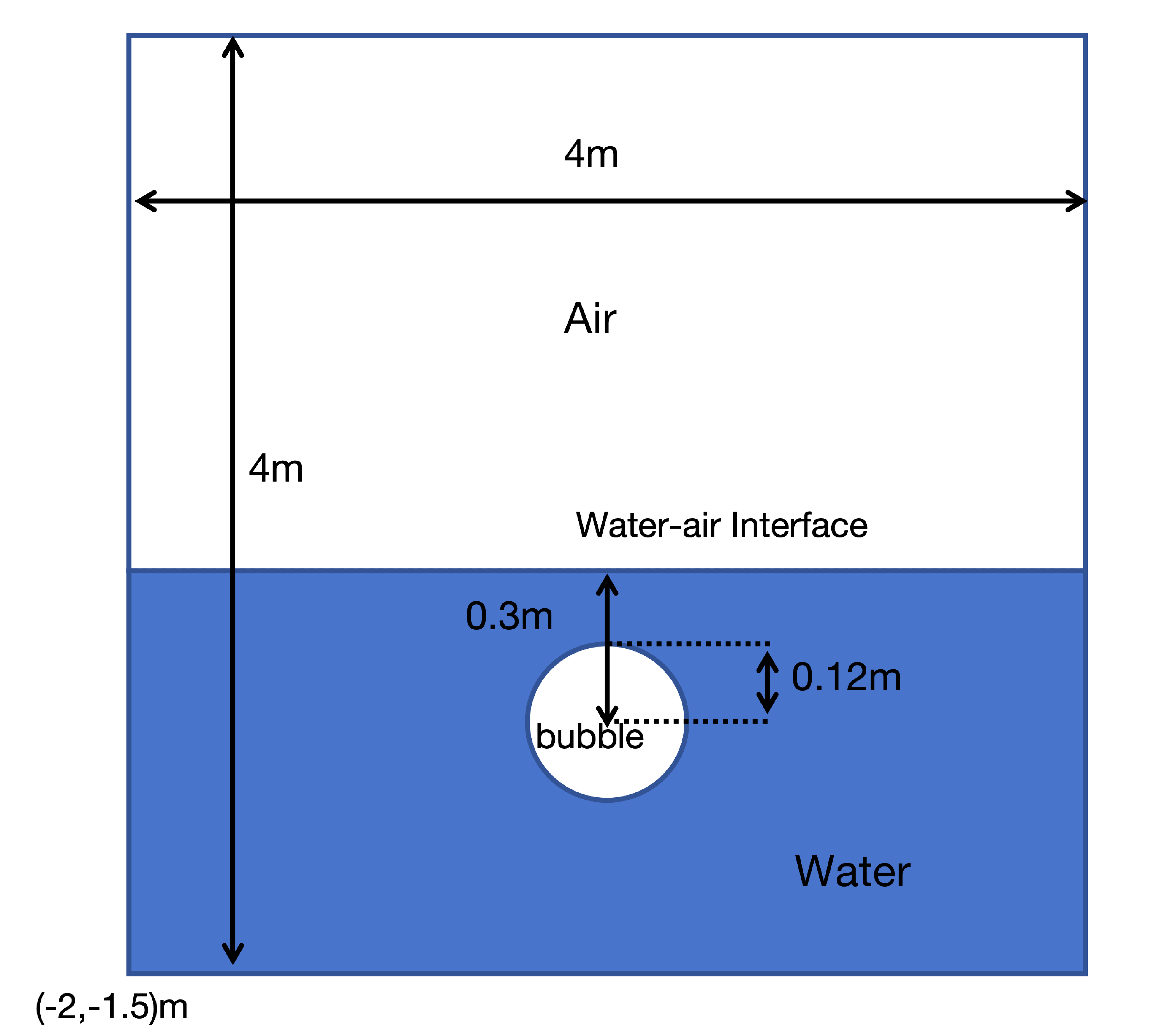}
    \caption{The schematic diagram of the computation region of example \ref{result:underwater:title}. }
    \label{result:example5:computation_domain}
\end{figure}

\subsubsection{Two-dimensional Underwater Explosion Test}\label{result:underwater:title}
We consider the underwater explosion problem which has been widely used to test numerical method \cite{RN7, RN6, RN2, RN19}. 
The initial high-pressure air cavity region interacts with the planar water-air interface above it. The computation domain is $[-2,2]\text{ m} \times [-1.5,2.5]$ m. The diameter of cylindrical air cavity is $0.24$ m which initially centered at $(0, -0.3)$ m with high pressure $10^{9}$ Pa and high density $1250\text{ kg}/\text{m}^{3}$. The planar water-air interface is located at $y = 0$. Both water and air mediums are under standard atmospheric conditions and stay stationary initially. Under this assumption, The density of pure water is $1000\text{ kg}/\text{m}^{3}$ and pure air is $1\text{ kg}/\text{m}^{3}$. In water medium, the volume fraction of air $\alpha_{1} = 10^{-8}$ and vice versa in air. Both water and air use stiffened gas EOS to describe their thermodynamic behavior. We use stiffened gas equation of state to model the thermodynamic behavior of corresponding liquid and gas where the $\gamma_{1} = 4.4, p_{\infty,1} = 6 \times 10^{8}$ Pa, $e_{\infty, 1} = 0$ for liquid phase and $\gamma_{2} = 1.4, p_{\infty, 2} = 0$ Pa$, e_{\infty, 2} = 0$ for gas phase respectively.
The transparent boundary condition is imposed on the left, right, and top boundaries, while the reflective boundary condition is applied on the bottom boundary. The mesh number is set to $600 \times 600$. 

\begin{figure}[ht!]
\centering
\subfloat{\includegraphics[width=0.31\textwidth]{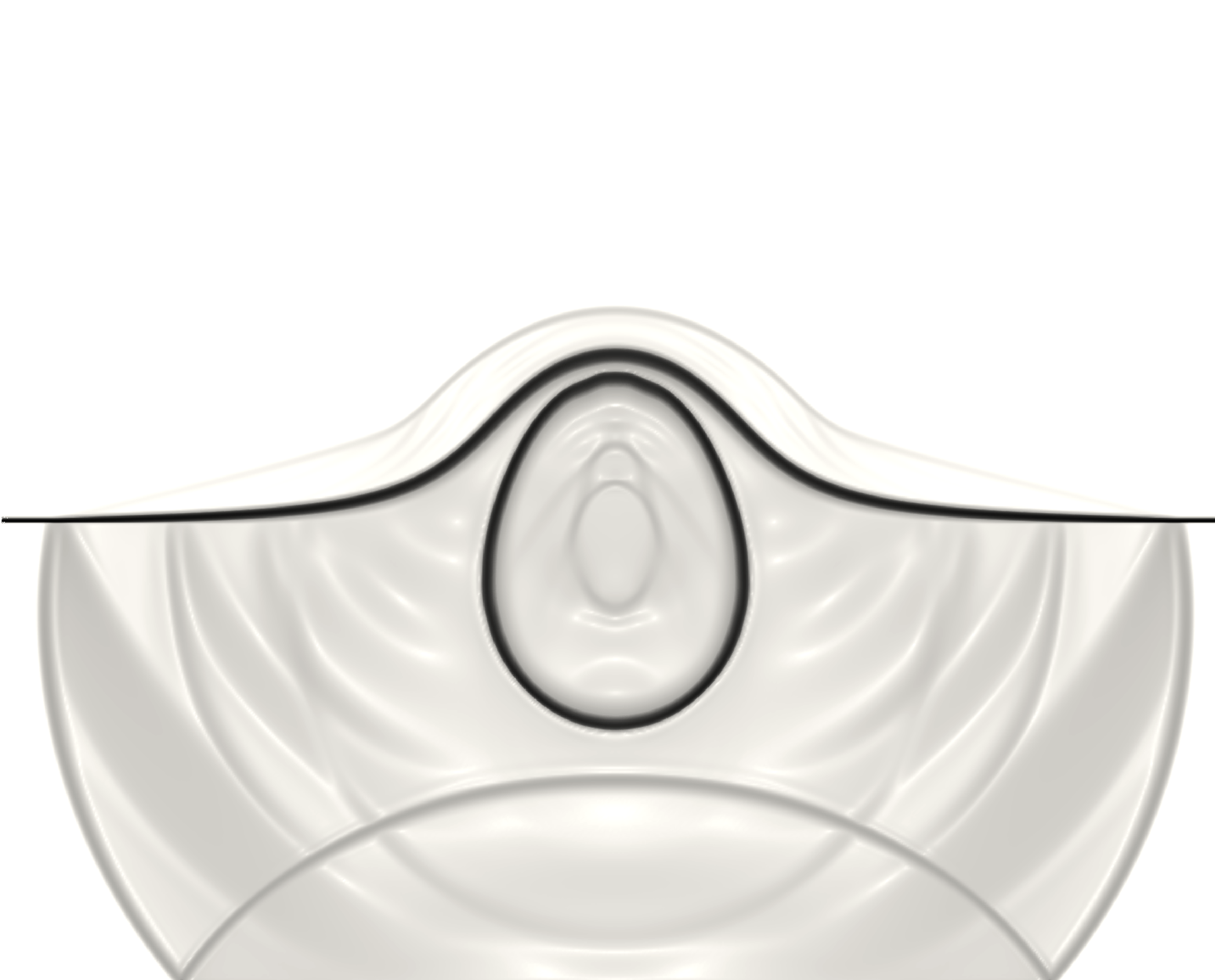}}
\subfloat{\includegraphics[width=0.31\textwidth]{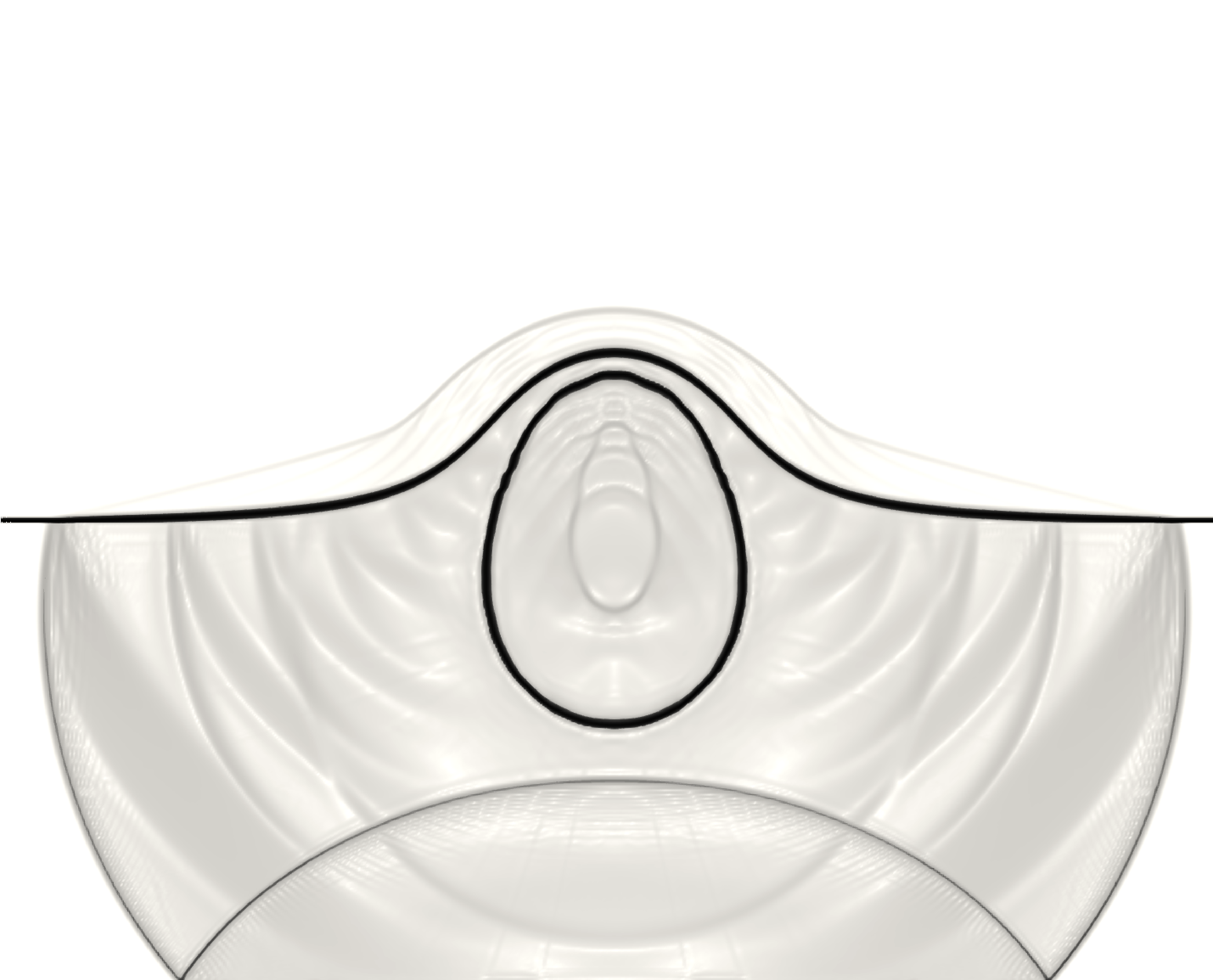}}
\subfloat{\includegraphics[width=0.31\textwidth]{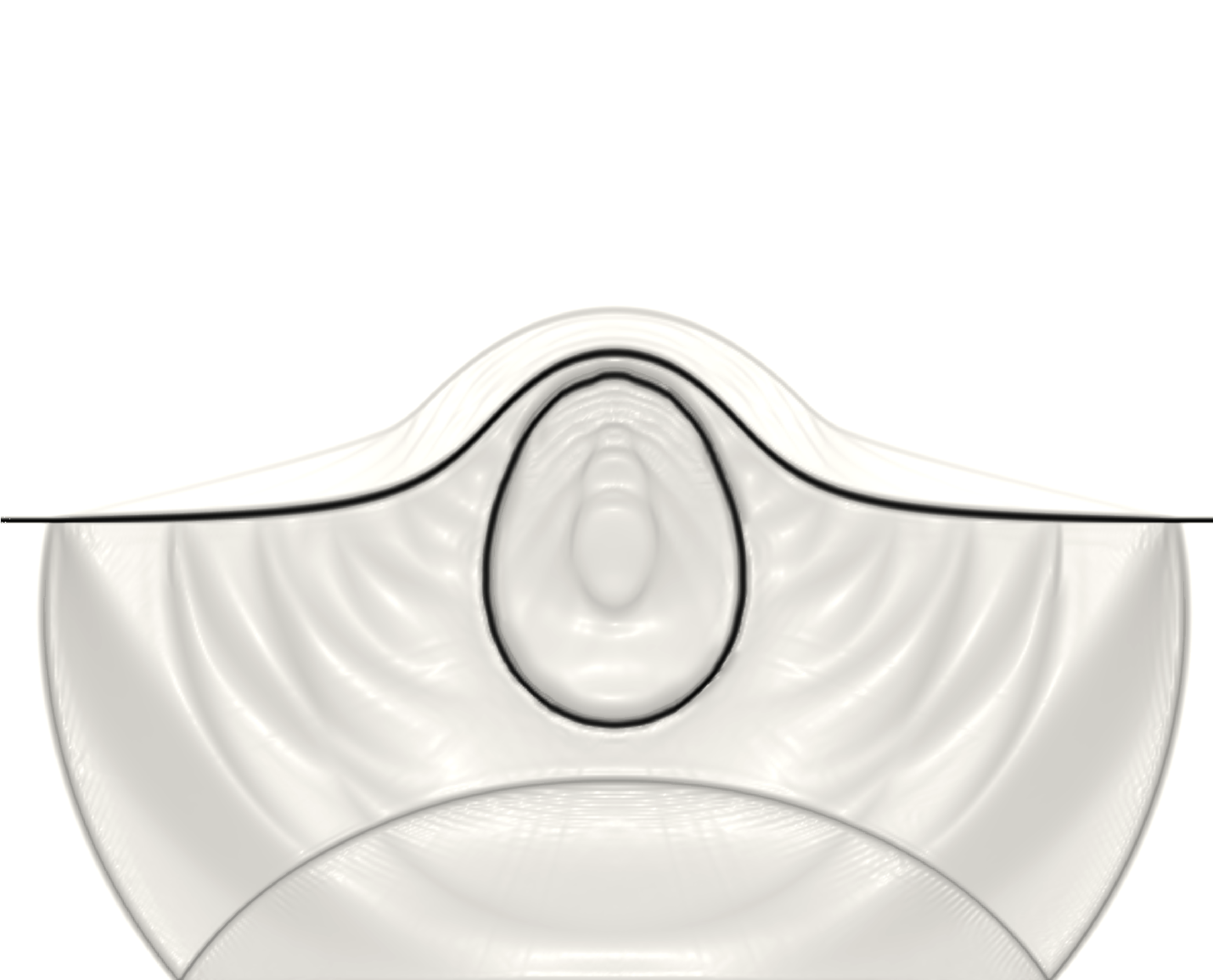}} \\
\subfloat{\includegraphics[width=0.31\textwidth]{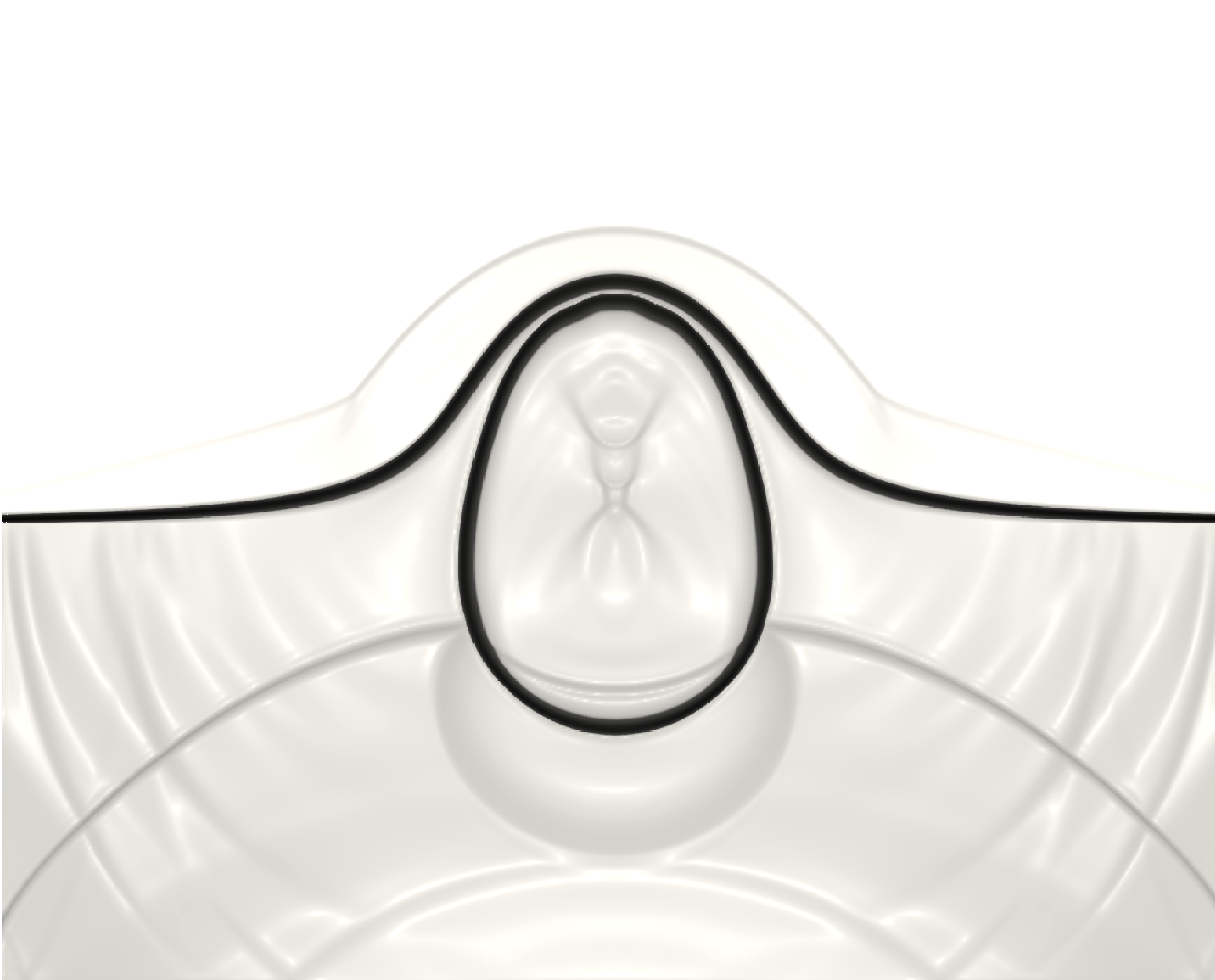}}
\subfloat{\includegraphics[width=0.31\textwidth]{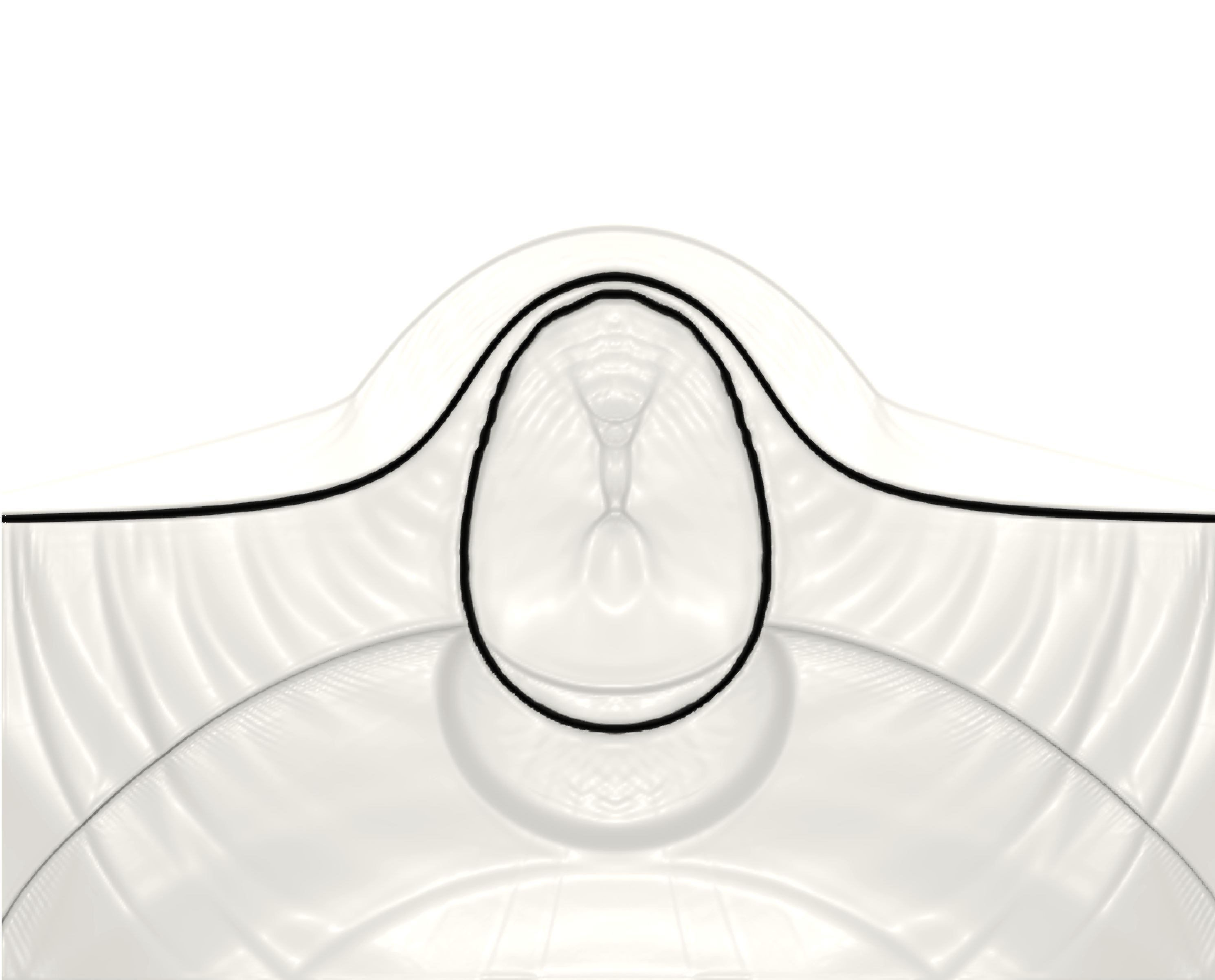}}
\subfloat{\includegraphics[width=0.31\textwidth]{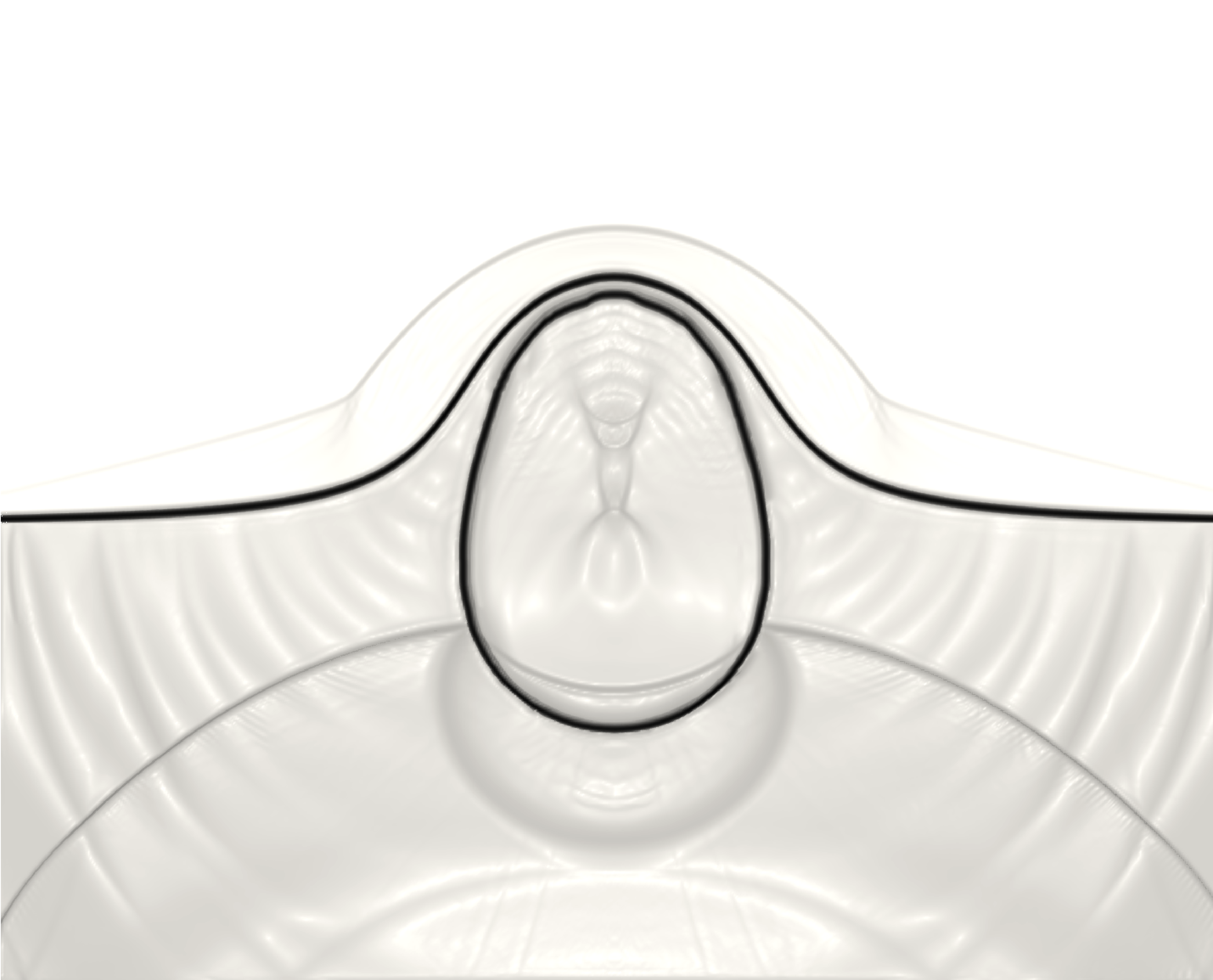}}\\
\subfloat{\includegraphics[width=0.31\textwidth]{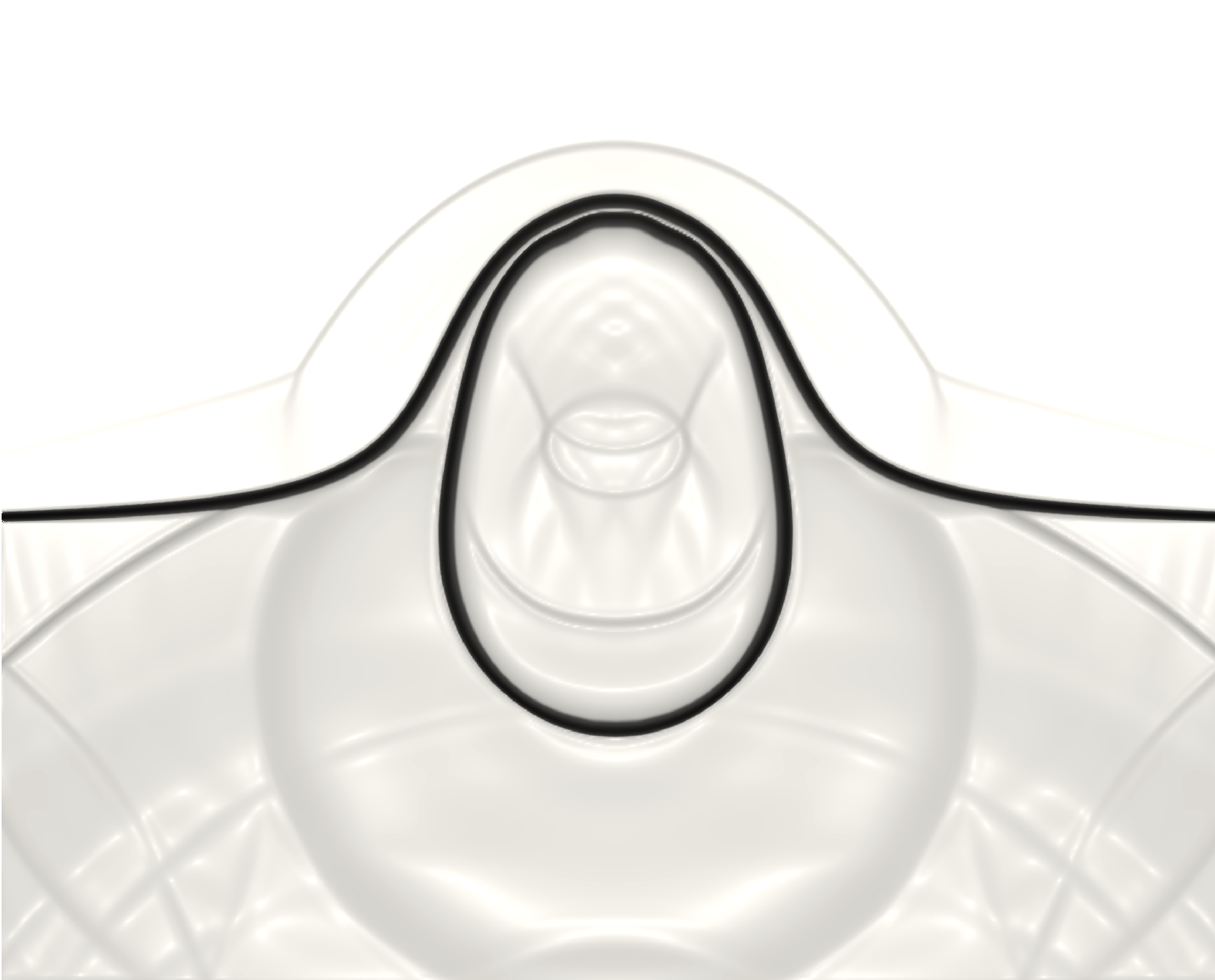}}
\subfloat{\includegraphics[width=0.31\textwidth]{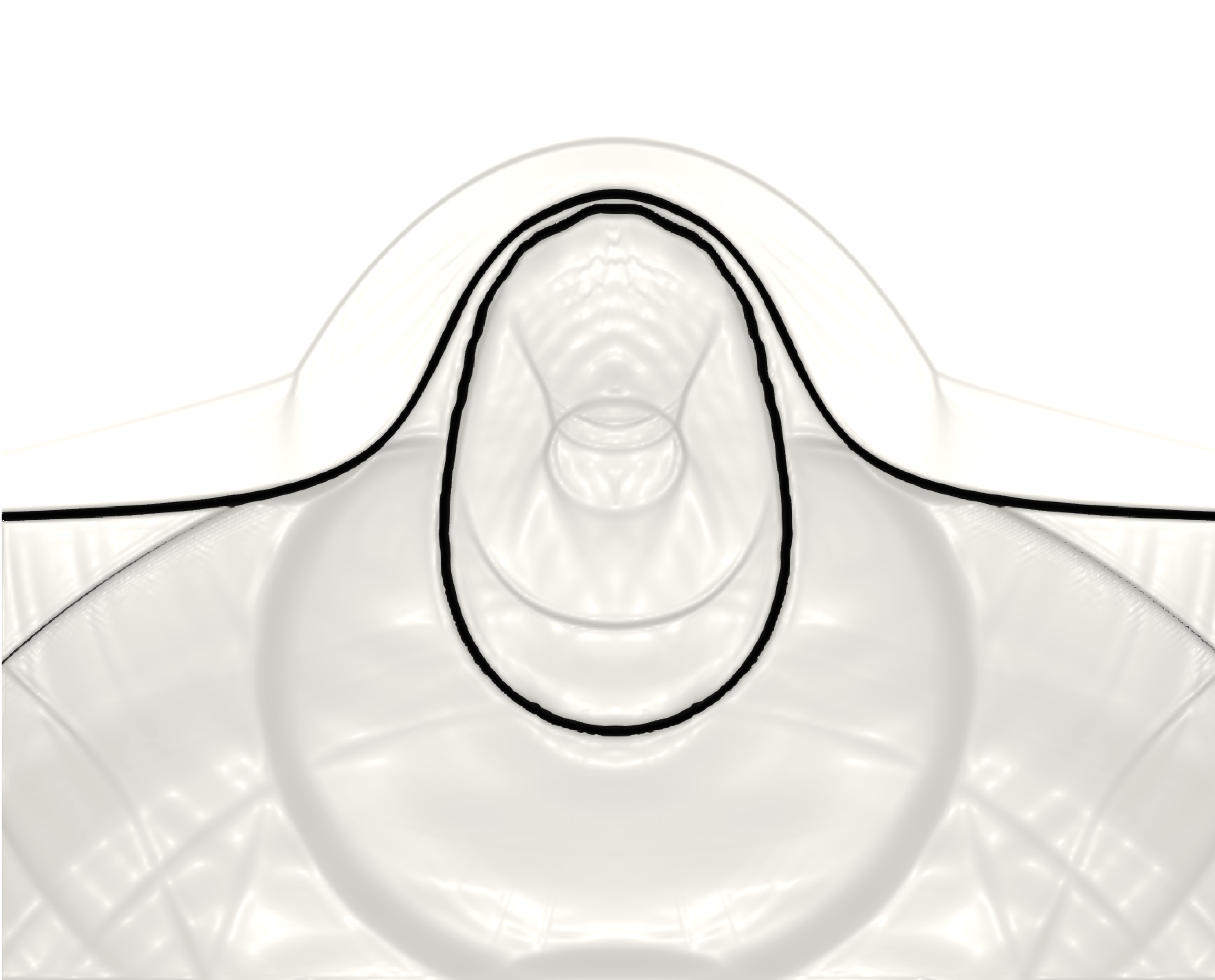}}
\subfloat{\includegraphics[width=0.31\textwidth]{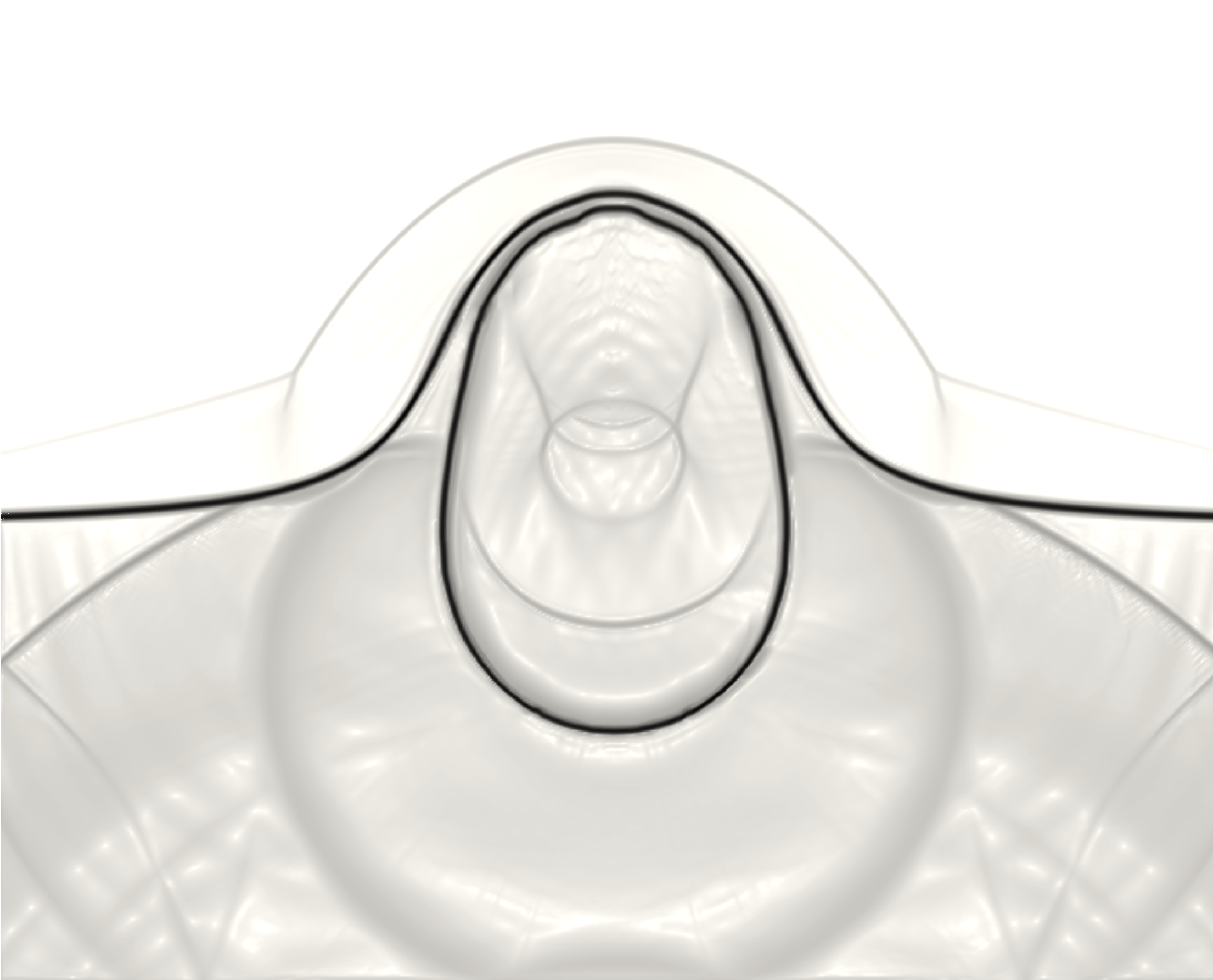}}\\
\subfloat{\includegraphics[width=0.31\textwidth]{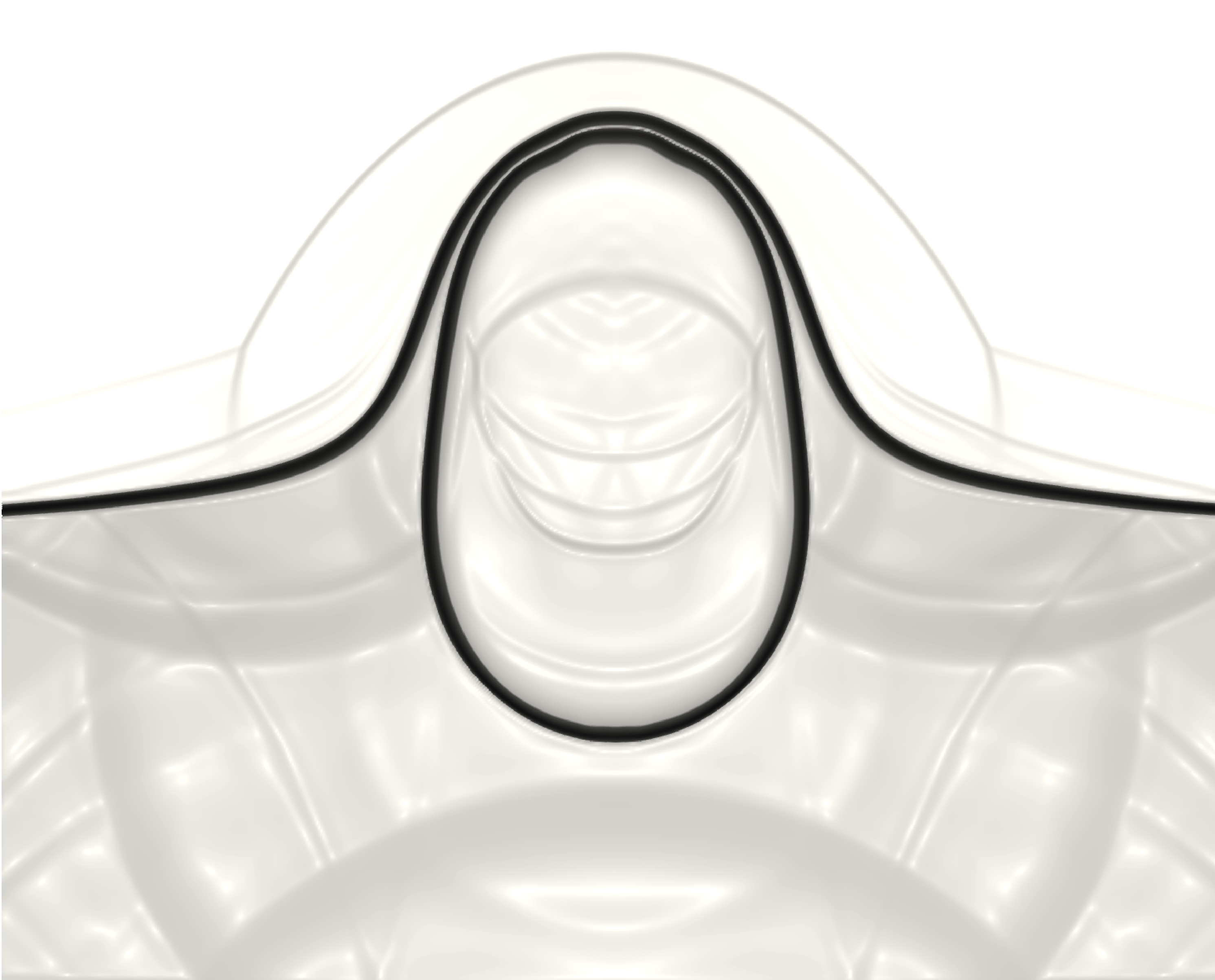}}
\subfloat{\includegraphics[width=0.31\textwidth]{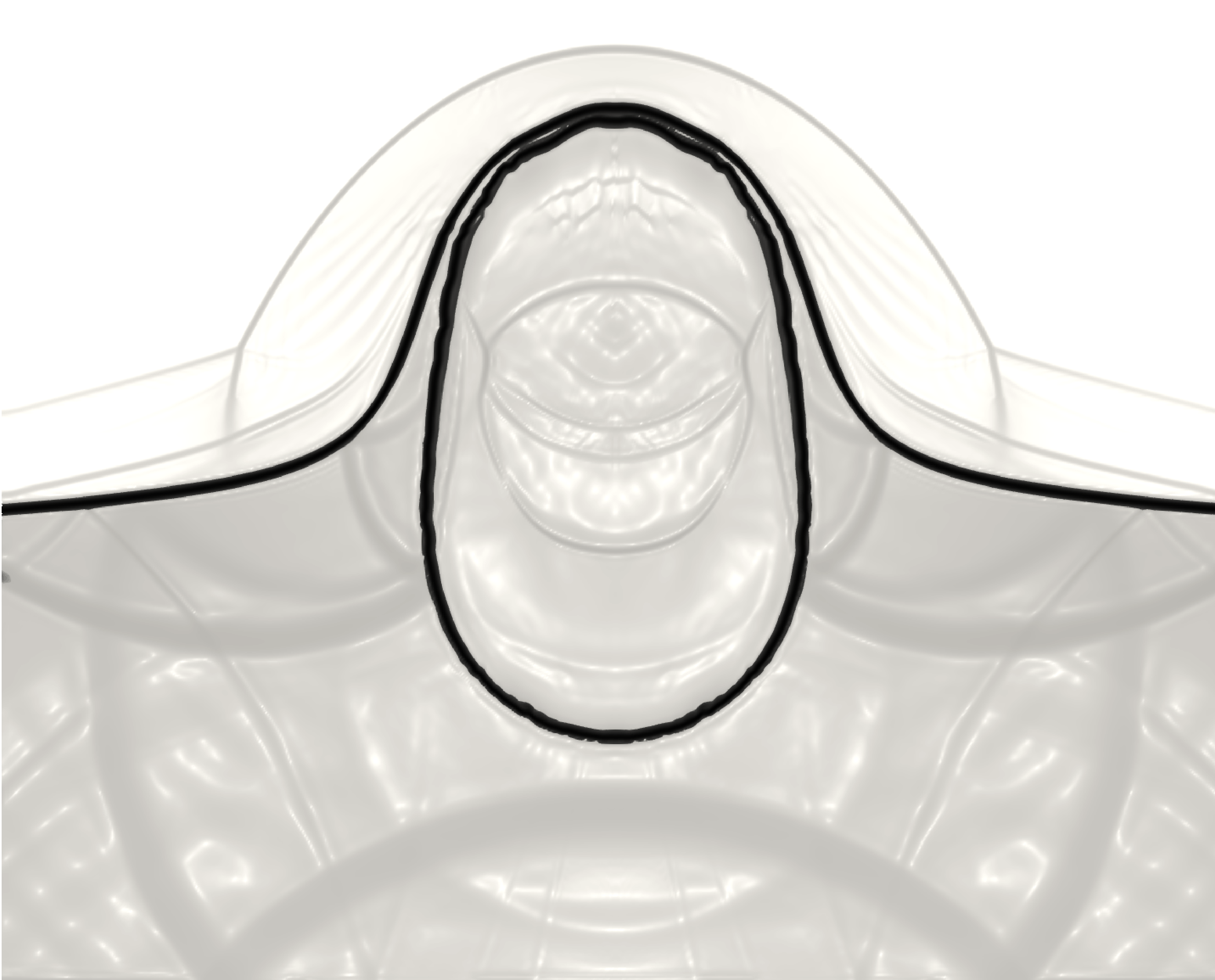}}
\subfloat{\includegraphics[width=0.31\textwidth]{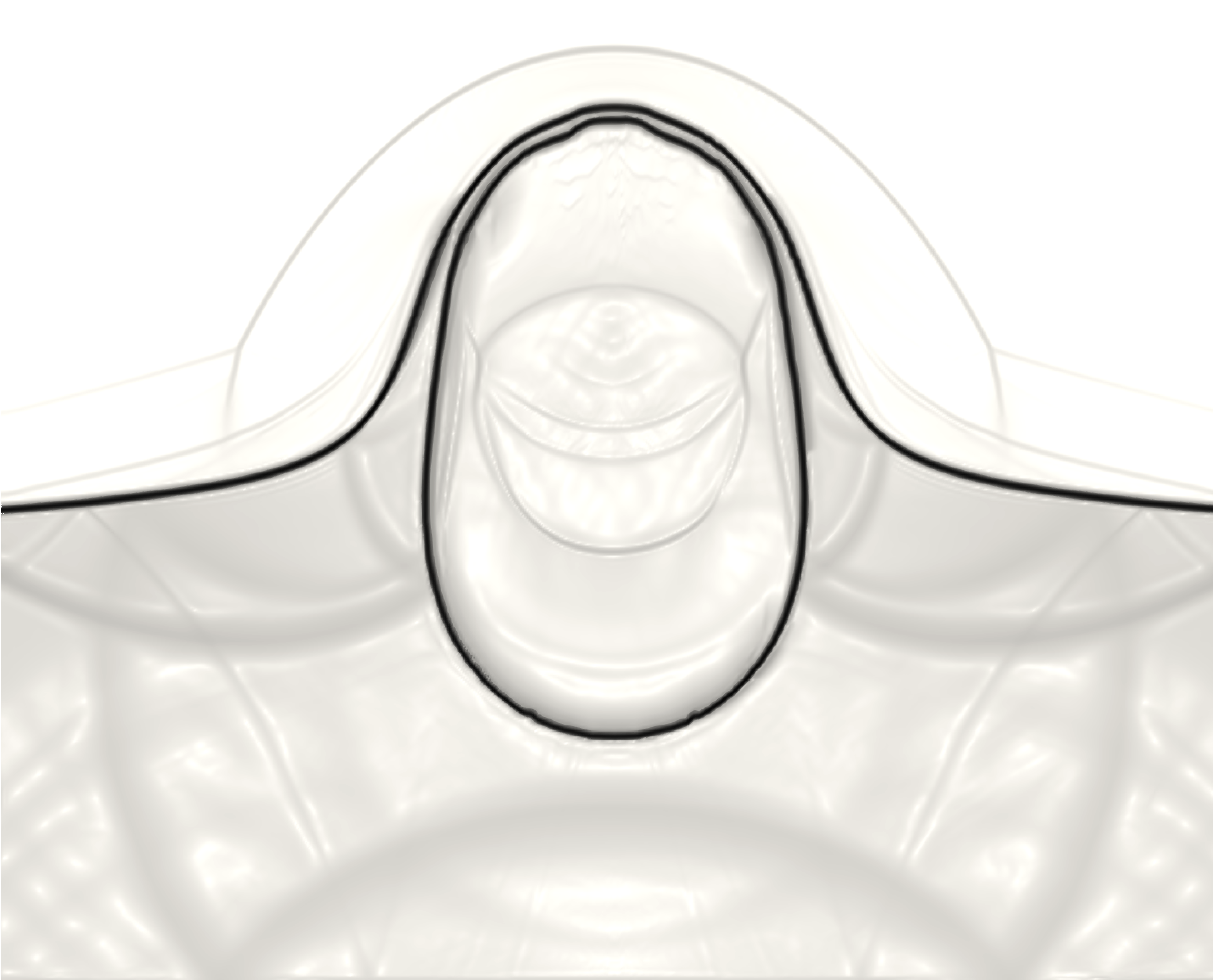}}
\caption{Schlieren figures of density for underwater explosion test at time t = $0.95$, $1.26$ , $1.58$ and $1.90$ ms from the top to bottom. Left column: the numerical results of the MUSCL scheme. Middle column: the numerical results of MUSCL-THINC-BVD scheme. Right column: the numerical results of the deepMTBVD. }
\label{result:example5:density}
\end{figure}

In Fig.~\ref{result:example5:density}, the numerical Schlieren figures for mixture density carried from different schemes at several instants are displayed. From the numerical results, all three schemes show circular shocks forming at the beginning and propagating outwards. At the same time, converging rarefaction waves are formed inside the air cavity. A weak shock is transmitted into the air since circular shock interacts with the water-air interface. The left and middle columns are the MUSCL scheme and MUSCL-THINC-BVD scheme respectively, and the far right is deepMTBVD. The MUSCL scheme can not prevent the water-air interface from smearing out. The MUSCL-THINC-BVD scheme captures the water-air interface sharply well, and the novel deepMTBVD scheme produces competitive results.

\begin{figure}[ht]
\centering
\subfloat[t = 0.20ms]{\includegraphics[width=0.31\textwidth]{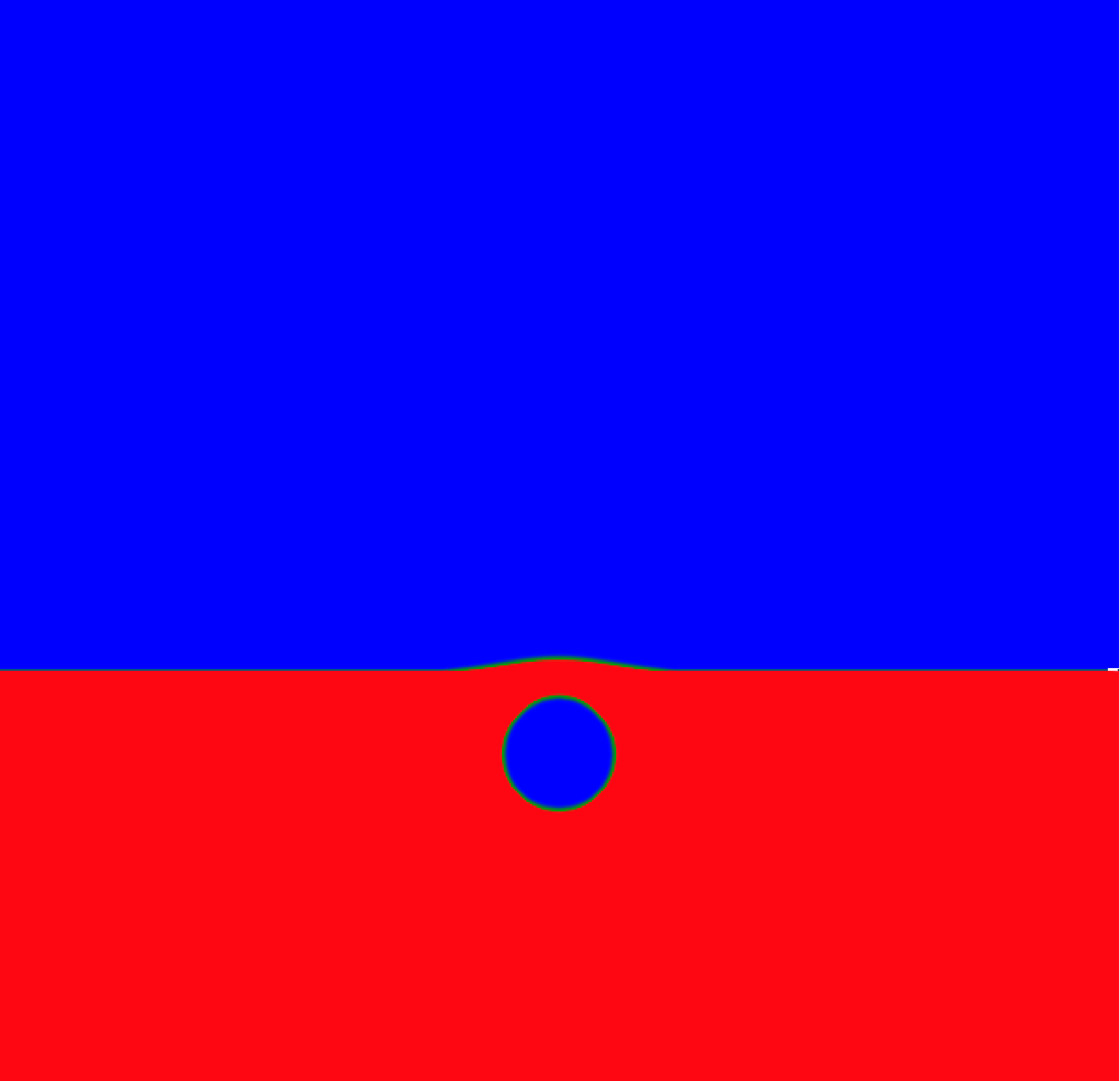}}
\subfloat[t = 0.79ms]{\includegraphics[width=0.31\textwidth]{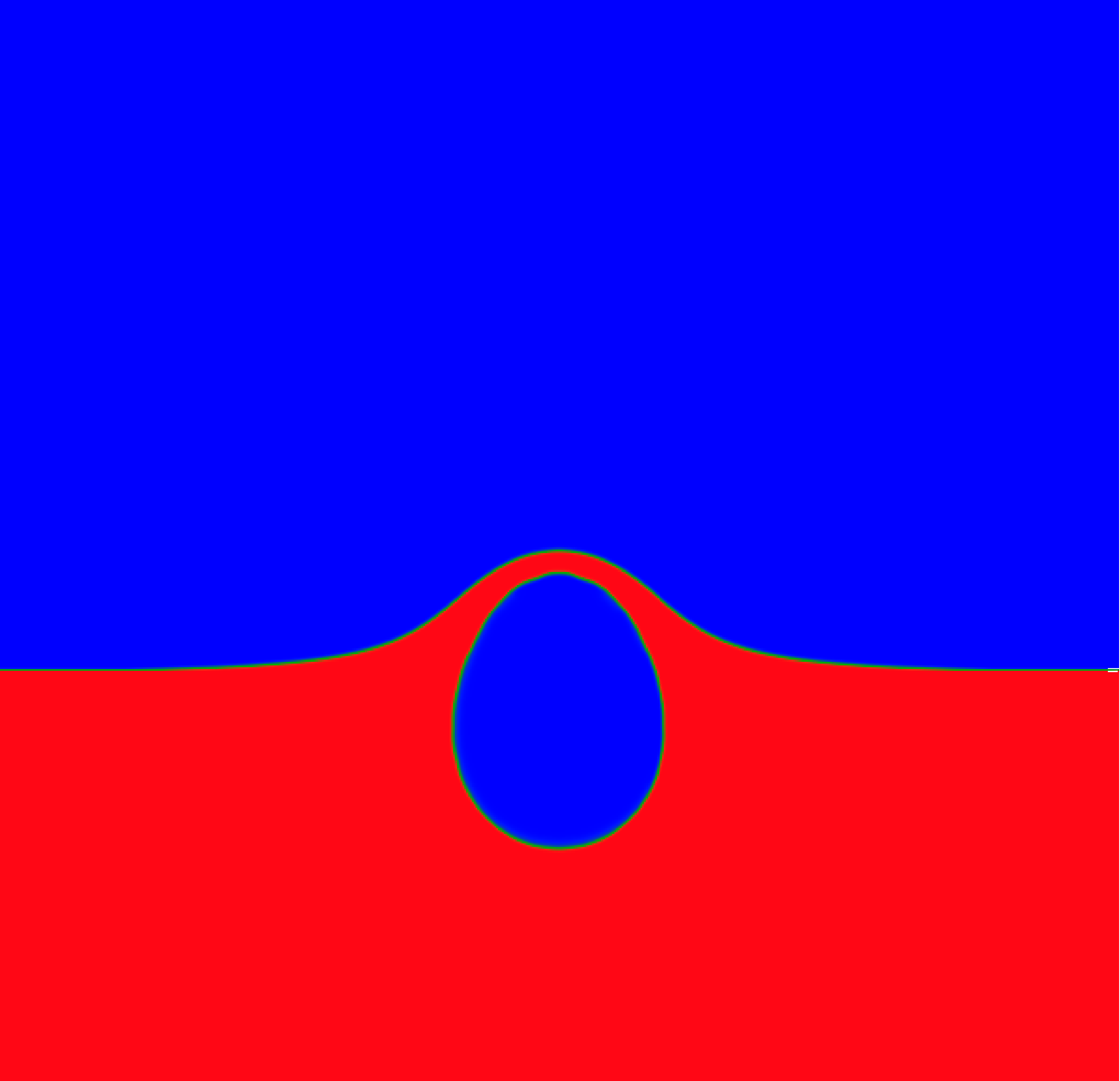}}
\subfloat[t = 0.95ms]{\includegraphics[width=0.31\textwidth]{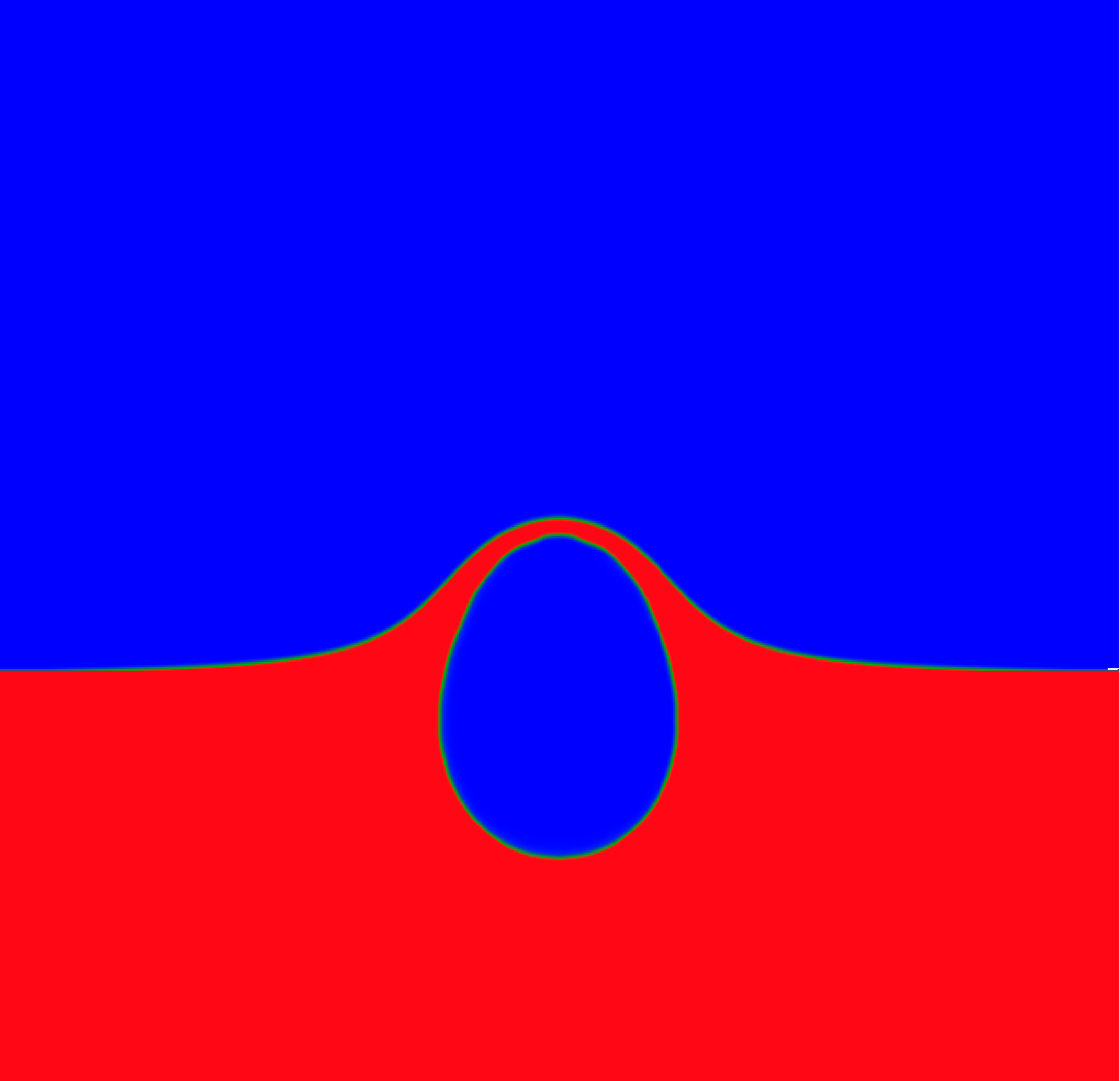}}\\
\subfloat[t = 1.26ms]{\includegraphics[width=0.31\textwidth]{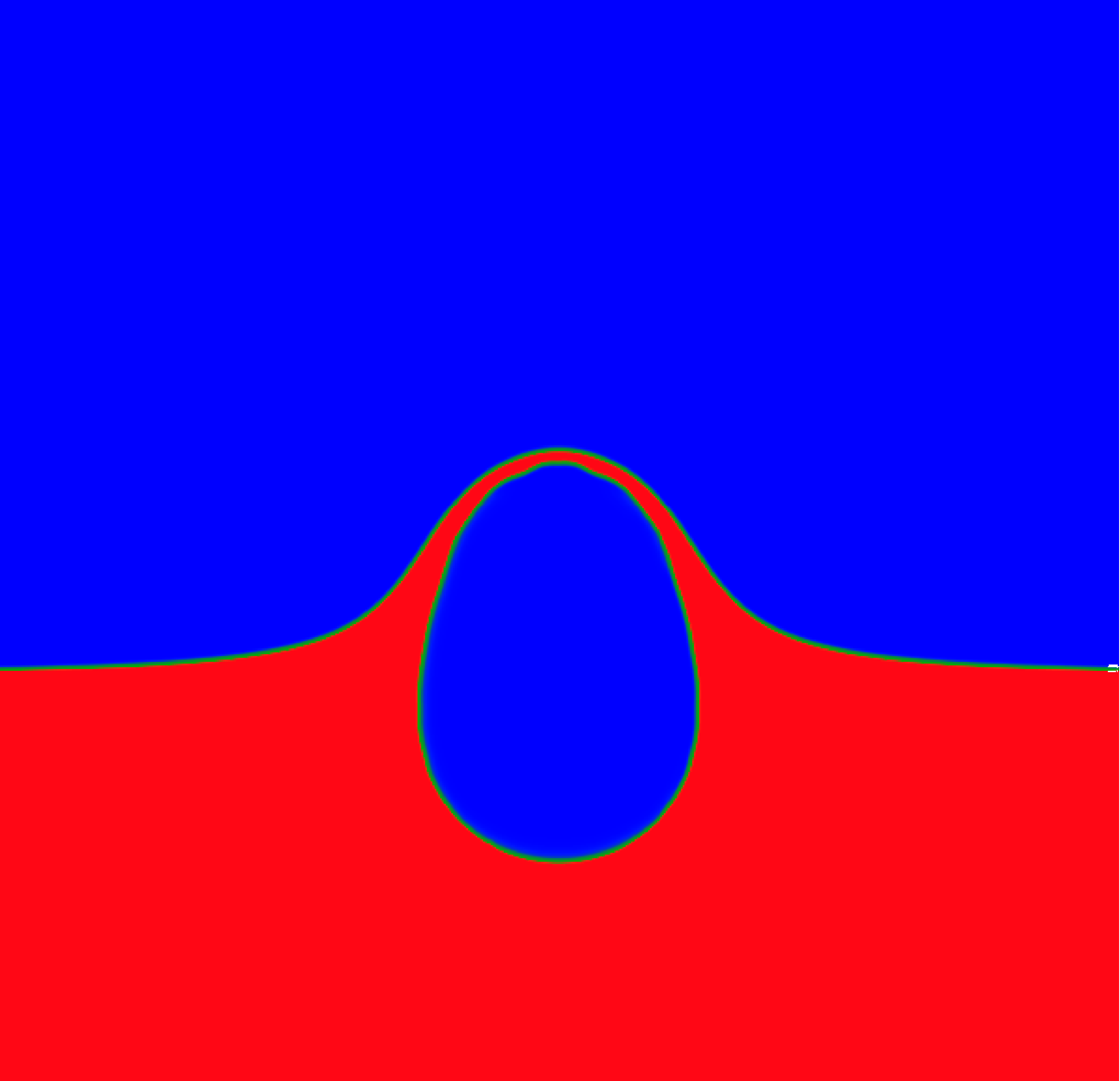}}
\subfloat[t = 1.58ms]{\includegraphics[width=0.31\textwidth]{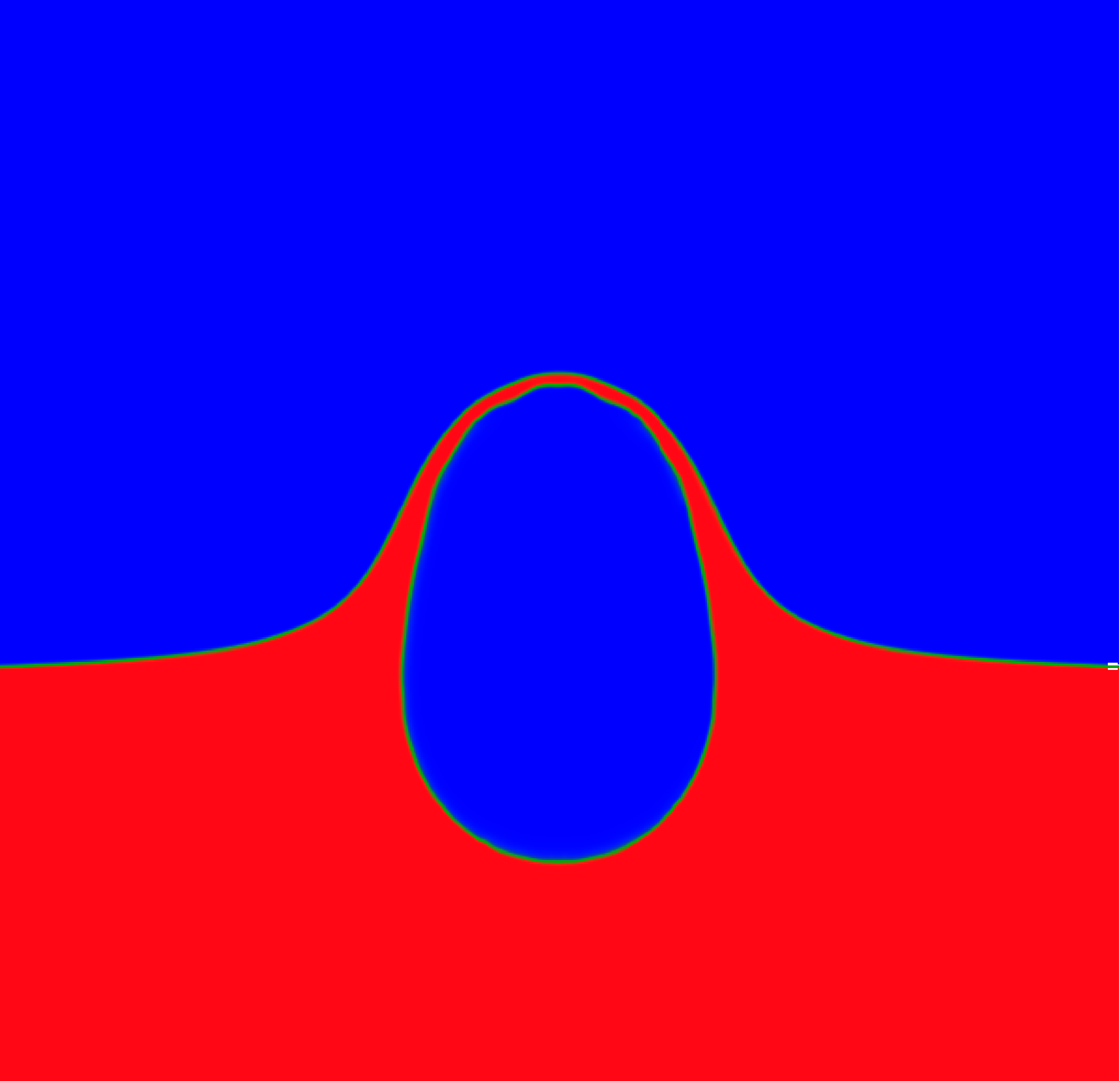}}
\subfloat[t = 1.90ms]{\includegraphics[width=0.31\textwidth]{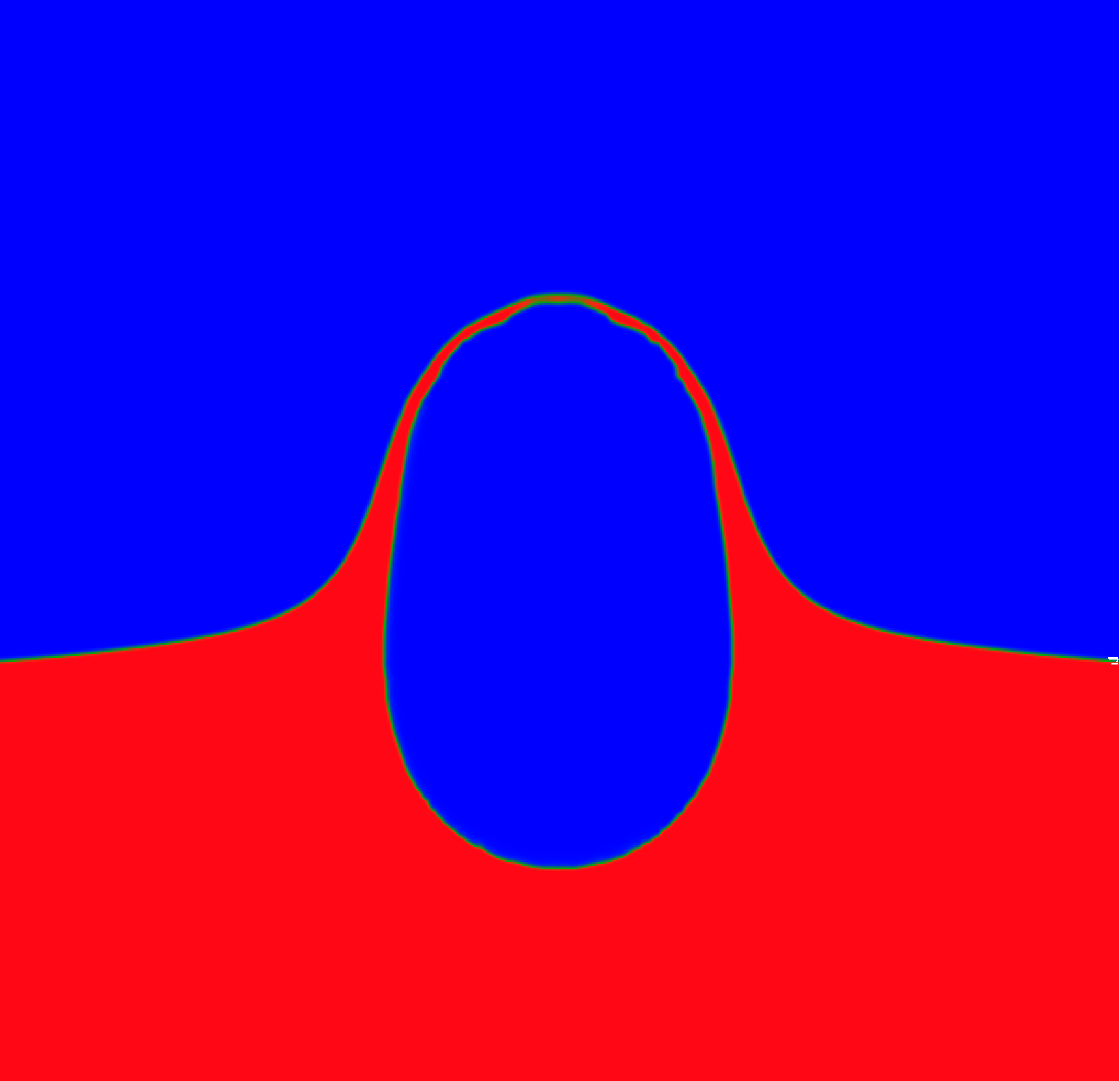}}
\caption{Volume fraction contour at different times $t=0.2, 0.79, 0.95, 1.26, 1.58$ and $1.9$ ms for underwater explosion test by the deepMTBVD on a uniform grid with 600$\times$600 cells.}
\label{result:example5:volume_fraction}
\end{figure}

The contour of volume fraction $\alpha_{1}$ is presented in Fig.~\ref{result:example5:volume_fraction} at several time instants. The air cavity gradually expands from a circle to an oval-like shape, which agrees with previous works \cite{RN11, RN2}. The circular shock is diffracted through the water-air interface and generates a water bridge between the air bubble and ambient air above the interface. It is well-known that resolving the thin water bridge sharply is quite challenging for other exiting methods \cite {RN7}. From Fig.~\ref{result:example5:volume_fraction}, we find that the deepMTBVD not only resolves the water bridge sharply and describes the changes from a circle-like bubble to an oval-like bubble.  

\begin{figure}[ht]
\centering
\subfloat[Compare density with Shukla's result \cite{RN6}]{\includegraphics[width=0.48\textwidth]{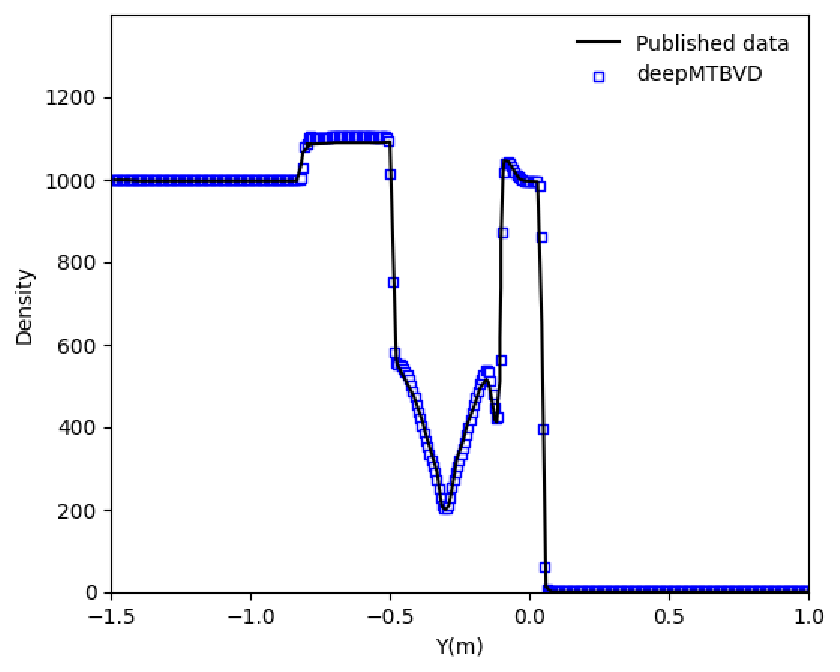}}   
\subfloat[Compare density with Hu's result \cite{RN16}]{\includegraphics[width=0.48\textwidth]{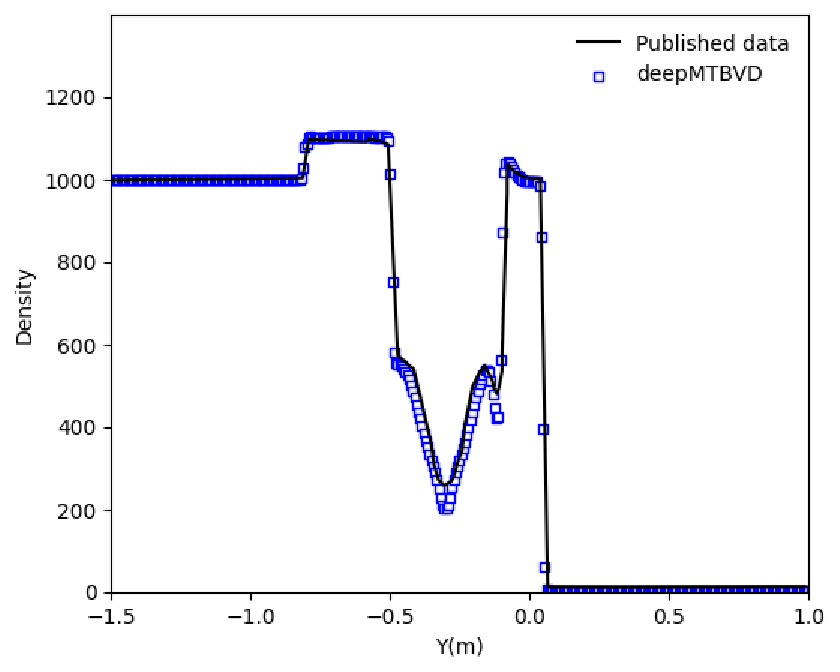}}
\caption{Density distribution along $x = 0$ cross-section for underwater explosion test at time t = 0.2ms compared with Shukla's (left)  and Hu's (right)  results. }
\label{result:example5:comparison_density}
\end{figure}

As a quantitative comparison, we plot the density field along the $x = 0$ cross-section in Fig.~\ref{result:example5:comparison_density}. Compared with other works by Shukla \cite{RN6} (left) and Hu \cite{RN16} (right), we find our results agree with those existing results in Fig.~\ref{result:example5:comparison_density}. Furthermore, we use coarser grid size $h = \frac{1}{150}$ than Shukla \cite{RN6}.

\subsubsection{Two-dimensional Shock–R22-cylinder Interaction} \label{result:R22:title}


As a well-known benchmark test \cite{RN11, RN18, RN23, RN24}, we consider the interaction between a shock wave of
Mach 1.22 in air and a cylindrical bubble of refrigerant-22 (R22) gas. As Fig.~11 in physical experiments \cite{RN18}, the R22 bubble is deformed due to moving post-shock waves. \rewb{According to the definition of vorticity provided in Eqn.~(1) of \cite{RN23}, the deformation arises from the interaction between the refracted and diffracted waves at the bubble boundary. The resulting vortical structures are generated by the baroclinic force as the shock wave traverses the R22 bubble, forming a vortex chain along the material interface. However, if the numerical scheme exhibits high dissipation, these vortices may be significantly suppressed or diminished unless a high-resolution computational mesh is employed.}
\begin{figure}[htbp]
    \centering
    \includegraphics[width=0.7\textwidth]{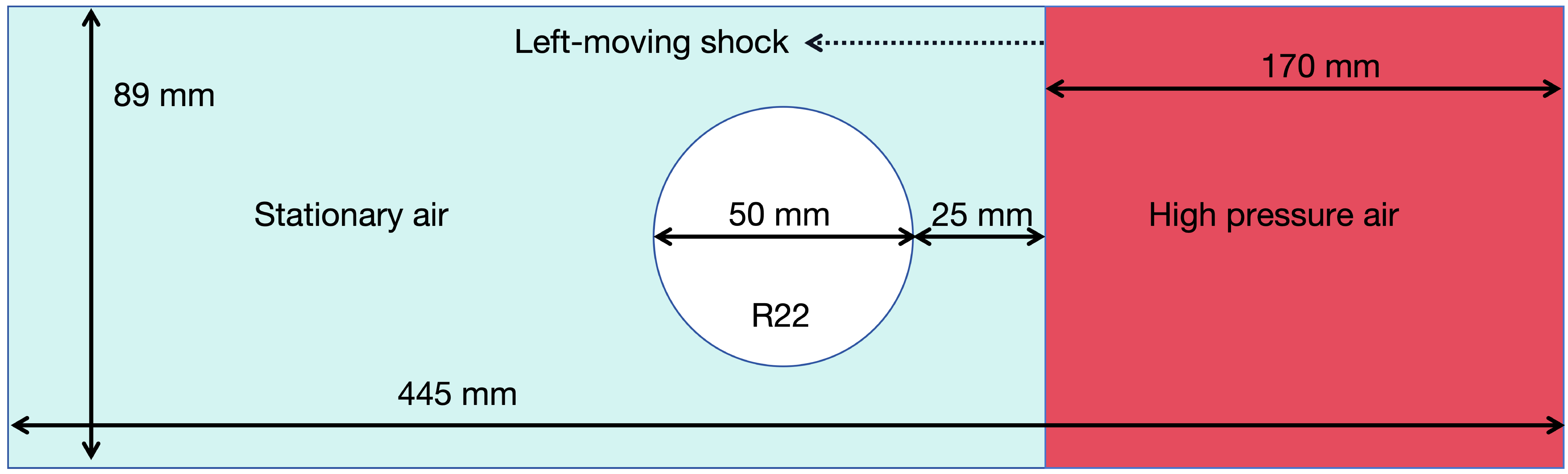}
    \caption{The schematic diagram for computational domain of R22-shock interaction in Example 6.}
    \label{result:example6:computational_domain}
\end{figure}
Initially, the computation region is set as shown in Fig.~\ref{result:example6:computational_domain}. 
There is a planar leftward-moving Mach 1.22 shock wave in air located at $x=275$ mm, moving towards a stationary R22 bubble centered at $(225, 44.5)$ mm with radius $r = 25$ mm. Both R22 (material 1) and air(material 2) are assumed to be perfect ideal gas with $\gamma_{1} = 1.249, p_{\infty, 1} = 0, e_{\infty, 1} = 0$ and $\gamma_{2} = 1.4, p_{\infty, 2} = 0, e_{\infty, 2} = 0$, respectively. Inside the R22 gas bubble, the  state variables are
\begin{equation*}
    (\rho_{1}, \rho_{2}, u_{1}, u_{2}, p, \alpha_{1}) = (3.869 \text{ kg}/\text{m}^{3}, 1.225 \text{ kg}/\text{m}^{3}, 0.0, 0.0, 1.01325 \times 10^{5} \text{ Pa}, 1.0 - \epsilon),
\end{equation*}
while outside the bubble, the parameters in the pre-shock are 
\begin{equation*}
    (\rho_{1}, \rho_{2}, u_{1}, u_{2}, p, \alpha_{1}) = (3.869 \text{ kg}/\text{m}^{3}, 1.225 \text{ kg}/\text{m}^{3}, 0.0, 0.0, 1.01325 \times 10^{5} \text{ Pa}, \epsilon),
\end{equation*}
and the corresponding parameters in the post-shock are
\begin{equation*}
    (\rho_{1}, \rho_{2}, u_{1}, u_{2}, p, \alpha_{1}) = (3.869 \text{ kg}/\text{m}^{3}, 1.686 \text{ kg}/\text{m}^{3}, -113.5\text{ m}/\text{s}, 0.0, 1.59 \times 10^{5} \text{ Pa}, \epsilon),
\end{equation*}
where $\epsilon = 10^{-8}$. In order to resolve the interface clearly, the mesh size is $\Delta x= \Delta y = \frac{1}{8}$ mm which corresponds to a grid-resolution of 400 cells across the bubble diameter \cite{RN2}. Inflow and outflow conditions are imposed on the left and right boundaries respectively. Reflective wall boundary conditions are implemented to the top and bottom boundaries. 

\begin{figure}[htbp!]
\centering
\subfloat[$t = 247 \mu s$]{\includegraphics[width=0.49\textwidth]{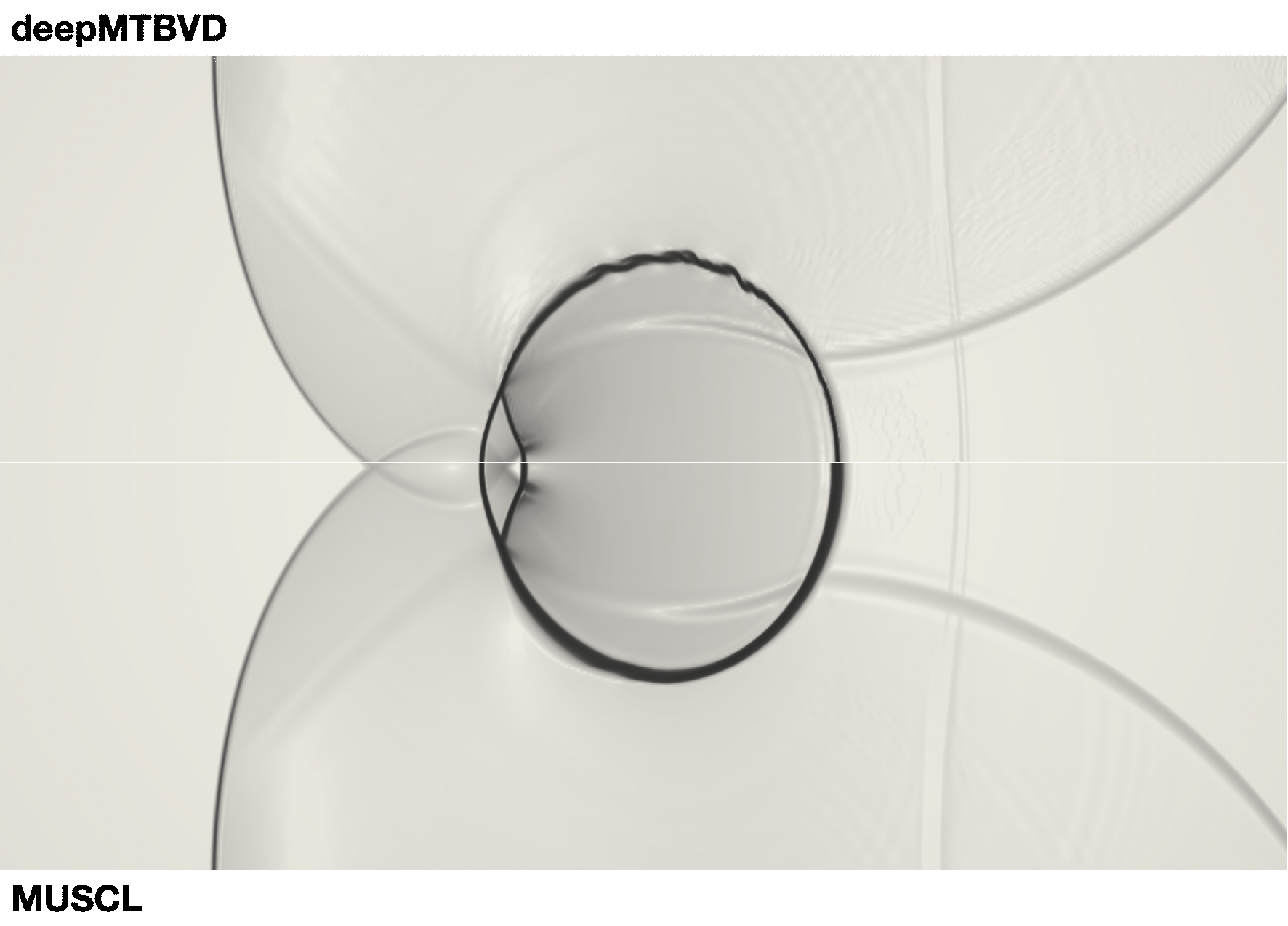}}
\subfloat[$t = 247 \mu s$]{\includegraphics[width=0.49\textwidth]{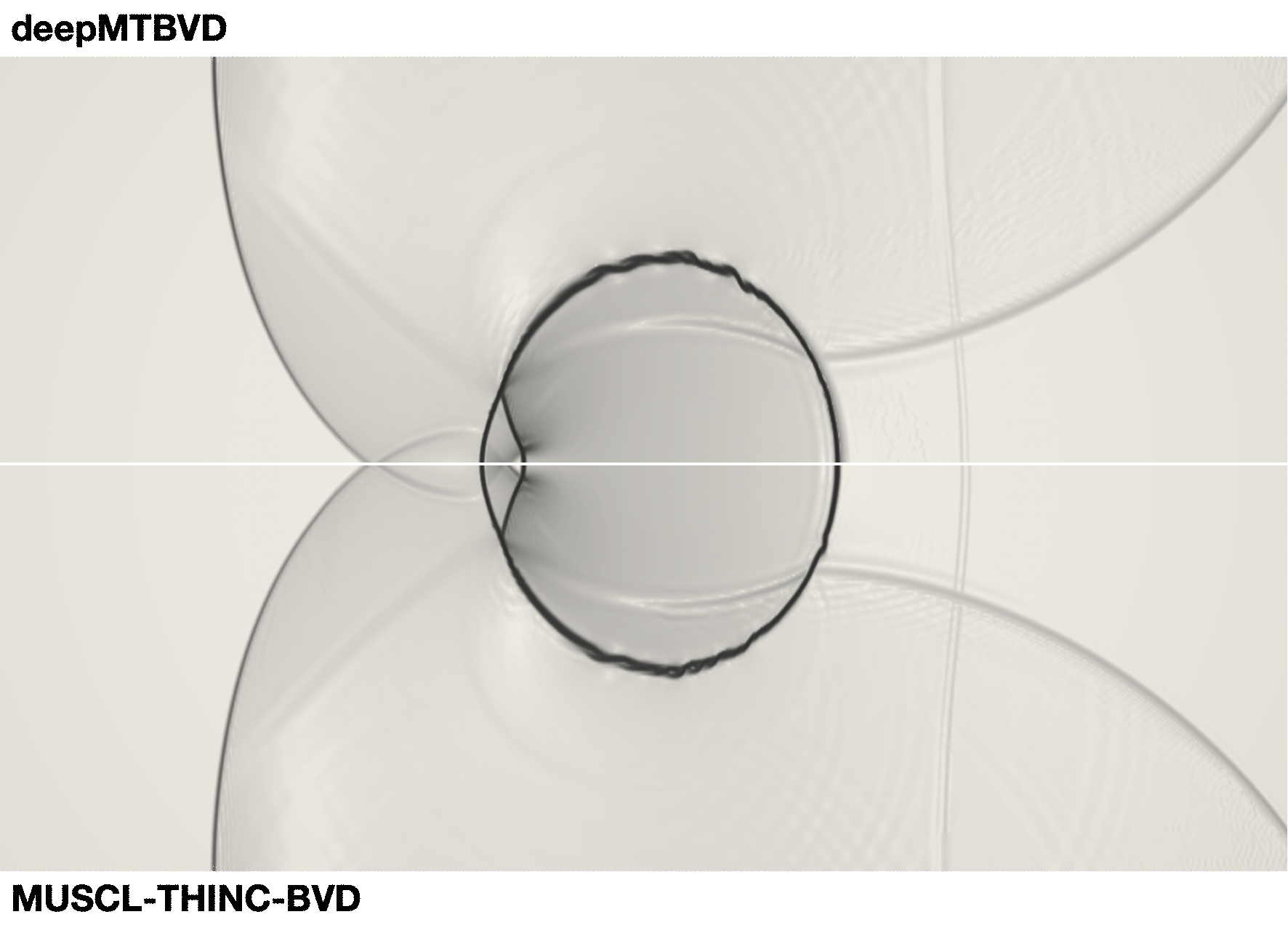}}\\
\subfloat[$t = 293 \mu s$]{\includegraphics[width=0.49\textwidth]{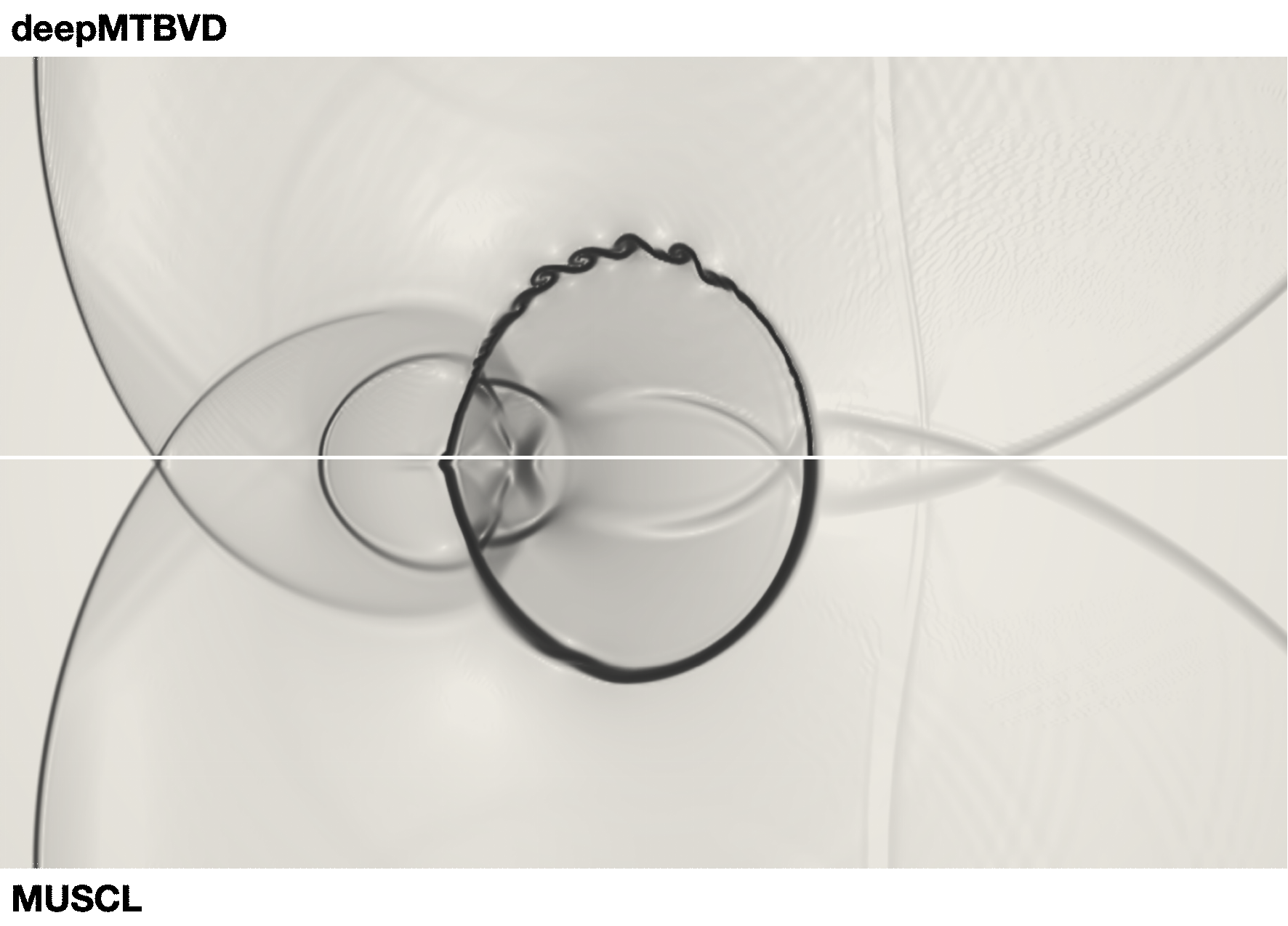}}
\subfloat[$t = 293 \mu s$]{\includegraphics[width=0.49\textwidth]{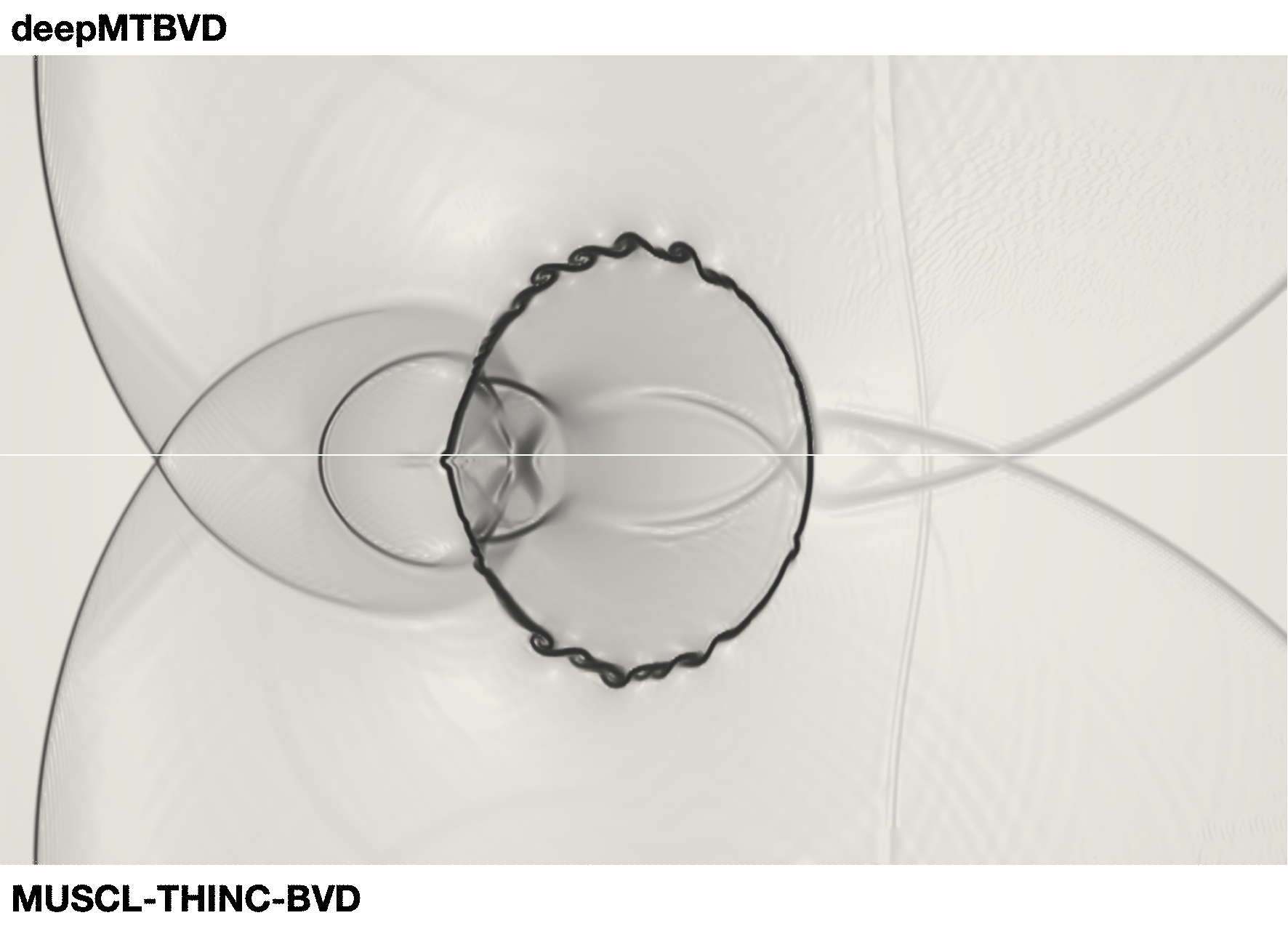}}\\
\subfloat[$t = 318 \mu s$]{\includegraphics[width=0.49\textwidth]{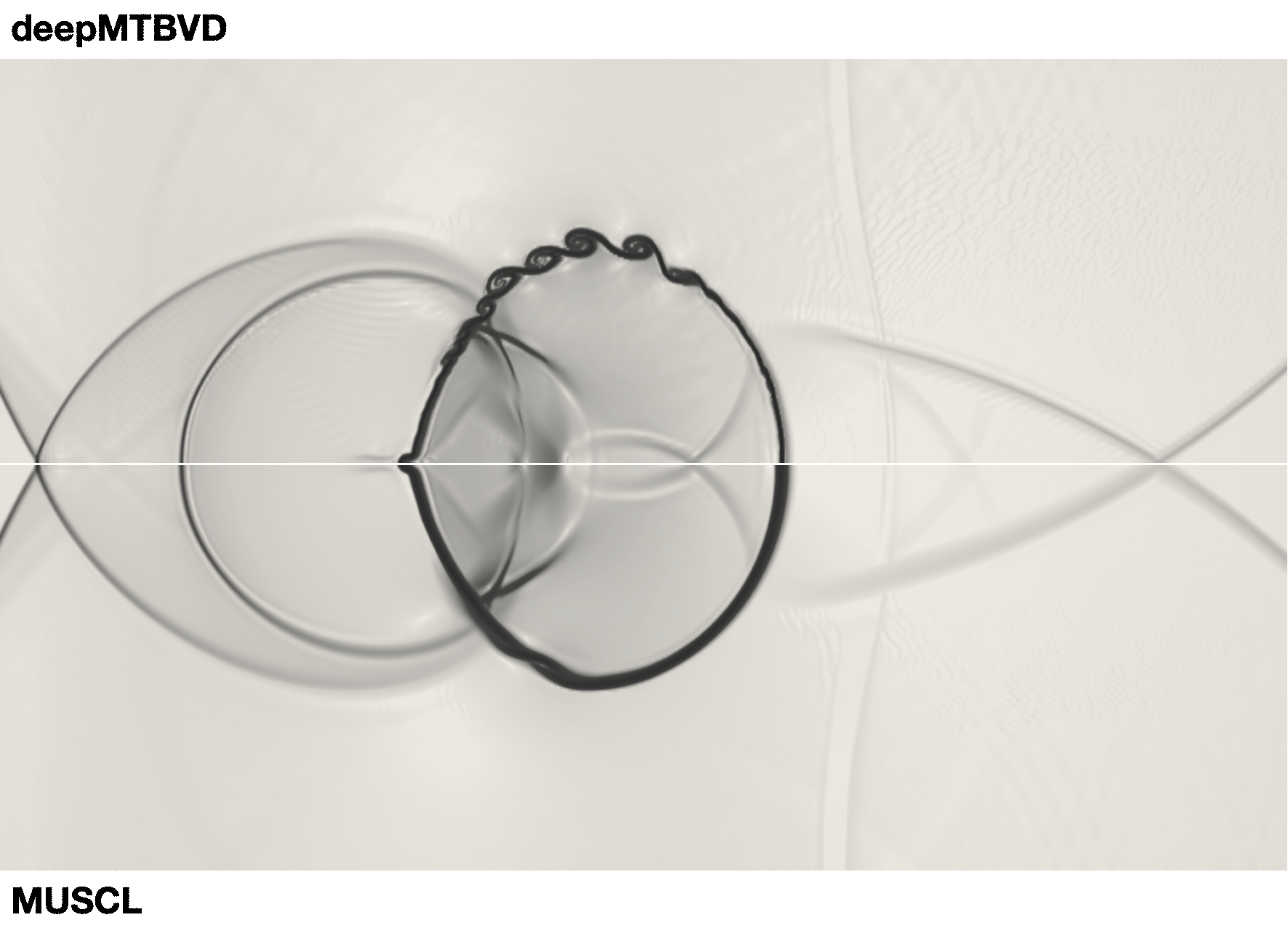}}
\subfloat[$t = 318 \mu s$]{\includegraphics[width=0.49\textwidth]{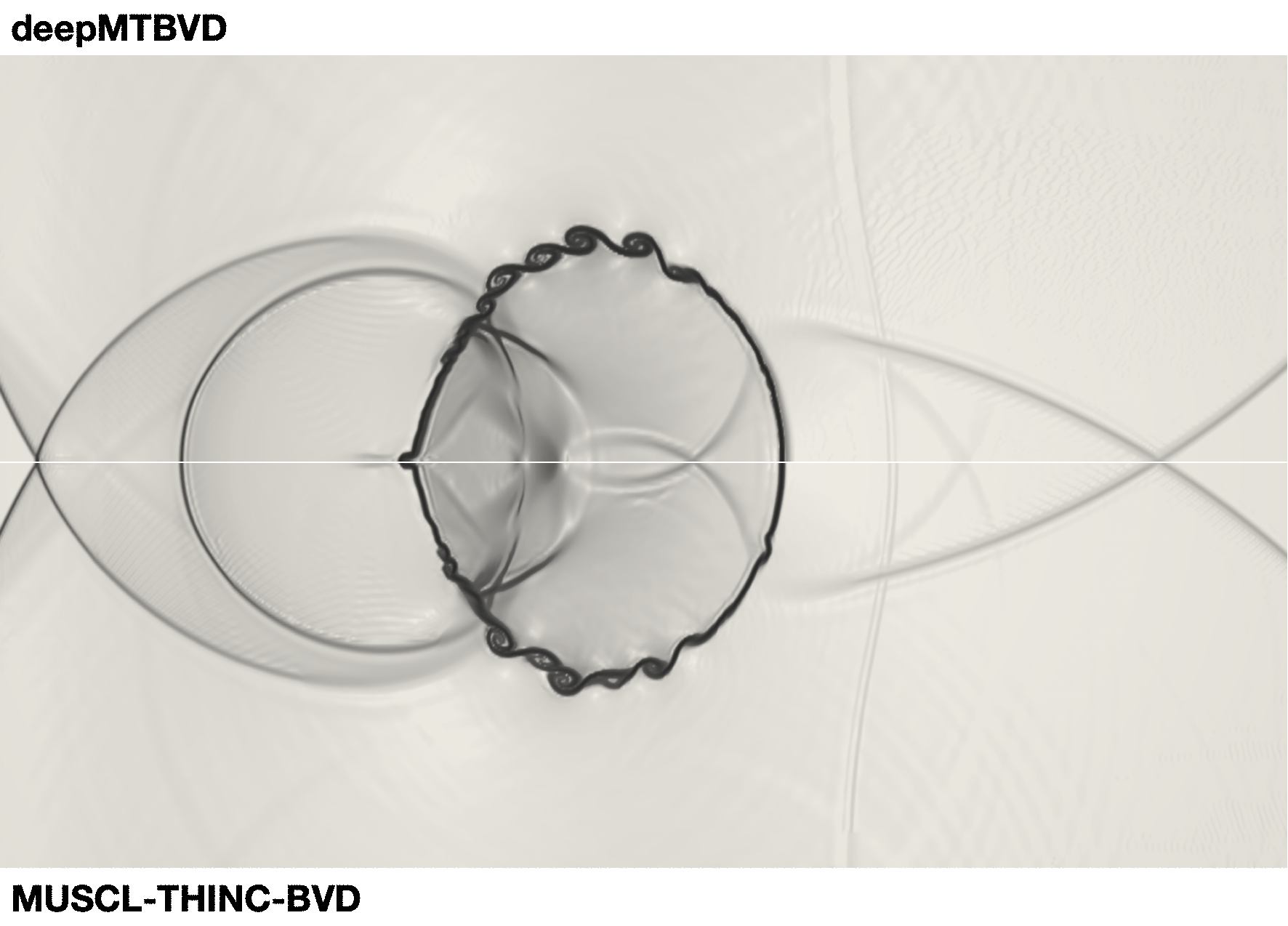}}\\
\caption{Numerical schlieren for the interaction of the air shock and R22 cylinder. The figures on each row represent the same moment. Left column: the results of the deepMTBVD (upper half) and MUSCL scheme (lower half). Right column: the results of the deepMTBVD (upper half) and MUSCL-THINC-BVD scheme (lower half).}
\label{result:example6:density}
\end{figure}

\begin{figure}[htbp!]
\centering
\subfloat[$t = 364 \mu s$]{\includegraphics[width=0.49\textwidth]{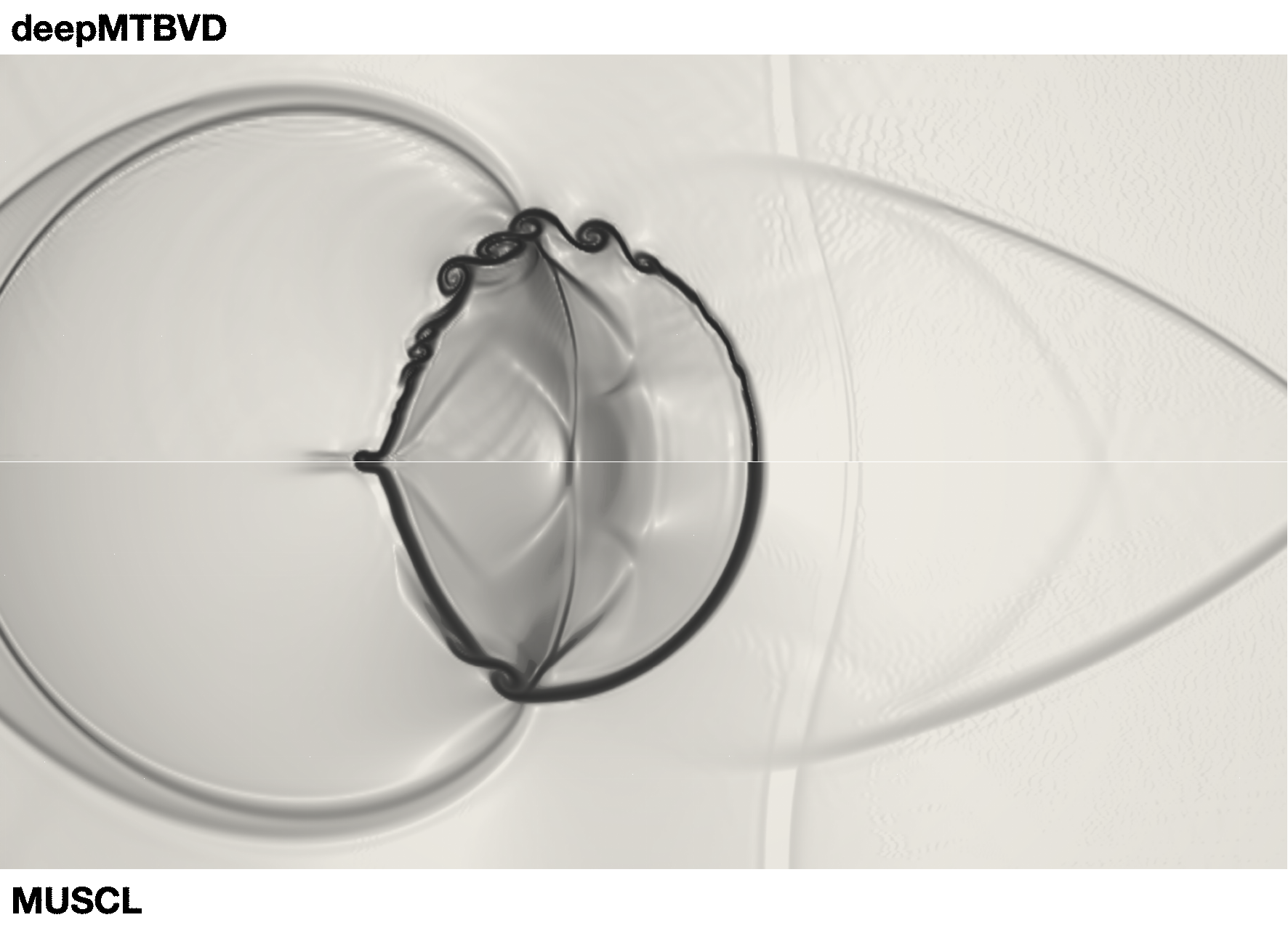}}
\subfloat[$t = 364 \mu s$]{\includegraphics[width=0.49\textwidth]{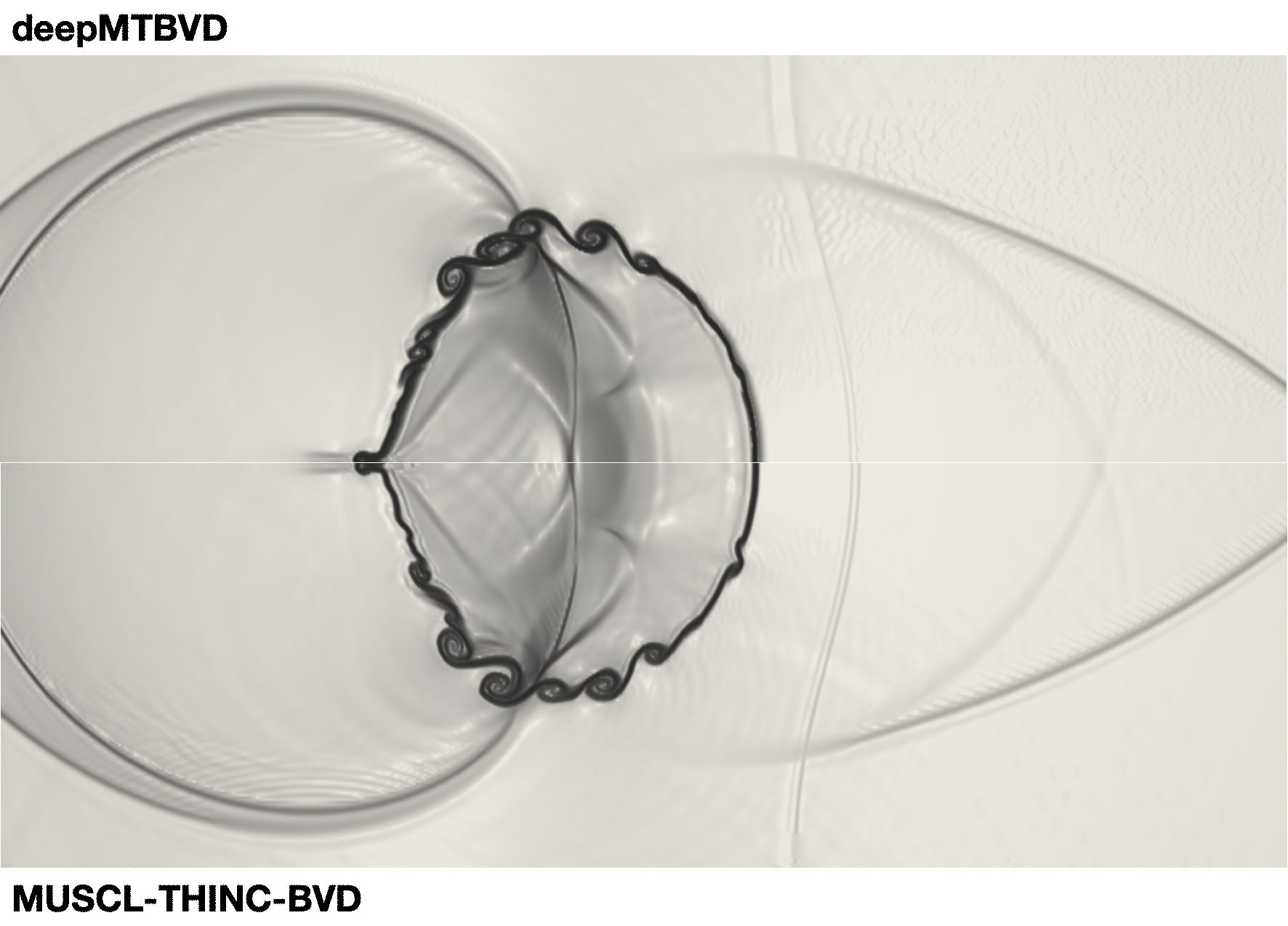}}\\
\subfloat[$t = 417 \mu s$]{\includegraphics[width=0.49\textwidth]{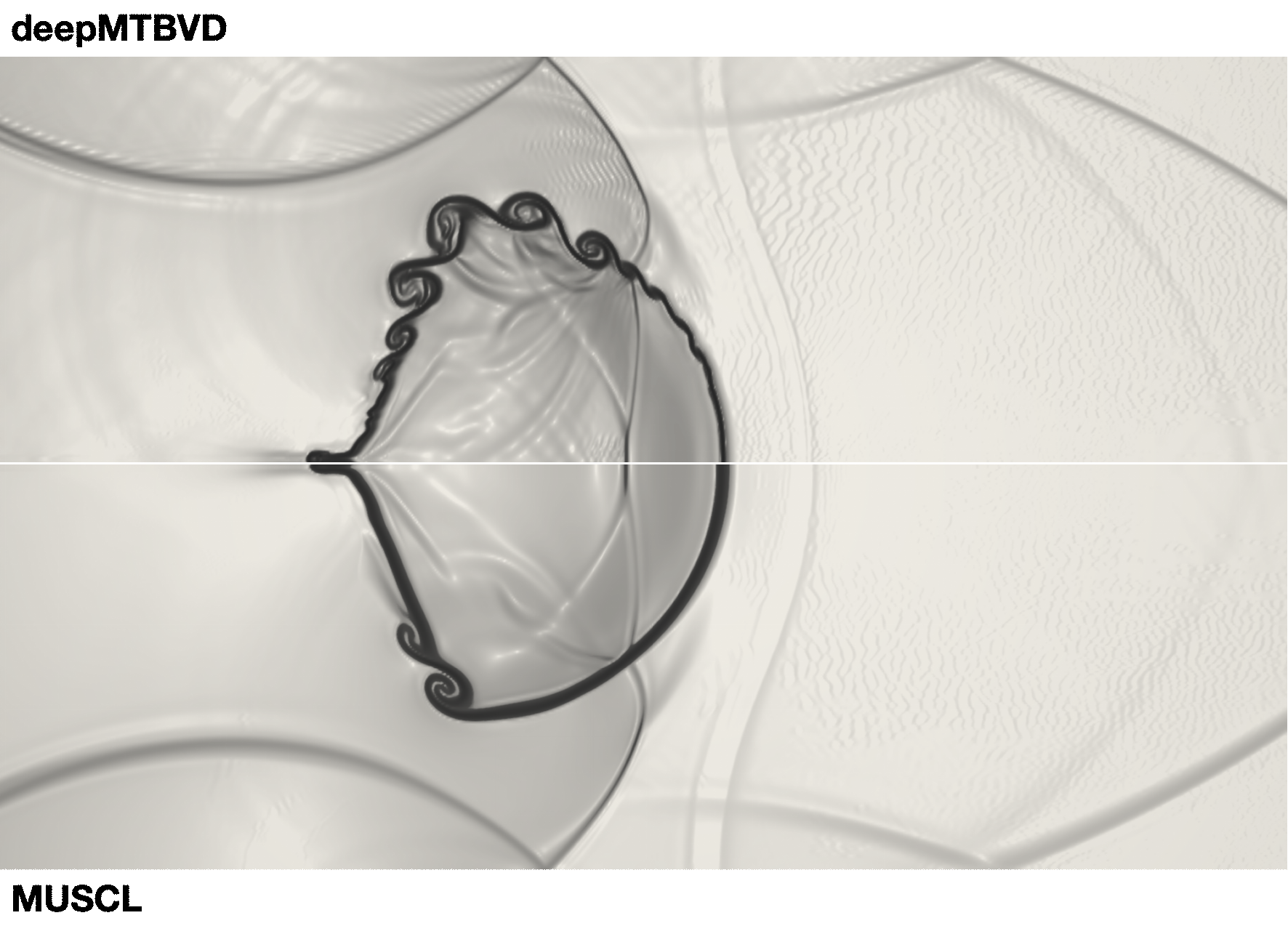}}
\subfloat[$t = 417 \mu s$]{\includegraphics[width=0.49\textwidth]{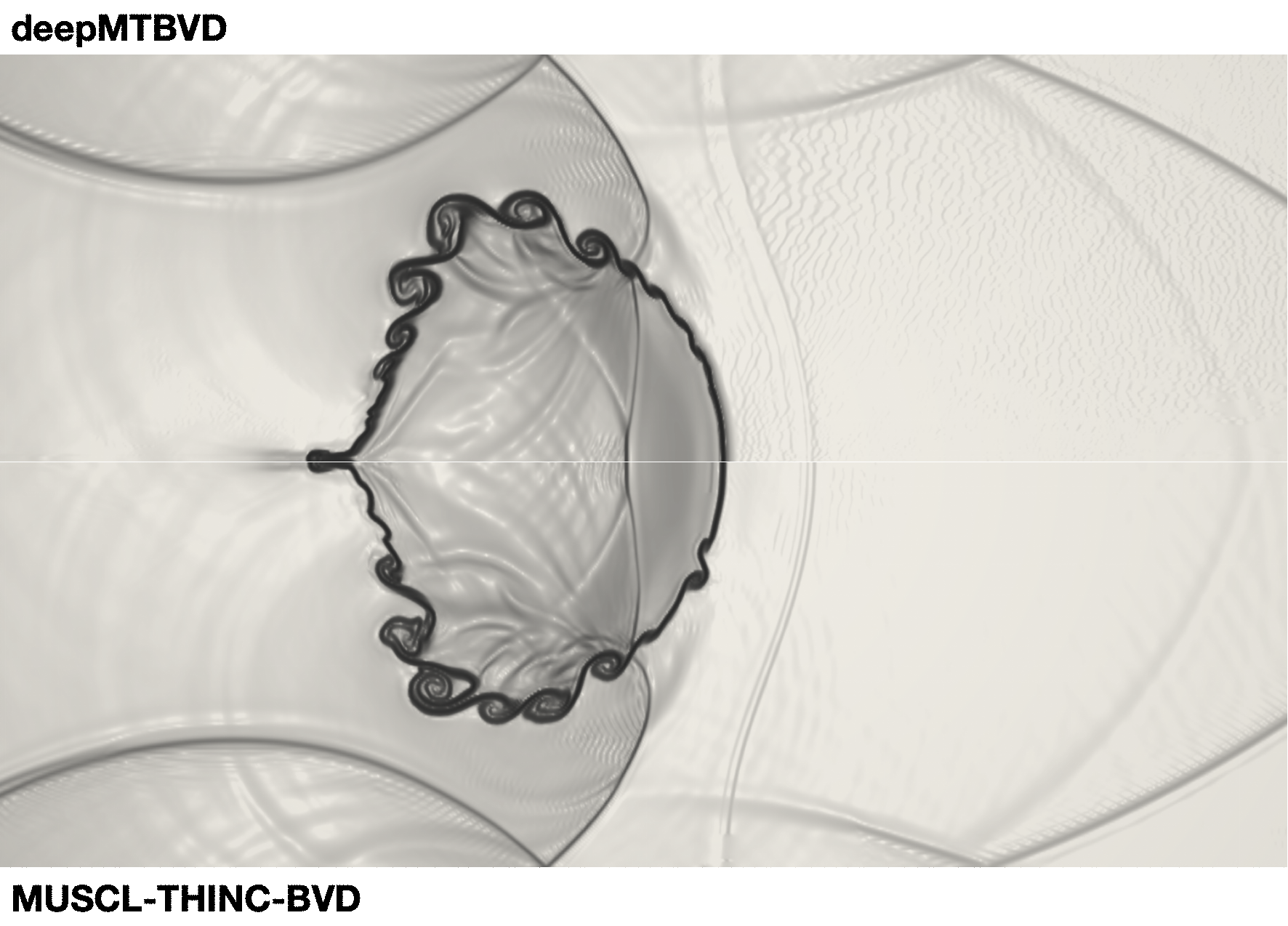}}\\
\subfloat[$t = 463 \mu s$]{\includegraphics[width=0.49\textwidth]{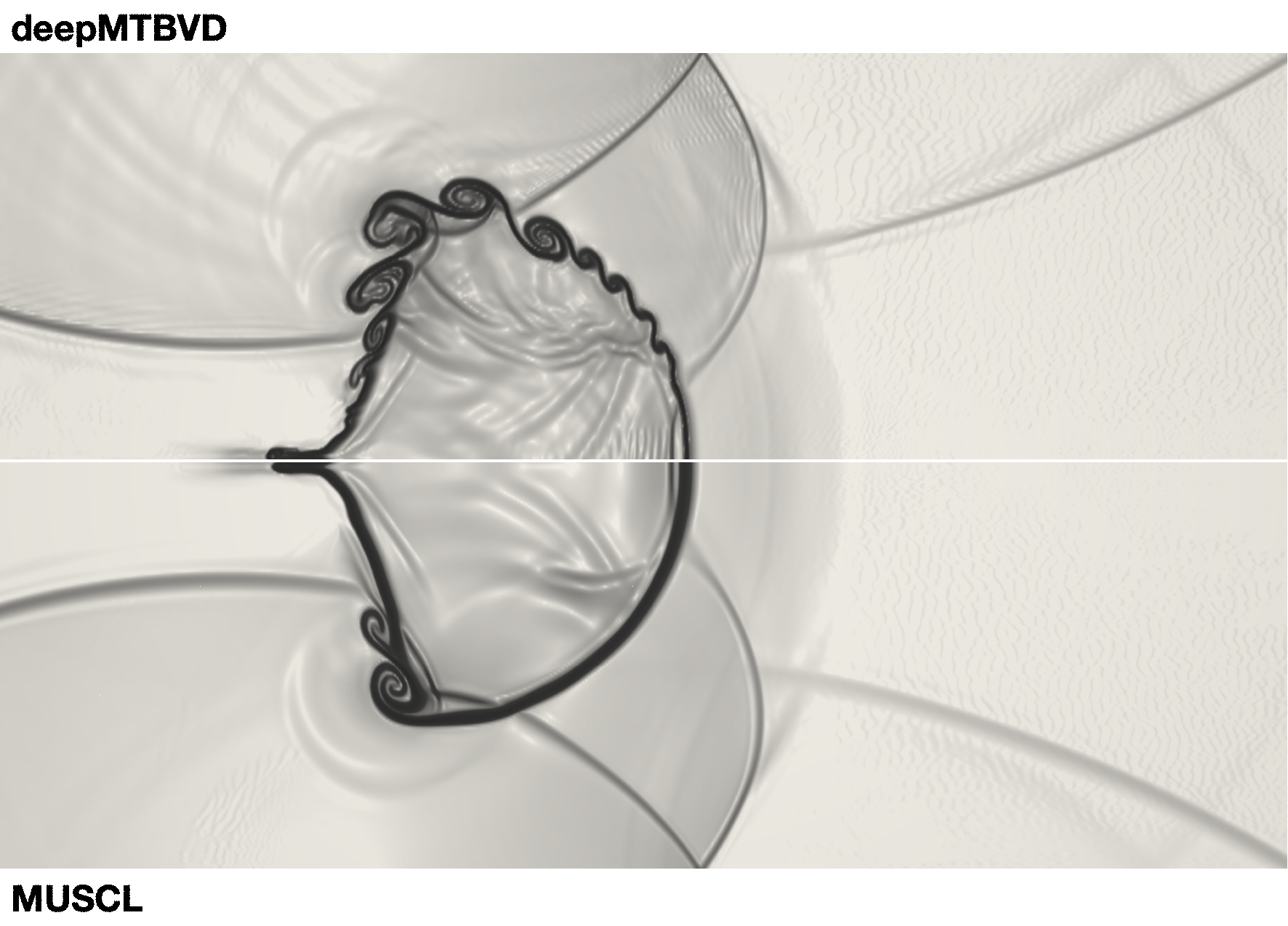}}
\subfloat[$t = 463 \mu s$]{\includegraphics[width=0.49\textwidth]{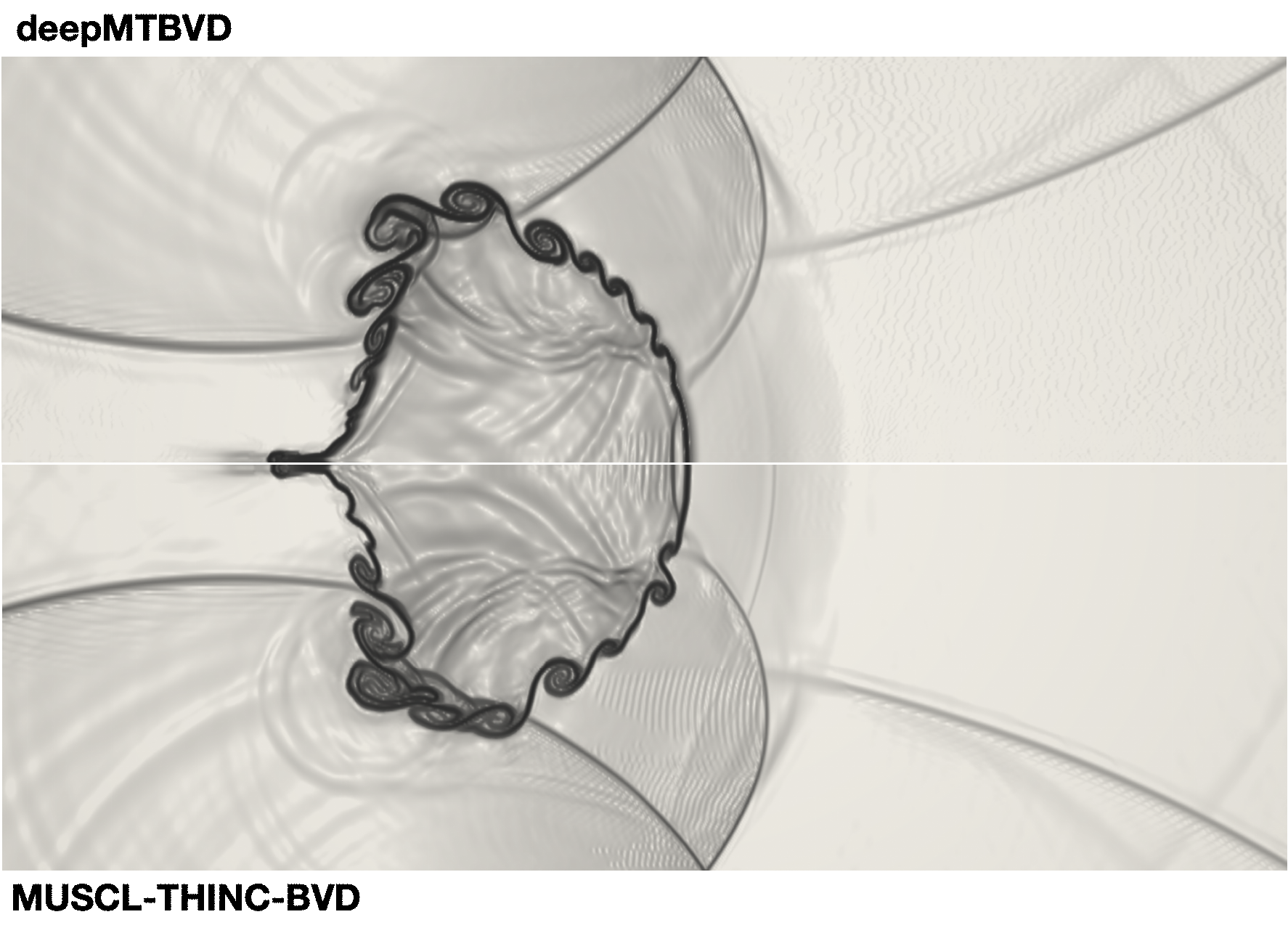}}\\
\caption{Continue of Fig.~$\ref{result:example6:density}$}
\label{result:example6:density:continue}
\end{figure}

\begin{figure}[ht!]
    \centering
    \subfloat[x-axis: MUSCL-THINC-BVD]{\includegraphics[width=0.49\linewidth]{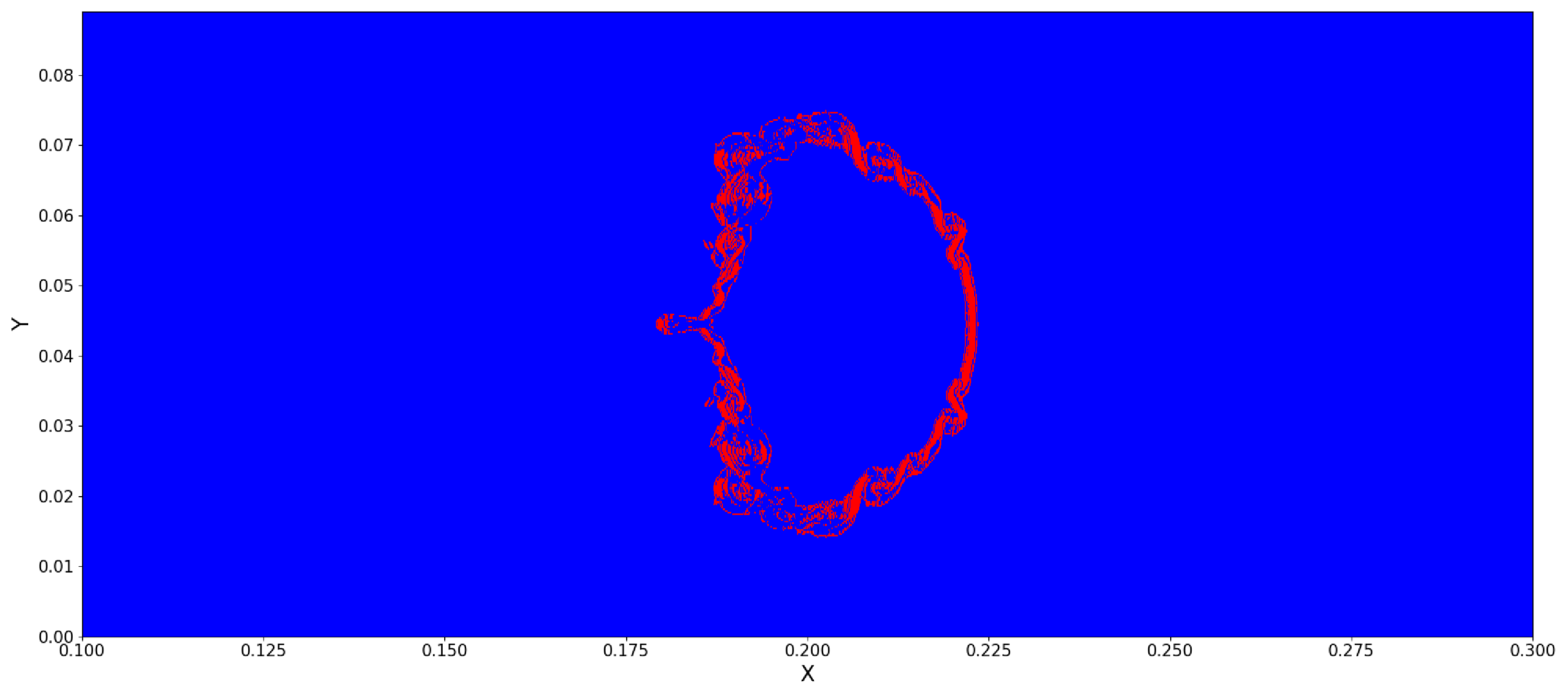}}
    \subfloat[x-axis: deepMTBVD]{\includegraphics[width=0.49\linewidth]{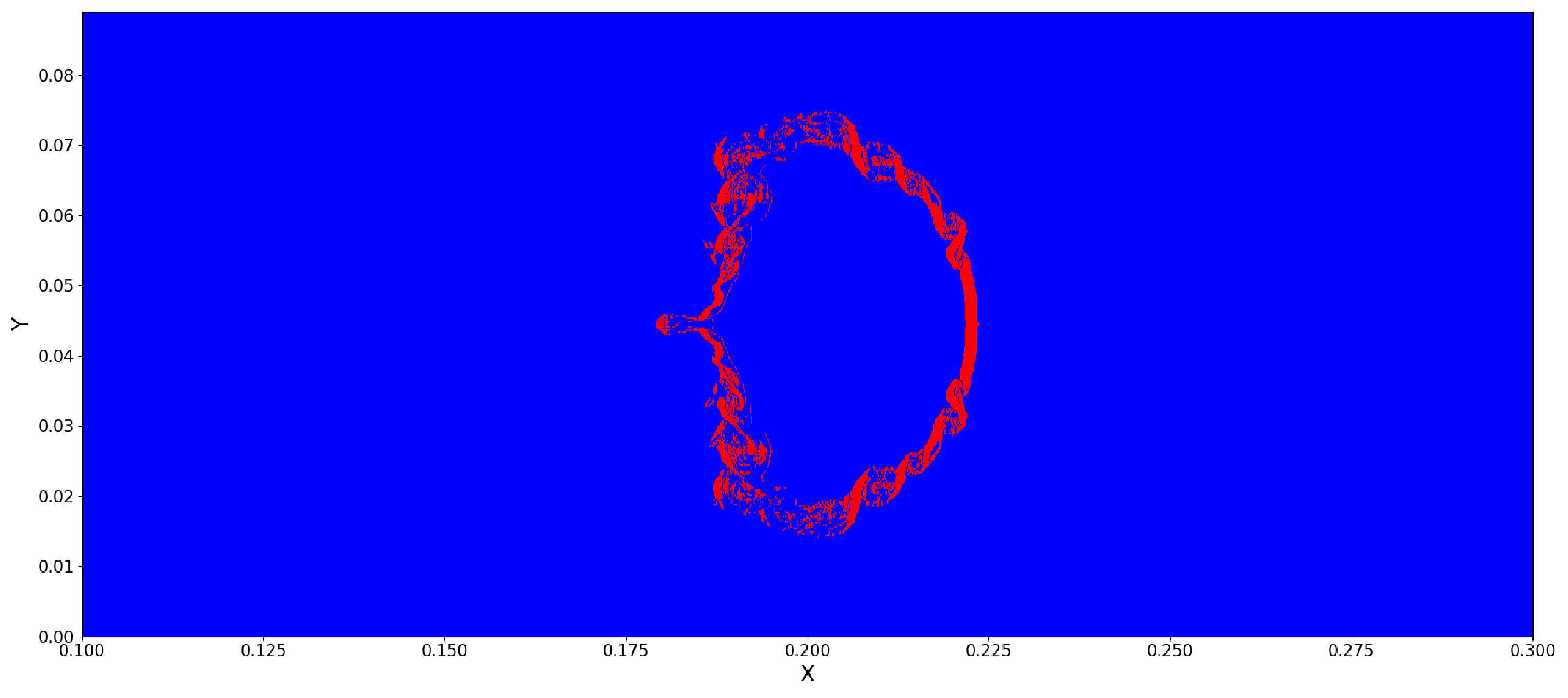}}\\
    \subfloat[y-axis: MUSCL-THINC-BVD]{\includegraphics[width=0.49\linewidth]{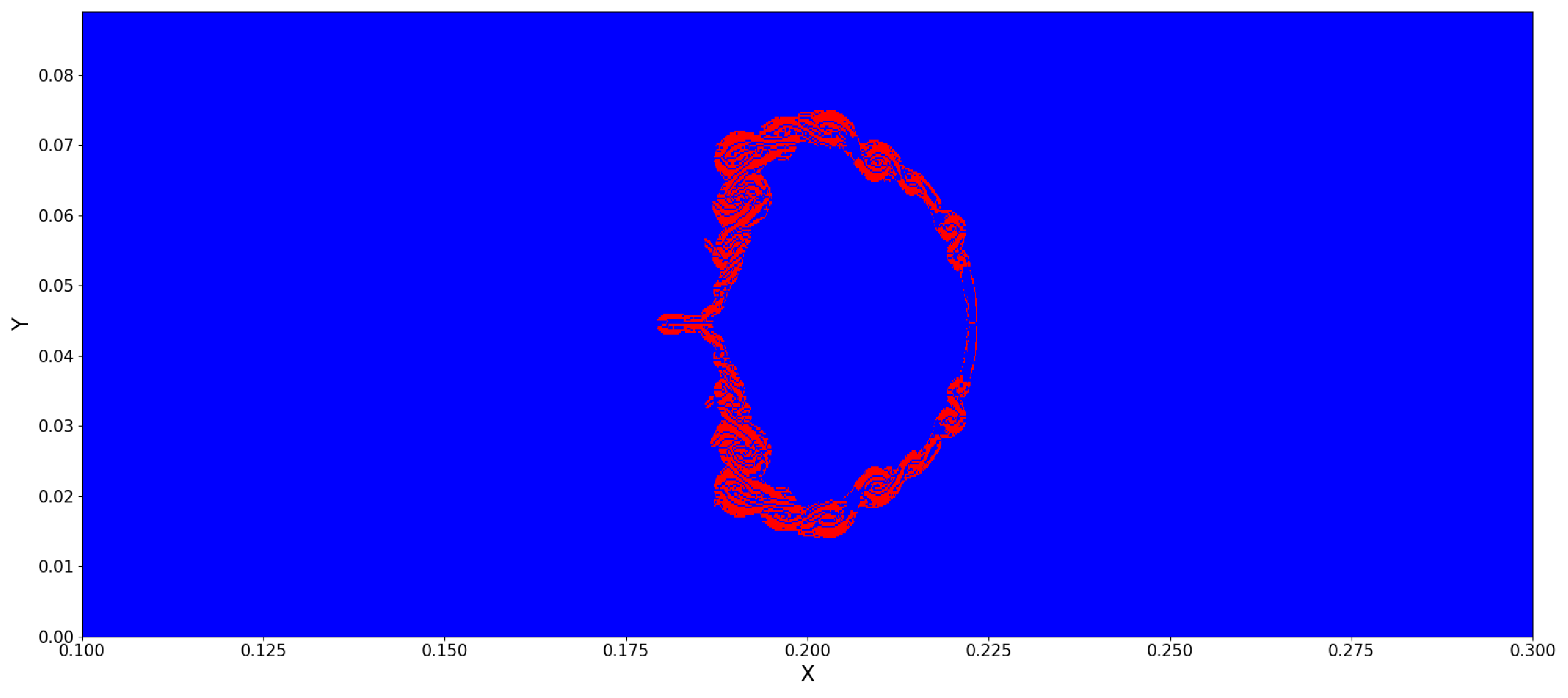}}
    \subfloat[y-axis: deepMTBVD]{\includegraphics[width=0.49\linewidth]{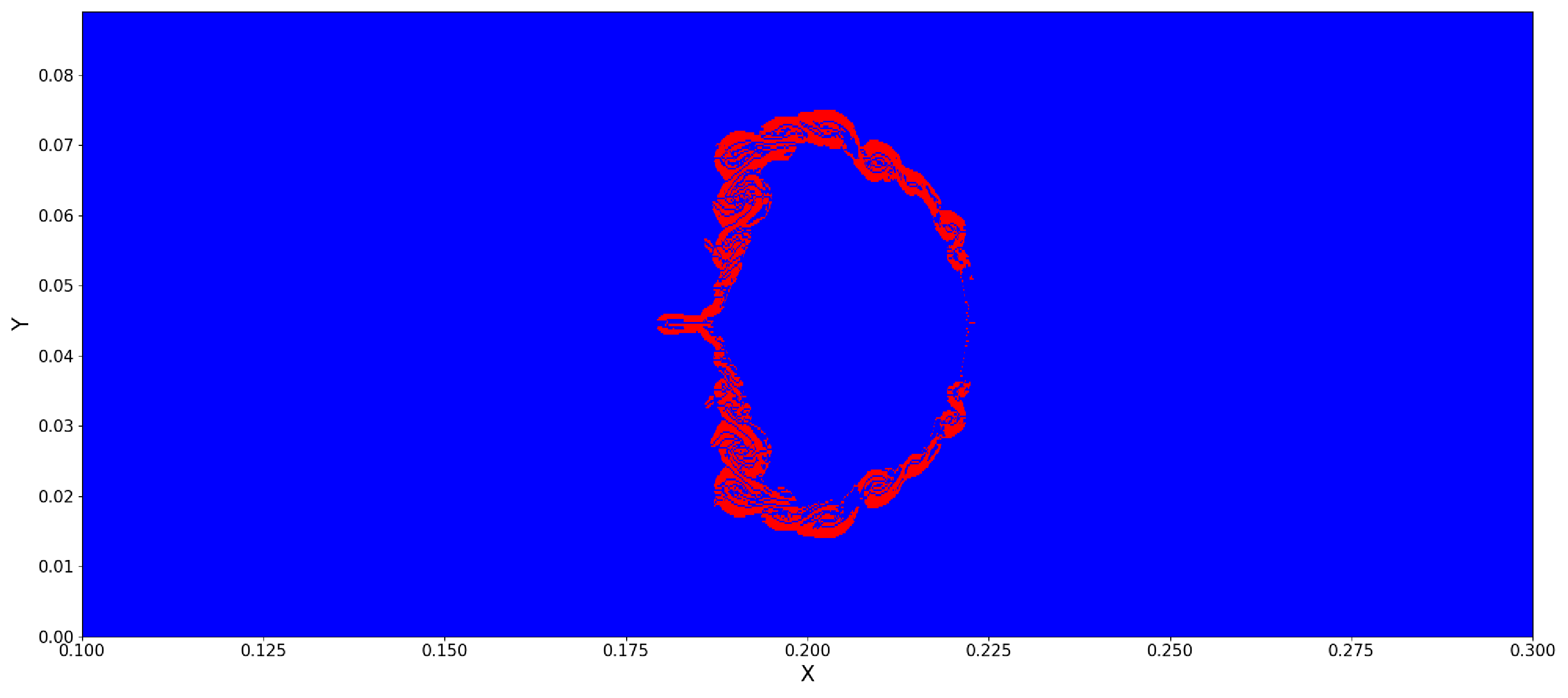}}
    \caption{\hms{The cell where is selected as THINC scheme of volume $\alpha_{1}$ along the x-axis (upper left and upper right) and y-axis (lower left and lower right) through the MUSCL-THINC-BVD scheme and the deepMTBVD scheme. ``Red” denotes the cell reconstructed by the THINC scheme, while the ``blue” represents the MUSCL scheme.}}
    \label{result:example6:celltype}
\end{figure}

Numerical Schlieren figures of the density gradient, $\log(1 + |\nabla \rho|)$, by three types of reconstruction schemes at different time instants are shown in Fig.~\ref{result:example6:density} and Fig.~\ref{result:example6:density:continue}. Each figure has two columns, and each panel of the left column corresponds to deepMTBVD (upper half) and MUSCL (lower half), while the right column corresponds to deepMTBVD (upper half) and MUSCL-THINC-BVD (lower half). 
It is noted that the deepMTBVD scheme resolves the material interface sharply and reproduces fine vortex structures around the R22 bubble interface. Examining the right column of Fig.~\ref{result:example6:density} and Fig.~\ref{result:example6:density:continue} reveals that both schemes involving THINC reconstruction keep the sharpness of the material interface. Although not perfectly matched, the results of deepMTBVD are comparable to that of the MUSCL-THINC-BVD scheme. Both of them resolve flow structures well and capture the reflected and transmitted shock waves. \hms{We present the selected reconstruction scheme for the cells with a quantity of \(\alpha_{1}\) along both the x-axis (upper) and y-axis (lower), utilizing the MUSCL-THINC-BVD scheme and the deepMTBVD scheme, as illustrated in Fig.~\ref{result:example6:celltype}. The figure indicates that the number of cells classified as THINC by the deepMTBVD scheme is lower than that identified by the MTBVD scheme. However, the two schemes reproduce very similar results overall.  From this, we demonstrate that both methods effectively capture discontinuities and subtle vortex structures at the material interface, yielding consistent simulation results.}

\section{Conclusion Remarks}\label{conclusion_remarks}
In this work, we introduce a new paradigm for constructing high-resolution hybrid schemes for compressible flows. This approach generates training data based on BVD schemes for supervised learning. It employs artificial neural networks to create an indicator that pre-selects each cell's most suitable reconstruction scheme, achieving the lowest global numerical dissipation. The numerical schemes under this paradigm are more computationally efficient than similar schemes within the BVD framework, as each cell only requires constructing a single interpolation function. We apply this paradigm to the MUSCL-THINC-BVD scheme, resulting in the new deepMTBVD scheme, and test its performance using multiple classical test cases for compressible single-phase and two-phase flows. The numerical results show that the new deepMTBVD scheme performs as well as or better than the original MUSCL-THINC-BVD scheme for numerical calculations of compressible flows while reducing computational time by up to 40\%. It can capture sharp, discontinuous flow structures such as shock waves and contact discontinuities, as well as intricate flow structures like shear vortices and Kelvin-Helmholtz vortices.
Furthermore, regarding the accuracy of the scheme, the proposed deepMTBVD scheme is second-order while the MUSCL-THINC-BVD scheme is only first-order. Therefore, the deepMTBVD scheme shows higher accuracy with a lower computational time. Under the paradigm used to construct the deepMTBVD scheme, a series of low-dissipation numerical schemes can also be developed, which we intend to explore further in future research.
\section*{Acknowledgments} 
This work is supported by the Fundamental Research Funds for the Central Universities. It is also partially supported by the National Key R$\&$D Program of China (Project No. 2020YFA0712000), the Shanghai Science and Technology Innovation Action Plan in Basic Research Area (Project No. 22JC1401700), the Strategic Priority Research Program of Chinese Academy of Sciences (Grant No. XDA25010405) and the National Natural Science Foundation of China (Grant No. DMS-11771290).  Lidong Cheng's work is supported by the Postdoctoral Fellowship Program of CPSF (Grant No. GZC20231592). Minsheng Huang would give special thanks to Prof. Zheng Ma at Shanghai Jiao Tong University for inspirational discussions.
\vspace{.3cm}

\noindent \textbf{Appendix A. The proof of the Conclusion.~\ref{conclusion:1}}\label{appendix:A}
\setcounter{conclusion}{0}
\begin{conclusion}
    \rewb{Assume that $\bar{u}_{i-1}, \bar{u}_{i}, \bar{u}_{i+1}$ are strictly monotone increasing (decreasing), 
    then the reconstruction function $\hat{u}^{s}(x)$ in cell $I_{i}$ satisfies
    $$
        \frac{\hat{u}^{s}(x) - \bar{u}_{\min}}{\bar{u}_{\max}-\bar{u}_{\min}} \in (0, 1), \quad s \in \{M, T\}, \quad x \in [x_{i-1/2}, x_{i+1/2}],
    $$
    where $\bar{u}_{\min} = \min\left\{\Bar{u}_{i-1}, \Bar{u}_{i+1}\right\}, \bar{u}_{\max}=\max\left\{\Bar{u}_{i-1}, \Bar{u}_{i+1}\right\}$ are the same definition in \eqref{numerical:thinc}.}
\end{conclusion}

\begin{proof}
    \rewb{Assume the $\bar{u}_{i-1}, \bar{u}_{i}, \bar{u}_{i+1}$ are monotone increasing, which follows that $\bar{u}_{i-1} < \bar{u}_{i} < \bar{u}_{i+1}$. According to the TVD property in \cite{torobook}, we get the MUSCL scheme satisfies
    $$
    \begin{aligned}
        \min\{\bar{u}_{i-1}, \bar{u}_{i}\} \leq u^{MUSCL,R}_{i-1/2} \leq \max\{\bar{u}_{i-1}, \bar{u}_{i}\}, \\
        \min\{\bar{u}_{i+1}, \bar{u}_{i}\} \leq u^{MUSCL,L}_{i+1/2} \leq \max\{\bar{u}_{i+1}, \bar{u}_{i}\}.
    \end{aligned}
    $$
    Since MUSCL is a linear reconstruction scheme, the $u^{MUSCL}(x)$ attains the extreme value on the cell boundary. $u^{MUSCL}(x)$ satisfies the condition Eq. \eqref{conclusion:normal} due to $u^{MUSCL} \in (\bar{u}_{\min}, \bar{u}_{\max}).$\\
    As for the THINC scheme, we straightly get the same result by reformulating \eqref{numerical:thinc} as follows
    $$
    \frac{u^{THINC} - \bar{u}_{\min}}{\bar{u}_{\max} -\bar{u}_{\min}} =  \frac{1}{2}(1+\theta \tanh(\beta \frac{x-x_{i-1/2}}{x_{i+1/2} - x_{i-1/2}} - \Tilde{x}_{i}),
    $$
    in which the right-hand side lies in $(0,1)$. The decreasing case is similar. }
\end{proof}

\noindent \textbf{Appendix B. The proof of the Conclusion.~\ref{conclusion:2}}\label{appendix:B}
\begin{conclusion}
    \rewb{Assume the individual cell average values in stencil $S_{i}$ are not identical, then the target $I_{i}$ chooses the same scheme as the MUSCL-THINC-BVD scheme under the normalization as follows
    \begin{equation*}
        \Tilde{u}_{i} = \frac{\bar{u}_{i}-\Tilde{u}_{\min}}{\Tilde{u}_{\max}-\Tilde{u}_{\min}},
    \end{equation*}
    where $\Tilde{u}_{\max} = \max\left\{\bar{u}_{i-2}, \bar{u}_{i-1}, \bar{u}_{i}, \bar{u}_{i+1}, \bar{u}_{i+1} \right\}$, $\Tilde{u}_{\min} = \min\left\{\bar{u}_{i-2}, \bar{u}_{i-1}, \bar{u}_{i}, \bar{u}_{i+1}, \bar{u}_{i+1} \right\}.$}
\end{conclusion}

\begin{proof}
    \rewb{Since the individual cell average values in stencil $S_{i}$ are not identical, it follows that $\Tilde{u}_{\max} > \Tilde{u}_{\min}$, avoiding the zero denominators. Then we define the normalized boundary variation function $\widetilde{BV}(s_{1},s,s_{2})$ as 
    \begin{equation}
        \widetilde{BV}(s_{1}, s, s_{2}) := \left| \Tilde{u}^{s_{1}}_{i-1/2,L} - \Tilde{u}^{s}_{i-1/2,R}\right| + \left|\Tilde{u}^{s}_{i+1/2,L} - \Tilde{u}^{s_{2}}_{i+1/2,R}\right|, s_{1}, s, s_{2} \in \left\{M, T\right\}. \label{appendixb:nbv}
    \end{equation}
    Here, we adapt the transformation \eqref{conclusion2:normalize} into each term in the $BV(s_{1}, s, s_{2})$. $M$ stands for the MUSCL scheme and $T$ denotes the THINC scheme. It should be noted that 
    $$
    \widetilde{BV}(s_{1}, s, s_{2}) = \dfrac{BV(s_{1}, s, s_{2})}{\Tilde{u}_{\max}-\Tilde{u}_{\min}}.
    $$
    Without loss of generality, we consider that the $I_{i}$ selects the THINC scheme. Thus, we have 
    \begin{equation*}
        TBV_{i, \min}^{T} < TBV_{i,\min}^{M}.
    \end{equation*}
    According to the definition of $TBV_{i, \min}^{s}$ in \eqref{bvd:tbv}, we have 
    \begin{equation}
        \begin{aligned}
            \widetilde{TBV}_{i,\min}^{T} & = \min\left\{\widetilde{BV}(M, T, M), \widetilde{BV}(T, T, T), \widetilde{BV}(M, T, T), \widetilde{BV}(T, T, M)\right\} \\ 
            & = \frac{1}{\Tilde{u}_{\max}-\Tilde{u}_{\min}} \min\left\{BV(M, M, M), BV(T, M, T), BV(M, M, T), BV(T, M, M)\right\} \\
            & = \frac{TBV_{i, \min}^{M}}{\Tilde{u}_{\max}-\Tilde{u}_{\min}} < \frac{TBV_{i, \min}^{T}}{\Tilde{u}_{\max}-\Tilde{u}_{\min}} = \widetilde{TBV}_{i, \min}^{M},
        \end{aligned}
    \end{equation}
    which completes the proof. The case that cell $I_{i}$ selects the MUSCL scheme is similar.}
\end{proof}

\noindent \textbf{Appendix C. The selection of the weights in the loss function.}\label{appendix:C}
\rewb{
\begin{table}[ht!]
\centering
\begin{tabular}{|c|c|c|c|c|c|c|c|}
\hline
$\omega_{0}:\omega_{1}$  &1:2 &1:5 &1:10 &1:20 &1:50 &1:100 & BCE \\ 
\hline
Accuracy & 0.968 &0.975	&0.963 &0.948 &0.923 &0.901	&0.981\\
\hline
FPR & 0.0118 &0.0046 &0.0044 &0.0038 &0.0024 &0.0024 &0.013\\
\hline
\end{tabular}
\caption{The different weights ratios used in the loss function \eqref{numerical:focal:loss}. }
\label{table:loss:compare}
\end{table}
}
 
\rewb{Refer to \cite{lin2018focal}, $\eta = 2$ gets the best performance, which is chosen as the default parameter in this paper. Here, we only compare various weight ratios under the same data and number of training epochs for the loss function \eqref{numerical:focal:loss} and binary cross entropy loss (BCE), as summarized in Table~\ref{table:loss:compare}. The data presented in the table above are the average of several experimental results. It is essential for $\omega_{0}$ to be smaller than $\omega_{1}$ because misclassifying the THINC scheme as the MUSCL scheme is acceptable, whereas reverse misclassification should be minimized. Therefore, we selected specific cases to examine the influence of different weight ratios. In Table~\ref{table:loss:compare}, ``Accuracy" represents the proportion of correctly classified instances. ``FPR" (False Positive Rate) refers to the proportion of the MUSCL scheme misclassified as the THINC scheme, where a smaller value is preferred and is of great importance in comparison. As shown in the Table.~\ref{table:loss:compare}, we find that the weighted loss function in \eqref{numerical:focal:loss} presents lower FPR than the ``BCE" loss function. Furthermore, the weight ratio of $\omega_{0}:\omega_{1} = 1:5$ achieves the most acceptable FPR with the highest accuracy when compared with other ratios. }

\noindent \textbf{Appendix D. The state space for the Riemann data.}\label{appendix:D}
\begin{table}[ht!]
    \centering
    \begin{tabular}{|c|c|c|c|c|c|c|}   
       \hline
       \text{Case}  & $\rho_{L}$ & $u_{L}$ & $p_{L}$ & $\rho_{R}$ & $u_{R}$ & $p_{R}$\\ 
        \hline 
        1 & 1.0 & u & 1.0& 0.125 & 0.0 & 1.0 \\ 
        \hline 
        2 & 1.0 & u & 1000.0 & 1.0 & 0.0 & 0.01 \\
        \hline
        3 & 3.857143 & 2.629369 & 10.333333 & 1 + 0.2 * sin(50x - 25) & 0.0 & 1.0 \\ 
        \hline 
        4 & 1.0 & u & 1000.0 & 1.0 & 0.0 & 100 \\ 
        \hline 
        5 & 1.0 & u & 100 & 1.0 & 0.0 & 0.01 \\
        \hline 
        6 & 0.445 & 0.698 & 3.528 & 0.5 & 0.0 & 0.571 \\
        \hline 
    \end{tabular}
    \caption{The Riemann state defined for generating the data.}
    \label{table:reimann_state}
\end{table}

\rewb{Table~\ref{table:reimann_state} presents the initial conditions of the Riemann problem used for data generation. In this context, cases 1, 2, 3, 4, and 5 will produce rarefaction waves, contact discontinuities, and shock waves as time progresses. Specifically, case 3 is designed for the scheme to learn a smooth yet fluctuating profile, while case 6 is designed for capturing the discontinuity and the shock waves. It is important to note that in cases 1, 2, 4, 5, we have the flexibility to choose random values for $u_{L}$ without affecting the resulting wave types. For practical purposes, we choose the fixed velocity $u=0$, which already generates a high-quality data set. To prevent the data from being pre-trained in actual numerical examples, we utilize only ten time steps (30 sub-time steps) of the SSP-RK3 method to curate the generated data. We vary grid sizes and maintain a small Courant number of 0.1. Since all data will be normalized to the range $[0,1]$, we focus on the magnitudes of the values, opting for different scales of physical quantities. In the MUSCL-THINC-BVD scheme, all physical variables are integrated into the same algorithm \eqref{bvd:mtbvd}. Similarly, we compile all physical variables as part of the data collection to create a universal surrogate model. 
}

\begin{figure}[htb!]
    \centering
    \includegraphics[width=\linewidth]{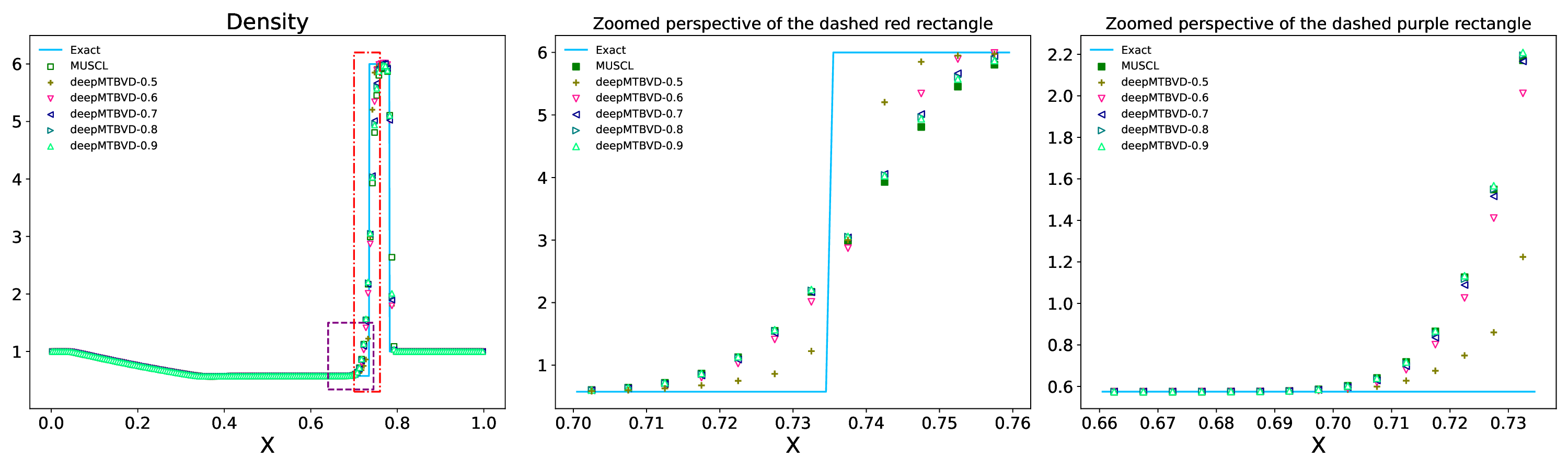}
    \caption{Numerical results of increasing the $\kappa_{\text{ref}}$ in Example.~\ref{strong_lax}. Left: the density profile computed from different \kaparef ranging from 0.5 to 0.9. Middle: the zoomed perspective of the dashed red rectangle. Right: the zoomed perspective of the dashed purple rectangle.}
    \label{result:example2:increasing:kappa}
\end{figure}
\begin{figure}[htb!]
    \centering
    \includegraphics[width=\linewidth]{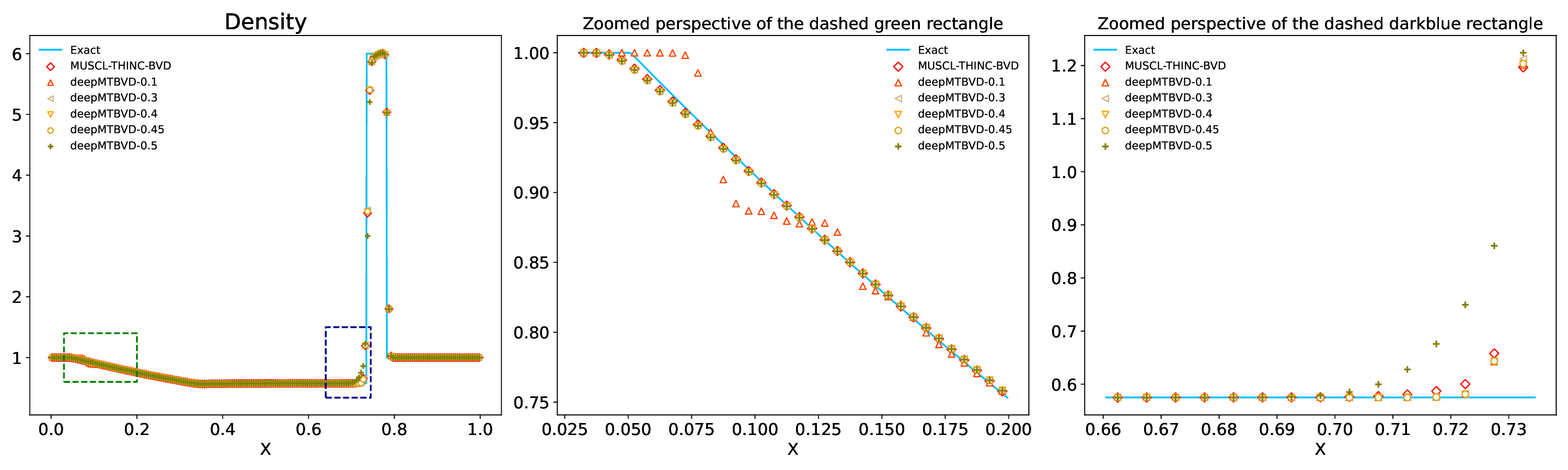}
    \caption{\hms{Numerical results of decreasing the \kaparef in Example.~\ref{strong_lax}. Left: the density profile computed from different \kaparef ranging from 0.1 to 0.5. Middle: the zoomed perspective of the dashed green rectangle. Right: the zoomed perspective of the dashed dark blue rectangle.}}
    \label{result:example2:decreasing:kappa}
\end{figure}

\noindent \textbf{Appendix E. The influence of the \kaparef.}\label{appendix:E}

\hms{To further emphasize the importance of \kaparef, we vary the \kaparef values from 0.1 to 0.9 in Example~\ref{strong_lax}. We categorize the results into two groups based on the values of \kaparef: Fig.~\ref{result:example2:increasing:kappa} shows results for \kaparef \(\geq 0.5\), while Fig.~\ref{result:example2:decreasing:kappa} shows results for \kaparef \(\leq 0.5\).}

\hms{In Fig.~\ref{result:example2:increasing:kappa}, the numerical solution shows a trend of increased diffusion as the value of \kaparef increases. As seen in the zoomed-in view of the dashed red rectangle, smaller values of \kaparef resolve the contact discontinuity with fewer mesh cells. Larger values of \kaparef tend to approximate the MUSCL scheme, as shown in the zoomed-in view of the dashed purple rectangle. This indicates that more cells are classified under the MUSCL scheme, and the numerical results approach those of the MUSCL scheme as \kaparef nears 1.0. When \kaparef = 1.0, all cells are classified according to the MUSCL scheme, thus simplifying the deepMTBVD scheme to the MUSCL scheme.}

\hms{For values of \kaparef \(\leq 0.5\), as shown in Fig.~\ref{result:example2:decreasing:kappa}, the numerical results exhibit the low-dissipative behavior characteristic of the MUSCL-THINC-BVD scheme. From the zoomed-in view of the dashed dark blue rectangle, we observe that as \kaparef approaches zero, the diffusive error decreases somewhat as more THINC cells are used compared to larger \kaparef values. However, when \kaparef falls below a certain threshold (empirically around 0.1), the deepMTBVD scheme starts to produce unsatisfactory results, introducing discontinuities in the rarefaction wave, as shown in the zoomed-in view of the dashed green rectangle. This occurs because smaller \kaparef values result in a higher number of cells classified as THINC, leading to more anti-dissipative error.}

\hms{It is important to note that when \kaparef is between 0.1 and 0.45, the numerical solution does not show significant improvement. In comparison, when the \kaparef is larger than 0.45, more MUSCL cells are introduced, which leads to more numerical dissipation. Additionally, the computational cost of the THINC scheme is higher than that of the MUSCL scheme. Therefore, we prefer to select configurations where performance is satisfactory and \kaparef is as large as possible to optimize efficiency.}
\bibliographystyle{unsrt}
\bibliography{ref}

\
\end{document}